\documentclass[12pt]{article}
\usepackage[margin=2 cm]{geometry}
\usepackage{comment}
\usepackage{graphicx}
\usepackage[font={small,it}]{caption}
\usepackage{mwe}
\usepackage[nottoc]{tocbibind}
\usepackage{amsmath,amssymb,extarrows,mathtools,graphicx,subfigure,setspace}
\usepackage{cite}
\usepackage{tikz}\usetikzlibrary{calc}
\usepackage{braket}
\usepackage{epsfig}
\usepackage[section]{placeins}
\usepackage{slashed}
\usepackage{color}
\usepackage{caption}
\usepackage{amsmath}
\usepackage{hyperref}
\makeatother

\newcommand{\be}{\begin{equation}}
\newcommand{\bea}{\begin{eqnarray}}
\newcommand{\eea}{\end{eqnarray}}
\newcommand{\ba}{\begin{array}}
\newcommand{\ea}{\end{array}}
\newcommand{\ee}{\end{equation}}
\newcommand{\bes}{\begin{equation*}}
\newcommand{\beas}{\begin{eqnarray*}}
\newcommand{\eeas}{\end{eqnarray*}}
\newcommand{\bas}{\begin{array*}}
\newcommand{\eas}{\end{array*}}
\newcommand{\ees}{\end{equation*}}

\numberwithin{equation}{section}

\begin{document}

\onehalfspacing
\vfill
\begin{titlepage}
\vspace{10mm}

\begin{center}

\vspace*{10mm}
\vspace*{1mm}
{\Large  \textbf{Correlations of mixed systems in confining backgrounds}} 
 \vspace*{1cm}
 
{$\text{Mahdis Ghodrati}^{a}$},

\vspace*{8mm}
{ \textsl{
$^a $Asia Pacific Center for Theoretical Physics, Pohang 37673, Republic of Korea}} 
 \vspace*{0.4cm}

\textsl{e-mails: {\href{mahdis.ghodrati@apctp.org}{mahdis.ghodrati@apctp.org}}}
 \vspace*{2mm}

\vspace*{1.7cm}

\end{center}

\begin{abstract}
We show that the entanglement of purification and the critical distance between the two mixed systems is a powerful measure in probing the phase structures of QCD and confining backgrounds, as it can distinguish the scale of chiral symmetry breaking versus the scale of confinement/deconfinement phase transitions. For two symmetric strips with equal and finite width and infinite length, and in the background of several confining geometries, we numerically calculate the critical distance between them where the mutual information vanishes and show that this quantity can probe the very rich phase structures of these backgrounds. The geometries that we study here are AdS-soliton, Witten Sakai Sugimoto and deformed Sakai-Sugimoto, Witten-QCD, Klebanov-Strassler, Klebanov-Tseytlin, Klebanov-Witten, Maldacena-Nunez, Nunez-Legramandi metric, and Domain-Wall QCD model. For each background we also present the relation for the entanglement of purification. Finally, we show that the Crofton forms of these geometries also behave in a universal form where a `` well'' is being observed around the IR wall, and therefore for all confining backgrounds, the Crofton form would also be capable of distinguishing the confining versus conformal backgrounds as it is also a tool in the reconstruction of various bulk geometries.
 \end{abstract}

\end{titlepage}

\tableofcontents


\section{Introduction}

From the advent of Ryu-Takayanagi (RT) prescription \cite{Ryu:2006bv} which for calculating the entanglement entropy of field theories uses calculations of the area of extremal surface which is homologous to the boundary, the deep connections between quantum information and geometry specially in the context of AdS/CFT has been ground in. The other new dualities in this context such as computational complexity/volume of the subregions \cite{Alishahiha:2015rta,Brown:2015bva,Ghodrati:2017roz,Ghodrati:2018hss,Ghodrati:2019bzz,Zhou:2019jlh}, and mixed correlation measures/minimal wedge cross section \cite{Takayanagi:2017knl,Ghodrati:2019hnn,Ghodrati:2020vzm} further strengthened this connection. 

However, it worths to note that for the dual of the entanglement wedge cross section (EWCS) in the bulk, recently, several mixed correlation measures have been proposed in the literature. These are measures of varied combinations of quantum and classical correlations. They include entanglement of purification (EoP)\cite{Takayanagi:2017knl},  logarithmic negativity \cite{Kudler-Flam:2018qjo}, odd entropy \cite{Tamaoka:2018ned}, entanglement distillation \cite{Agon:2018lwq}, and reflected entropy \cite{Dutta:2019gen}. It is not still clear which of these measures exactly would be dual to the minimal wedge cross section, or if in the large $N$ limit, all or subset of them could work as the dual quantity, however, here we only calculate the critical distance $D_c$ for two strips, the mutual information and the minimal wedge cross section in the confining geometries without considering any specific dual quantity for EWCS.

As the calculations of the properties of geometrical objects in the bulk are much easier than the calculations of quantum measures in the boundary field theories and some of these geometrical objects can deeply probe the bulk and gather non-local information of it which would be dual to interesting quantum measures of the boundary, one would expect that these geometrical objects can also detect the rich phase structures of the exotic forms of matters such as quark gluon plasmas, strange metals, high temperature superconductors, superfluids, etc.   The motivation of this work is to use minimal wedge cross section or mutual information between two mixed systems in the background of confining geometries to gain the properties of the phase structures of QCD.

We show in this work that the minimal wedge cross section of two symmetric strips which as proposed in \cite{Takayanagi:2017knl} could be dual to the mixed correlation measures such as entanglement of purification between the two mixed state, can capture a lot of information about the phase structures of various confining models.  
 Specially this would be true for QCD and confining systems. Using the correlation measures of two symmetric mixed strips which would be easy to calculate, one could gain a lot about the phase structures of QCD and confining backgrounds, which is the motivation of this work. We found that in addition to the direct calculation of mutual information or minimal wedge cross section, in order to probe the phase structures of confining backgrounds, it would actually be easier and more preferable to calculate the critical distance between the two symmetric strips where the mutual information suddenly drops to zero. In addition, it could even detect the scale of chiral symmetry breaking and the confining-deconfining phase transitions which is the main result of this work.

For doing so, we use several top-down confining backgrounds here. These top-down, and ten-dimensional backgrounds use the fundamental models of string theory with different Dp-brane structures. For example, the Witten model of low-energy QCD is constructed through the type-IIA supergravity, using D4-D8 intersected branes. The boundary dual of this theory is a supersymmetric $4+1$ dimensional gauge theory which by higher temperatures dimensional reduction would lead to a non-supersymmetric QCD in $3+1$ dimension. This theory has a large number of colors and a large 't Hooft coupling. Extending this model, for instance, by adding flavor D8-branes, Sakai and Sugimoto could also explain the chiral symmetry breaking and its restoration at high temperatures by an interesting geometrical realization. The only free parameters in their model is the Kaluza-Klein mass parameter $M_{KK}$ or $u_{KK}$ and the 't Hooft coupling $\lambda$ at that scale. Another example is Klebanov-Strassler which contains $N$ regular and $M$ fractional D3 branes. Here for each metric we discuss briefly about the D-brane structures of each background.

 Here, we study the behavior of several quantum information quantities such as entanglement entropy and mixed correlation measures such as EWCS/entanglement of purification (EoP) versus these free parameters, such as $u_{KK}$, to probe the phase structures of QCD in these backgrounds. We hope it could also help to find the corrections needed to approach real QCD models at finite $N_c$ and $\lambda$ that could be gained from type-IIA string theory. Note that specifically since the Sakai-Sugimoto model, both qualitatively and quantitatively is very close to real QCD, our results in principle should be close to the results of lattice QCD.

Additionally, in any holographic QCD model, one would have an IR wall where the geometry would be terminated in the holographic direction, or there would be some sorts of singularities there. In this work we also would like to check the effects of these soft or hard walls and cut-offs and the terminated radius $u_{KK}$ on the quantum information measures such as mutual information, entanglement of purification and negativity, and the phase diagrams of various confining backgrounds. Therefore, another aim of this work is to check how the topology of different confining geometries would affect QCD properties and if/how the new holographic quantum measures such as entanglement of purification or quantum negativity, or the critical distance between two mixed systems, (and later complexity or complexity of purification), could probe these topological properties or dually the characteristics of QCD. This would actually be in continuation of our previous work \cite{Ghodrati:2018hss}, which we used complexity to probe the phase structures of Gubser's models of QCD \cite{Gubser:2008ny}, dubbed $V_1$, $V_2$, $V_{\text{QCD}}$ and $V_{\text{IHQCD}}$. In the current work however, to construct the phase structure, we use new mixed correlations measures and new confining backgrounds, including AdS-soliton, Witten-Sakai-Sugimoto and deformed Sakai-Sugimoto, Witten-QCD, Klebanov-Strassler, Klebanov-Tseytlin, Klebanov-Witten, Maldacena-Nunez, domain-wall QCD and Nunez-Legramandi metric. So, first in section \ref{sec:generalize}, we show how to generalize the new quantum measures in confining backgrounds and then in section \ref{sec:critdQCD}, we present the results for each of our confining models.

Then, in section \ref{sec:Crofton}, we present the relations and plots for the behavior of Crofton-from (which is a quantity in studying the Kinematic space and bulk reconstruction \cite{Czech:2015qta}) for each of these confining geometries and show they all have a universal behavior where a well could be detected around the singularity while it would become constant at the bigger holographic distances.

We finish with conclusion in section \ref{sec:conclude}.

\section{Mixed correlation measures in confining backgrounds}\label{sec:generalize}
The generalization of entanglement entropy (EE) from conformal to non-conformal cases would be from 
\begin{gather}
S_A= \frac{1}{4 G_N^{(d+2)} } \int_\gamma d^d \sigma \sqrt{  G^{(d)}_{\text{ind}} },
\end{gather}
to 
\begin{gather}\label{eq:EEconfined}
S_A=\frac{1}{4G_N ^{(10)}} \int d^8 \sigma e^{-2 \Phi} \sqrt{G^{(8)} _{\text{ind}}}, 
\end{gather}
where $G_N^{(d+2)}$ is the $d+2$ dimensional Newton constant and $G_{\text{ind} }^{(d)}$ is the induced string frame metric on $\gamma$. This generalization would be due to the fact that, in general, in the non-conformal theories, the volume of the $8-d$ compact dimensions and also the dilaton would be \textit{non-constant}.

Note that, because of the symmetry, or isotropy, in out setup, where we have two equal strips, we expect that this formula works for the confining backgrounds since the only difference with the conformal case is the wall at the end of the geometry in the deep IR, in the large $u$, which does not break the symmetry in the $x$ direction. Therefore, in the integral of equation \ref{eq:EEconfined}, we only need to consider the variation of this dilaton field along the radial direction $u$, and therefore simply the term $e^\phi$ is multiplied by the induced metric along the minimal wedge cross section.

 \begin{figure}[ht!]
 \centering
  \includegraphics[width=3.7cm] {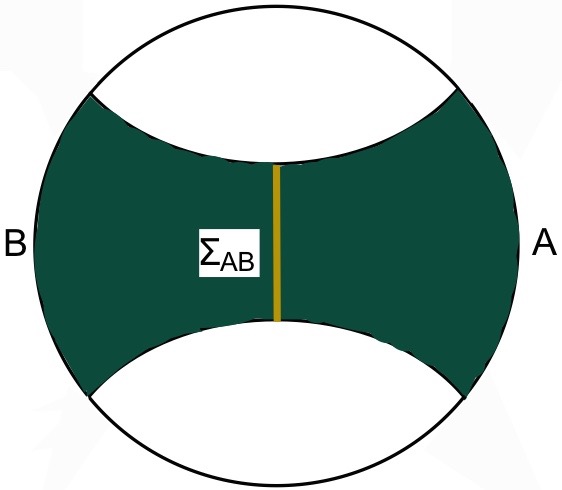} 
  \caption{The minimal wedge cross section for the two connected mixed systems of $A$ and $B$, i.e, $\Sigma_{AB}$ is shown in yellow while the whole entanglement wedge is shown in green.}
 \label{fig:wedge}
\end{figure}

We put the setup of figure \ref{fig:wedge} in the background of confining geometries (as in figure \ref{fig:AdSBHsol} which is for the AdS soliton case) and probe various exotic features of these top-down, supergravity and confining backgrounds and show that mixed quantum entanglement measures and correspondingly the geometrical objects such as minimal wedge cross section can detect the effects of singularities, walls, throats and exotic topologies of such backgrounds.

First, note that if we write the confining gravitational backgrounds in the string frame in the form of 
\begin{gather}
ds^2= \alpha( \rho) \lbrack \beta(\rho) d\rho^2 + dx^\mu dx_\mu \rbrack + g_{ij} d \theta^i d \theta^j,
\end{gather}
where $\rho$ is the holographic radial coordinate in the interval $\rho_\Lambda < \rho < \infty $, $x^\mu, \ \ (\mu=0,1,...,d)$ parameterize $\mathbb{R}^{d+1}$, and $\theta^i, \ \ (i= d+2, . . . ,9)$ are the $8-d$ internal directions.

The bulk minimal wedge cross section constructed from two mixed boundary regions, has been proposed in \cite{Takayanagi:2017knl} to be dual to the entanglement of purification between these two mixed systems of $A$ and $B$, as in figure \ref{fig:wedge}. For the simple Schwarzchild AdS black brane geometry and for two regions with width of $l$ and distance $D$, in \cite{Ghodrati:2019hnn}, this quantity has been found as
\begin{gather}
\Gamma=\int_{z_D}^{z_{2l+D}} \frac{dz}{ z^{d-1} \sqrt{1-\frac{z^d}{z_h^d} }},
\end{gather}
where $z_D$ and $z_{2l+D}$ are the two turning points.

This then should be generalized for the non-conformal and specifically for confining geometry similar to the EE case. In fact, for several Dp-brane geometries, in \cite{Lala:2020lcp}, several similar quantum information measures have been calculated.

Here we could generalize it as 
\begin{gather}\label{eq:generalizedeEW}
\Gamma=\int_{u_D}^{u_{2l+D} } du e^{-2 \Phi} \sqrt{-\gamma_{ab} },
\end{gather}
where $\gamma_{ab}$ is the determinant of the eight dimensional induced metric on $\Gamma$ that results from the ten dimensions by setting the two coordinates of time and radial to zero, i.e, $dt=dr=0$.

Note that they could be some constant factors behind the integral which we will set to one in our studied here. Also, $e^{-2\Phi}$ should be brought here, which is due to the fact that in non-conformal geometries, the dilaton field $\Phi$ and the volume of the compact dimensions are not constant and so similar to the entanglement entropy case, are needed to be considered for the formulation of the entanglement of purification as well. The dilaton field would be denoted by $\Phi$ and the RR and NS fluxes would complete the background which for each geometry is different.

One can define another quantity 
\begin{gather}
H(\rho) = e^{-4 \Phi } V^2_{\text{int}} \alpha^2,
\end{gather}
where $V_{\text{int}} = \int d \vec{\theta} \sqrt{\text{det} \lbrack g_{ij} \rbrack }$ is the volume of the internal manifold described by the coordinates $\vec{\theta}$.

Then, at any confining background, the entanglement entropy (EE) could be written as \cite{Kol:2014nqa}
\begin{gather}\label{eq:connected}
S_C(\rho_0) = \frac{V_{d-1} }{2 G_N^{(10)} } \int_{\rho_0}^\infty d\rho \sqrt{ \frac{\beta(\rho) H(\rho) }{1- \frac{H(\rho_0) }{H(\rho) } } },
\end{gather}
where $\rho_0$ is the minimum value of $\rho_0$ and the width of one interval as a function of $\rho_0$ would be
\begin{gather}\label{eq:L}
L(\rho_0) =2 \int_{\rho_0}^ \infty d\rho \sqrt{\frac{\beta(\rho) }{ \frac{H(\rho)}{H(\rho_0)}-1} },
\end{gather}
and the disconnected solution would be written as
\begin{gather}\label{eq:disconnected}
S_D (\rho_0) = \frac{V_{d-1} }{2 G_N^{(10)} } \int_{\rho_\Lambda}^\infty d\rho \sqrt{\beta(\rho) H(\rho)}.
\end{gather}

The difference between the connected and disconnected solution would be finite.

These quantities would be useful for pure states. When we have mixed system, the main quantity that should be examined is the mutual information, $I(A,B)= S_A + S_B - S_{AB}$, which needs these relations \ref{eq:connected}, \ref{eq:L} and \ref{eq:disconnected}. From the width of one interval $l$ or the distance between intervals $D$, the turning point for each region would be calculated and then the entanglement entropy for each region could be calculated. In addition, from the distance between the turning points, the minimal wedge cross section $\Gamma$ could be found.

\section{Critical distances in the confining backgrounds}\label{sec:critdQCD}

When the distance between two mixed systems becomes larger, at a specific critical distance $D_c$, the correlations, mutual information and entanglement of purification between them vanish or rather drop to one order of magnitude lower. 

For two equal strips with the width $l$ and distance $D$, the mutual information would be
\begin{gather}\label{eq:mutualrel}
I(D,l)= S_A+S_B - S_{AB}= 2S(l)-S(D)-S(2l+D),
\end{gather}
and the critical distance can be found by where it drops to zero, i.e, $I(D_c,l)=0$.

We show that this critical distance could give interesting new information about QCD phase structures as it in fact, would be a new scale in the theory.

The critical width for ``one strip'' could distinguish the confinement/deconfinement phase transition and it has been proposed that this critical width $L_{\text{crit}}$ is proportional to the inverse of confinement temperature, i.e, $L_{\text{crit} }= \mathcal{O} (\Lambda_{IR}^{-1})$, where $\Lambda_{IR}$ correspond to the singularity in each theory. However, here we can show that the critical distance between \textit{``two strips''} is a stronger quantity and can even distinguish the scale of chiral symmetry breaking in addition to  the confinement/deconfinement scale.

\begin{figure}[ht!]
 \centering
  \includegraphics[width=8.5cm] {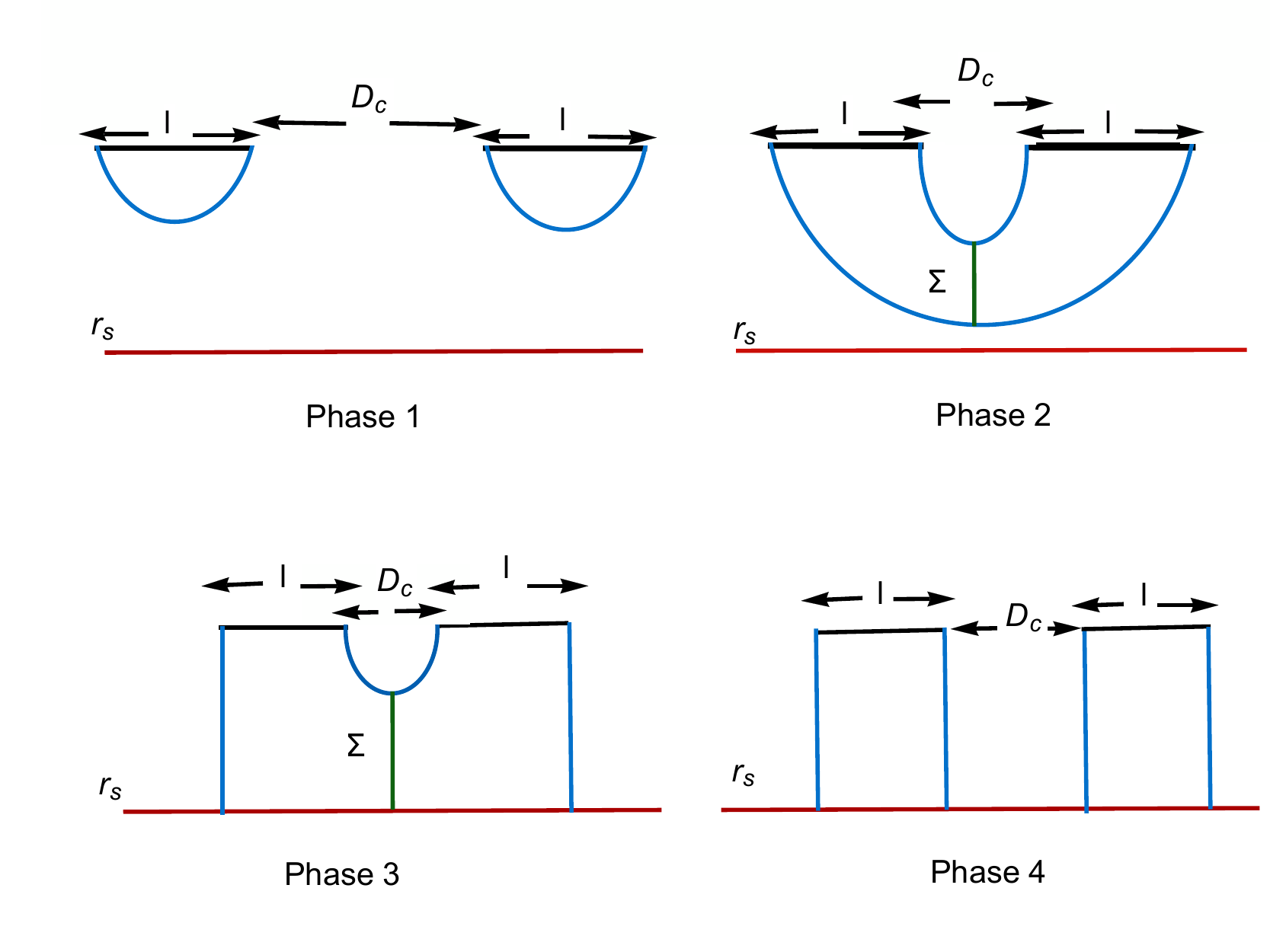} 
    \includegraphics[width=8.5 cm] {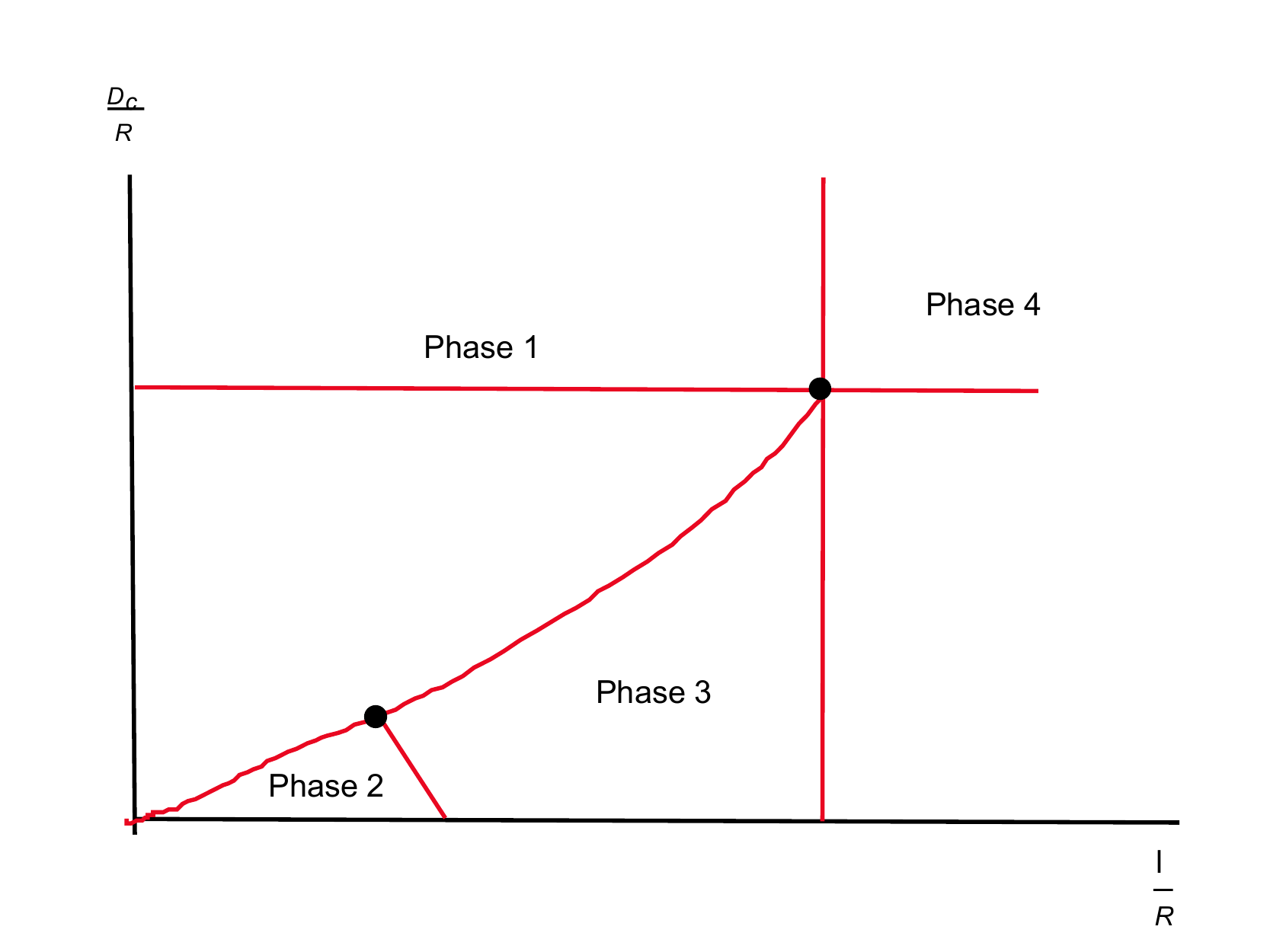} 
  \caption{The structure of phases in confining backgrounds as first presented in \cite{Jain:2020rbb}.}
 \label{fig:phasesQCDstrip}
\end{figure}

Similar to the work of \cite{Jain:2020rbb}, we find that the phase structure of confining background would be as shown in figure \ref{fig:phasesQCDstrip}. Note that, in each of these based on the relative sizes of $r_s$ or $u_{KK}$, $D_c$ and $l$, the entanglement wedge cross section becomes minimum between all other solutions and therefore that specific phase beocome dominant, and each time this happens we see a jump in the numerical solutions.

These jumps then could be used to distinguish the scale of chiral symmetry breaking versus confinement/deconfinement phase transitions. We present more numerical details about how this quantity would probe phase structures of QCD and could capture important scales in each theory.

\subsection{AdS soliton background}
The first background where we probe its phase structures using mixed correlations would be AdS soliton.  This geometry was first introduced in the work of Myers and Horowitz \cite{Horowitz:1998ha} where the boundary CFT side have supersymmetry-breaking boundary conditions and the bulk geometry have negative energy dual to the negative Casimir energy in the boundary gauge theory side. Therefore, the AdS-soliton would have the lowest negative boundary energy among all solutions with the same asymptotic boundary. This negative energy which could source and generate the AdS-soliton geometry has been calculated as $\frac{E}{L_x^{n-2} L_y} \Big |_{\text{BH}} =\frac{(n-1) r_0^n }{2 \ell^{n+1}} $, where $n$ is the dimension of the boundary, $n+1$ is the dimension of the bulk geometry, $L_x$ and $L_y$ represent the extension of the surface on $x$ and $y$ directions and $\ell$ is the radius of AdS spacetime. In the bulk it means that all massive and massless particles would repel each other and therefore it means AdS-soliton would actually anti-gravitates. The effects of this feature would then be carved in the structures of quantum states and on the behavior of quantum information measures, and bit threads structures.

In addition, since there are more phase transitions possible between AdS-black hole (BH) solutions and AdS-solition geometry compared with other cases like AdS-BH and thermal AdS geometry, it would be more interesting to study the behaviors of mixed quantum information measures like entanglement of purification (EoP), mutual information (MI), complexity of purification (CoP), negativity, etc and also tools such as bit threads and Wilson loops during such phase transitions.

We choose the following background for the AdS soliton in the Poincare coordinates \cite{Ishihara:2012jg}
\begin{gather}\label{eq:maingeometry}
ds^2=\frac{L^2_{\text{AdS}} }{z^2} \left ( \frac{dz^2}{f(z)} +f(z) d\theta^2-dt^2+dr^2+r^2 d\Omega_{d-3} \right),
\end{gather}
where $f(z)$ is a harmonic function and has a general form as 
\begin{gather}
f(z)= \left( 1- k_1 \frac{z}{z_0} \right ) \left (1-k_2 \frac{z}{z_0} \right ) (1+ \sum_{n=1} c_n z^n).
\end{gather}

This metric \cite{Ishihara:2012jg} has been found by double Wick rotating the asymptotically AdS space where the coefficients $c_n$ could be derived by the condition that the $1+\sum c_n z^n$ should not contain poles and zeros at $z=z_0$.

 \begin{figure}[ht!]
 \centering
  \includegraphics[width=6 cm] {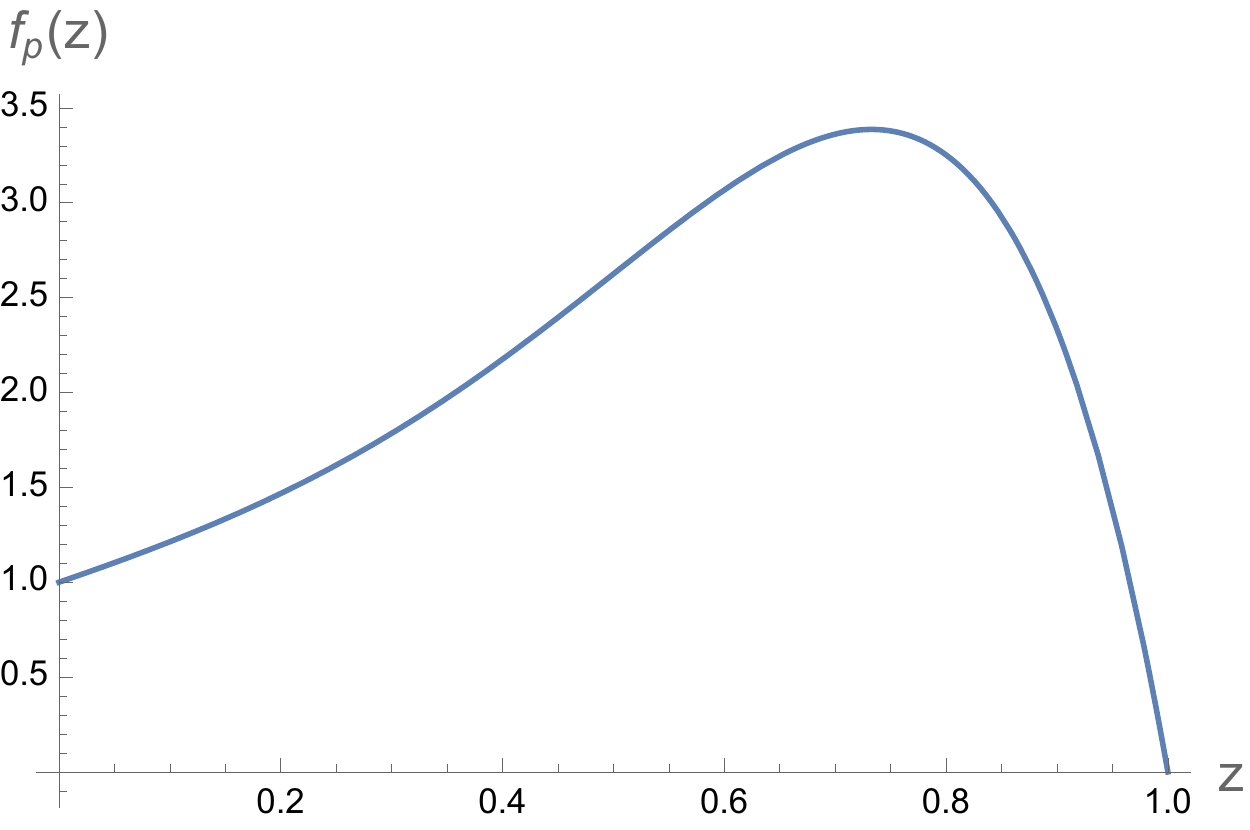} \hspace{1cm}
    \includegraphics[width=6 cm] {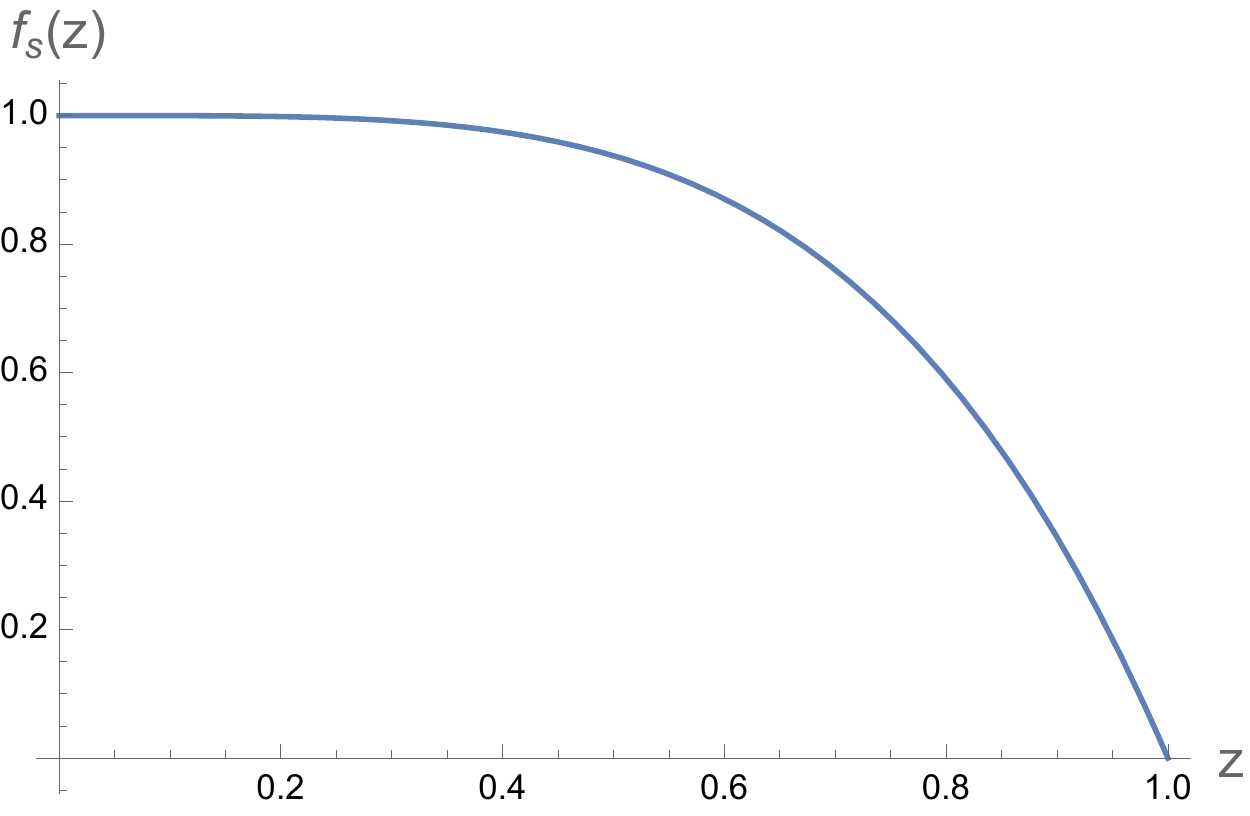} \hspace{1cm}
  \caption{The behavior of $f_p(z)=( 1- k_1 \frac{z}{z_0}  )  (1-k_2 \frac{z}{z_0}  ) (1+ \sum_{n=1} c_n z^n)$ which uses the perturbative sum and arbitrary values of $c_i$ as $k_1=1, k_2=-2, c_1=1, c_2=2, c_3=3, c_4=4, c_5=5$, which is used for the general case of soliton is shown in the left part, and the function for the simplest AdS soliton case, $f_s(p)=1- (\frac{z}{z_0})^{8-d}$, $d=4$, which we use here for calculations, is shown in the right part.}
 \label{fig:f}
\end{figure}

By changing $k_1$ and $k_2$, different IR behaviors could be yielded, but not for any $k_1$ and $k_2$, this ansatz would become a solution of Einstein theory. In order to get a soliton AdS geometry one of the $k_1$ or $k_2$ should be negative, and the simplest AdS soliton would be with $k_1=1$ and $k_2=-1$. The behavior of $f(z)$ for the general case and simplest case are shown in figure \ref{fig:f}. Here we take $ f(z)= 1- (\frac{z}{z_0})^{8-d}$ as for the simplest AdS soliton.

It worths to mention that other complicated soliton solutions can also be obtained by double Wick rotating the charged or hairy scalar AdS black hole solutions. For example, by double Wick rotating charged $\text{AdS}_5$, the function $f$ would be found as $f(z)=1-m z^4 + q^2 z^6$, which is related to a boundary theory with non-zero magnetic flux or current density condensate.

Now, to study entanglement and other correlation measures, first, the width of the strip versus turning point $z_t$, for AdS soliton, could be found as
\begin{gather}
L(z_0)= 2 \int_{z_0}^\infty \frac{dz}{ \left ( 1- \left(\frac{z}{z_0} \right)^{8-d}  \right ) \left ( \left(\frac{z_0}{z} \right)^{2d-4} \left (1-  \left(\frac{z}{z_0} \right)^{8-d}\right )   -1  \right )  },  
\end{gather}
and the entanglement entropy for the connected solution in this background would be
\begin{gather}
S_C(z_t)= c_1 \int_{z_0}^\infty   \frac{dz}{z^{d-2} \sqrt{1- \frac{ \left( \frac{z}{z_0}\right)^{2d-4}  } {1- \left(\frac{z}{z_0}\right)^{8-d} } } },
\end{gather}
and for the disconnected part one gets
\begin{gather}
S_D= -c_2 \int_{z_\Lambda}^\infty \frac{dz}{z} ,
\end{gather}
where $c_1$ and $c_2$ are just two constants that are the combination of the parameters $G_N^{(10)}$, $d$, $L$, $\pi$ and  $V_{d-3}$, and are not important in our numerical studies here.

\begin{figure}[ht!]
 \centering
  \includegraphics[width=7.5cm] {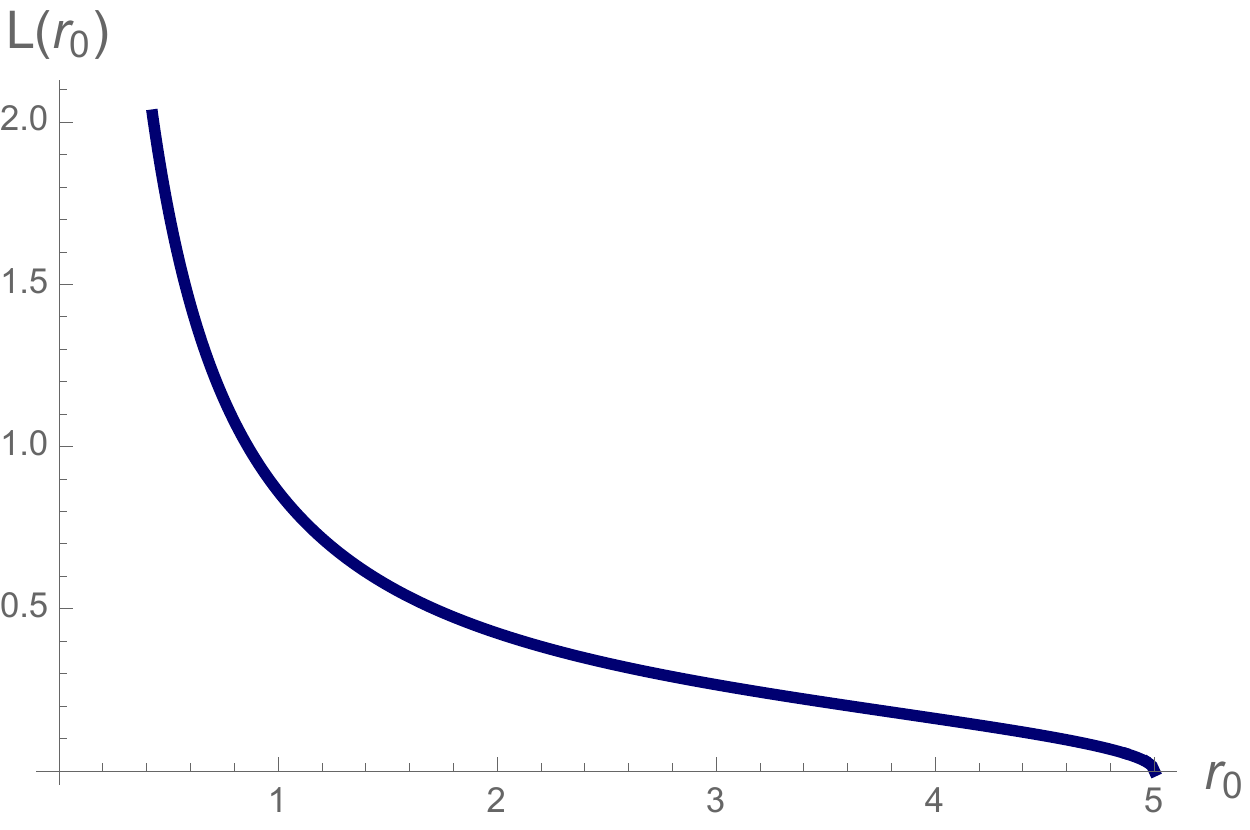} 
    \includegraphics[width=7.5 cm] {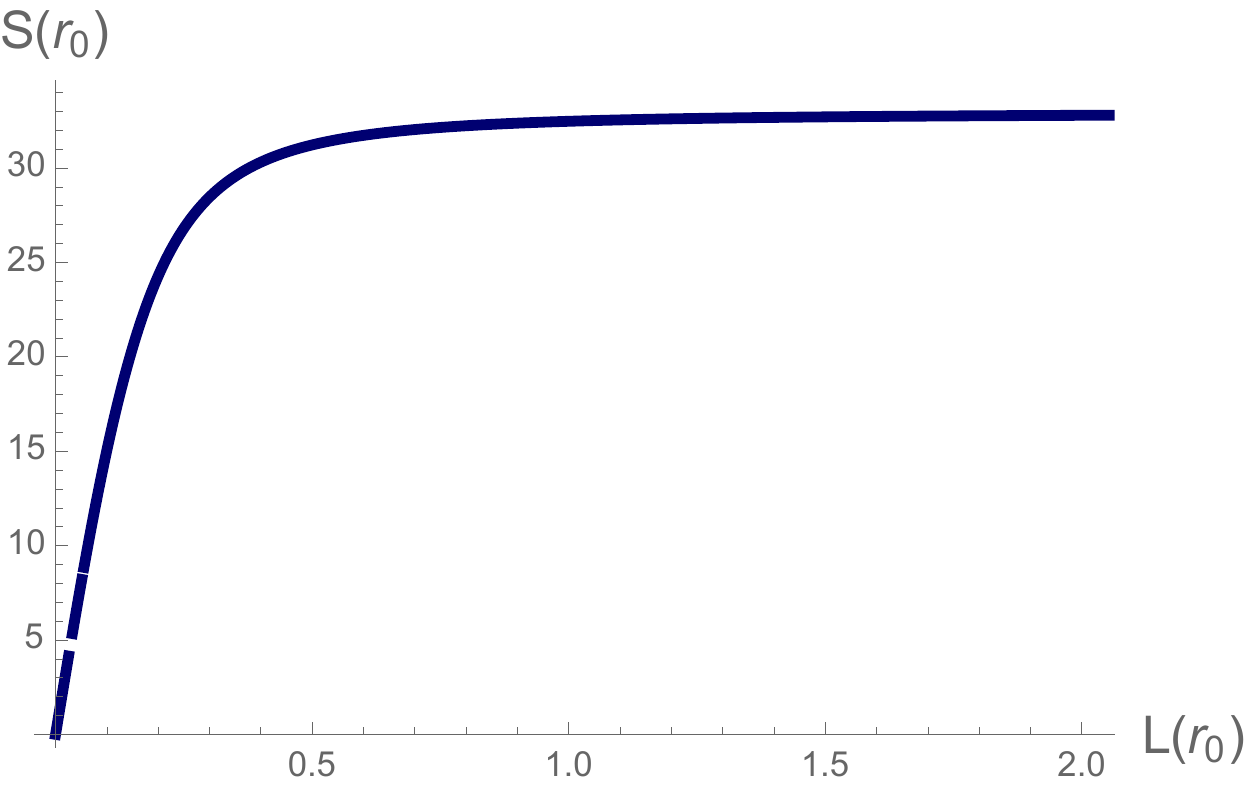} 
  \caption{The relationship between width of a strip versus turning point $\tau_0$ (in the left), and entanglement entropy versus $L$ (in the right) in the AdS-soliton background.}
 \label{fig:Stripinsoliton}
\end{figure}

The behavior of these quantities are shown in figure \ref{fig:Stripinsoliton}. The noises observed in these phase diagrams could be due to various phenomena in the system such as presence of bound-entangled states, creation and annihilation of EPR pairs, etc.

\begin{figure}[ht!]
 \centering
  \includegraphics[width=8.5cm] {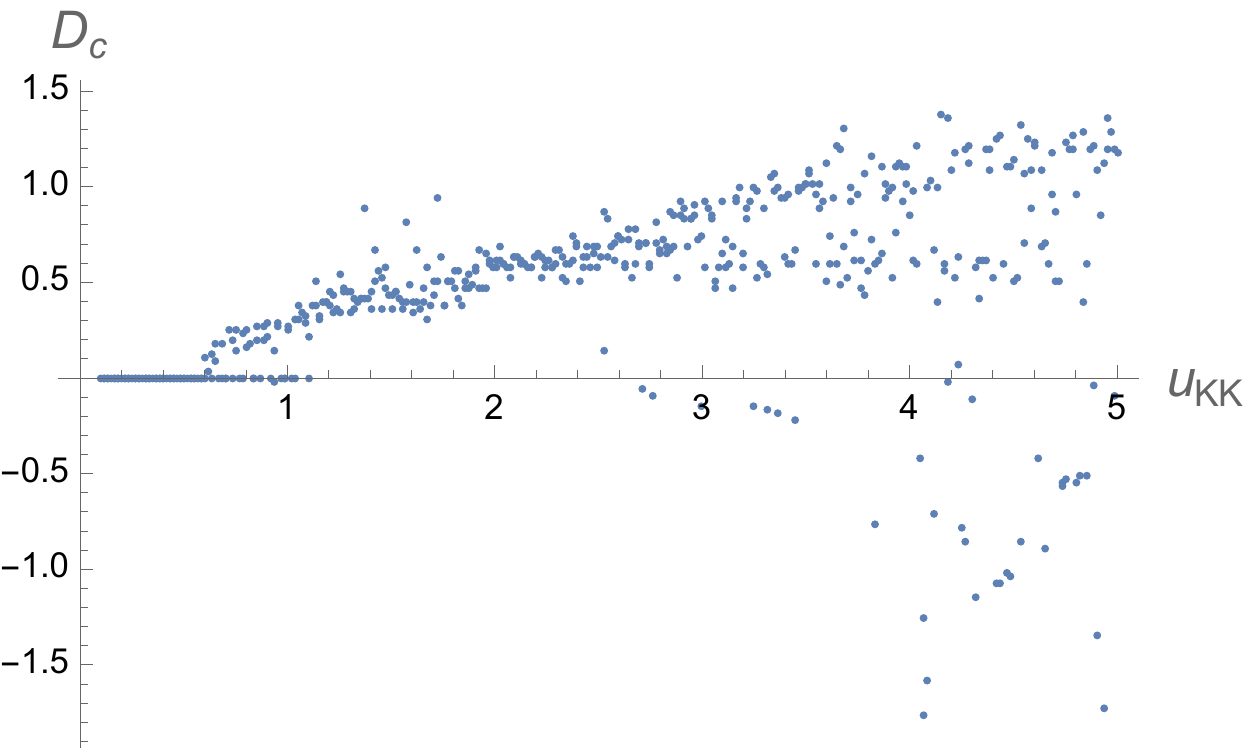} 
  \caption{The behavior of $D_c$ versus $u_{KK}$ in AdS-soliton background with the metric \ref{eq:maingeometry} .}
 \label{fig:datapointsoliton}
\end{figure}

The behavior of $D_c$ versus $u_{KK}$ in AdS-soliton background is shown in figure \ref{fig:datapointsoliton}.

For the AdS-soliton metric in the form of \cite{Reynolds:2017jfs}
\begin{gather}
ds^2= \frac{r^2}{\ell^2}  \left \lbrack - dt^2 + \left ( 1- \frac{r_+^d}{r^d} \right ) d \chi^2 + d\vec{x}^2 \right \rbrack + \left ( 1- \frac{r_+^d}{r^d} \right )^{-1} \frac{\ell^2}{r^2} dr^2,
\label{eq:metricAdS2}
\end{gather}
where there are $d-2$ components of $\vec{x}$, the function for $L(r_0)$ and $S(r_0)$ would be
\begin{gather}
L(r_0)= 2 \int_{r_0}^ \infty dr \sqrt{\frac{ \frac{\ell^4}{r^4} \left(1-\frac{r_+^d}{r^d} \right)^{-1}  }  {\left(\frac{r}{r_0} \right)^{2d-6} \frac{1-\left(\frac{r_+}{r^d} \right)^d}{1- \left(\frac{r_+}{r_0} \right)^d } } },\nonumber\\
S_C(r_0)= \frac{V_d}{2G_N} \int_{r_0}^\infty dr \frac{ \left (\frac{r}{\ell}\right )^{d-5} }{\sqrt{1- \left(\frac{r_0}{r} \right)^{2d-6} \frac{1-\left(\frac{r_+}{r_0} \right)^d}{1- \left (\frac{r_+}{r}\right)^d} }},
\end{gather}

\begin{figure}[ht!]
 \centering
  \includegraphics[width=5.5cm] {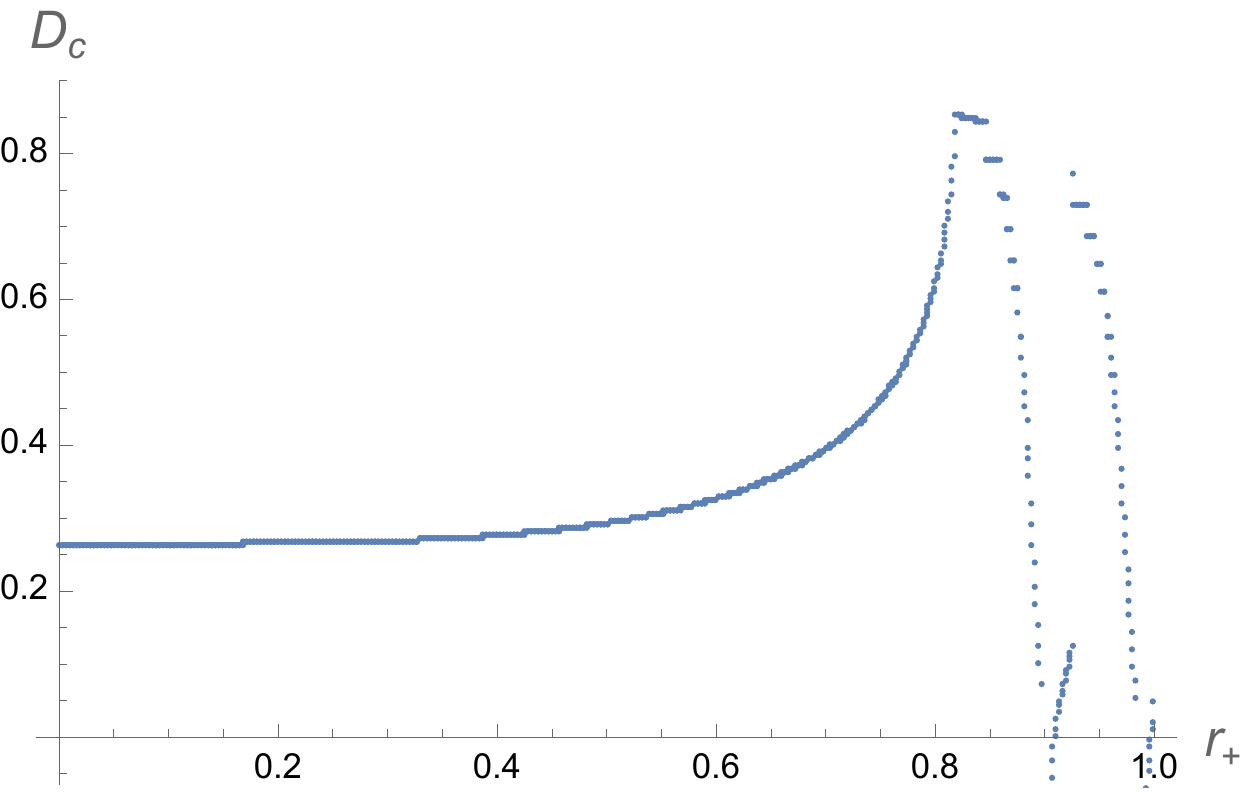} 
    \includegraphics[width=5.5cm] {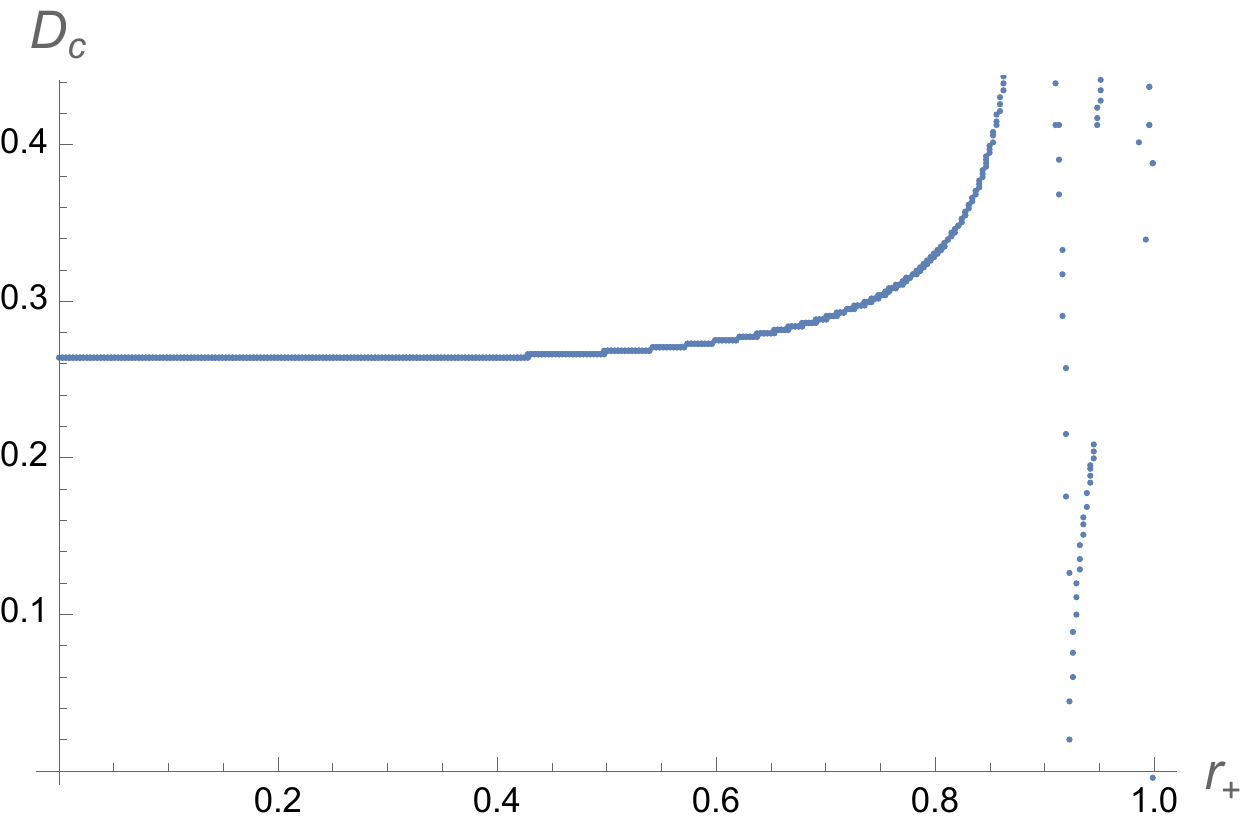} 
        \includegraphics[width=5.5cm] {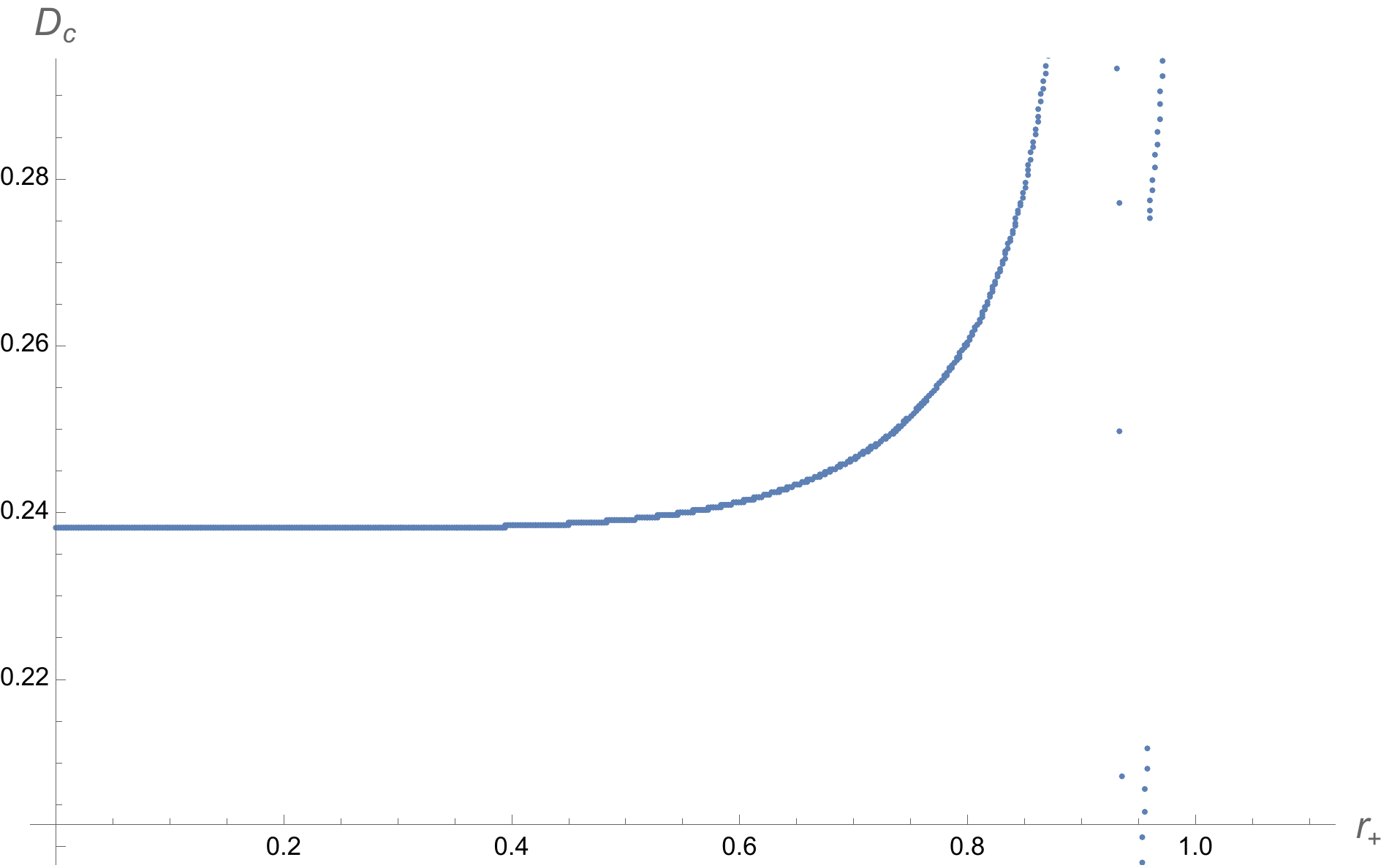} 
  \caption{The behavior of $D_c$ versus $r_{+}$ in the AdS-soliton background with the metric \ref{eq:metricAdS2}, for the case of $d=4$ in the left, $d=5$ in the center, and $d=6$ in the right part.}
 \label{fig:solitondata3}
\end{figure}

\begin{figure}[ht!]
 \centering
  \includegraphics[width=8cm] {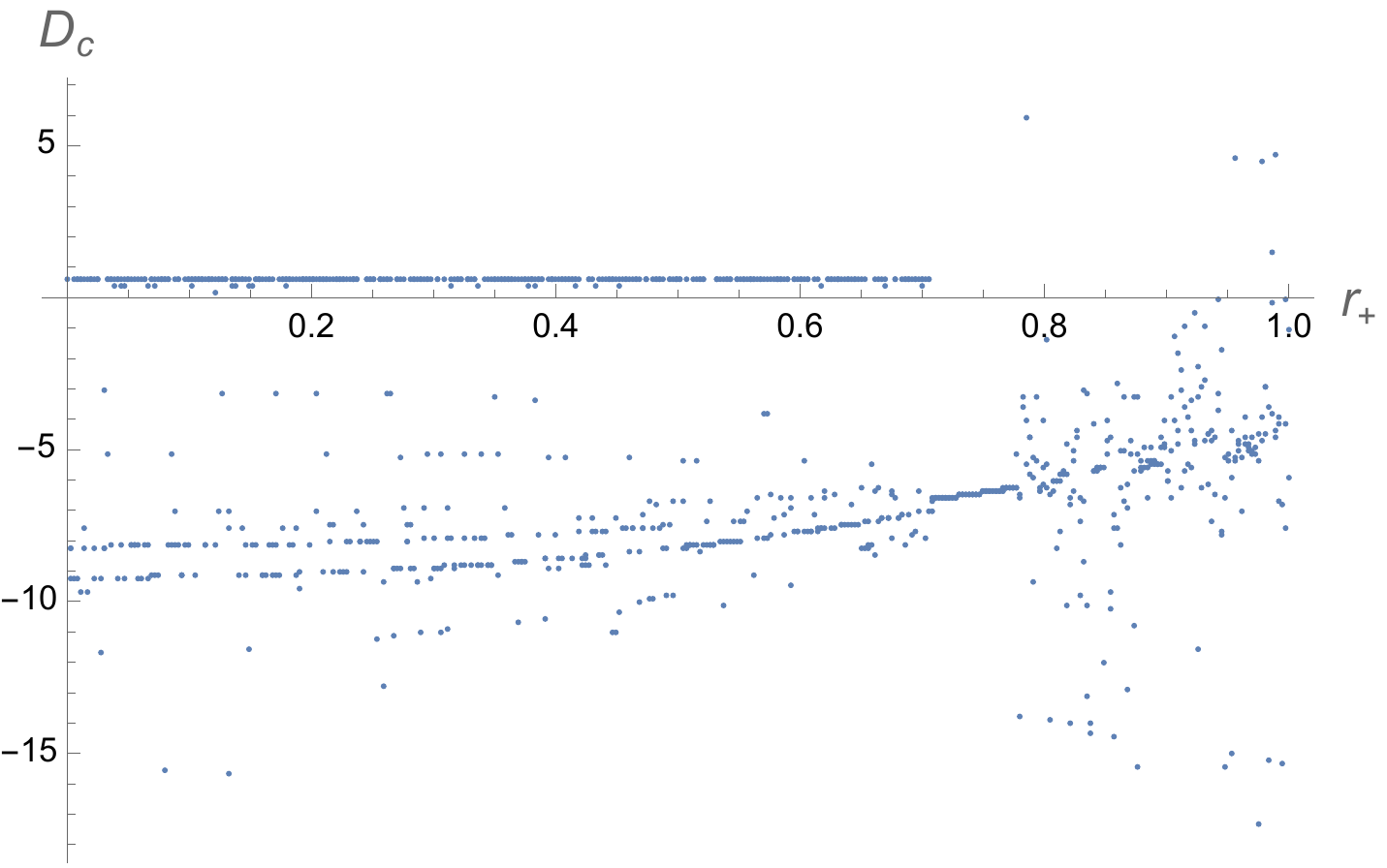} 
  \caption{$D_c$ versus $r_{+}$ in the AdS-soliton background with the metric \ref{eq:metricAdS2}, for the case of $d=2$.}
 \label{fig:solitond2}
\end{figure}

For the case of $d=4$, $d=5$ and $d=6$, the behavior of $D_c$ versus $r_+$ is shown in figures \ref{fig:solitondata3}, where the four phases are clear. The phases become less distinct when the dimension increases, for $d \ge 4$. The case of $d=2$ also is shown in figure \ref{fig:solitond2}.

Note that the previous discussions are for \textbf{one} strip and for \textbf{pure} states. Now we move to the case for \textbf{two} strips and \textbf{mixed} states, where the mutual information and critical distance, $D_c$ between the two strips are being used to probe the phase space. This critical distance $D_c$ can be calculated by setting the mutual information \ref{eq:mutualrel} to zero, while we use the relation \ref{eq:L}, for calculating the turning points from the widths of each region and then we use the relation \ref{eq:connected} for calculating the entanglement entropies.

\begin{figure}[ht!]
 \centering
  \includegraphics[width=6.5 cm] {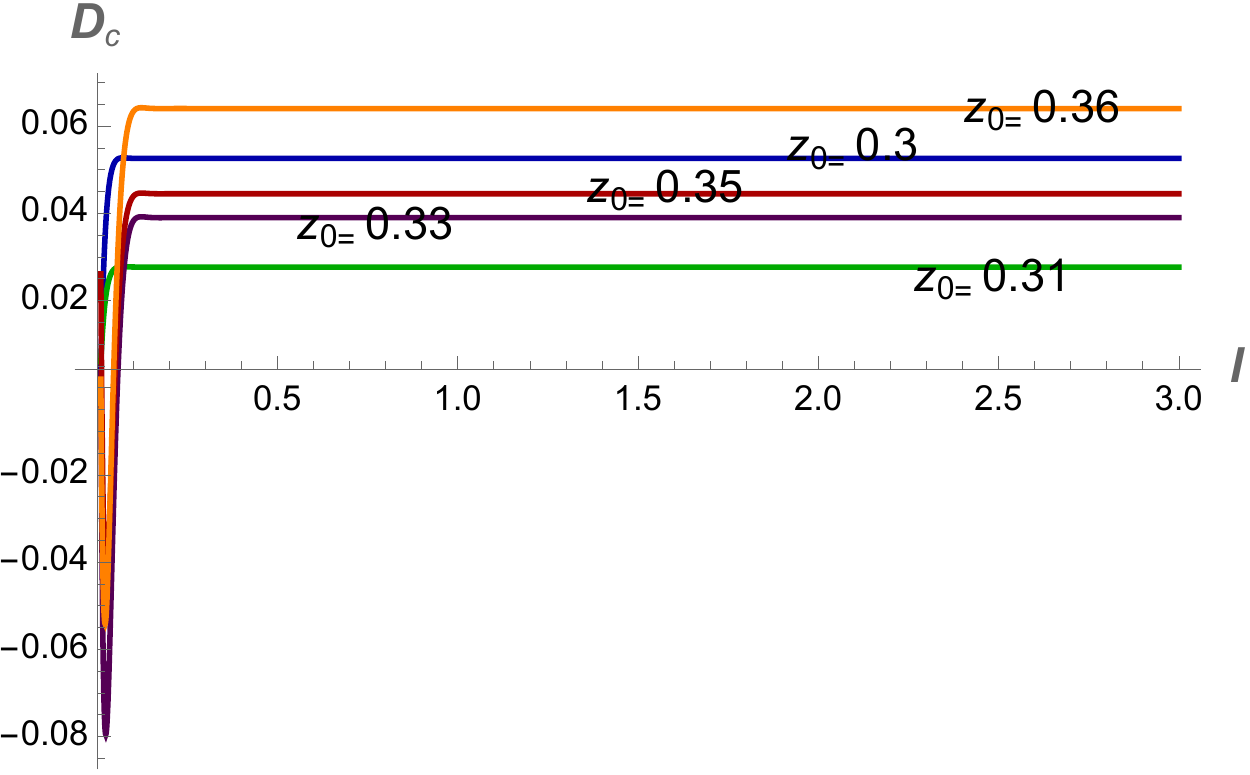} 
    \includegraphics[width=7.5 cm] {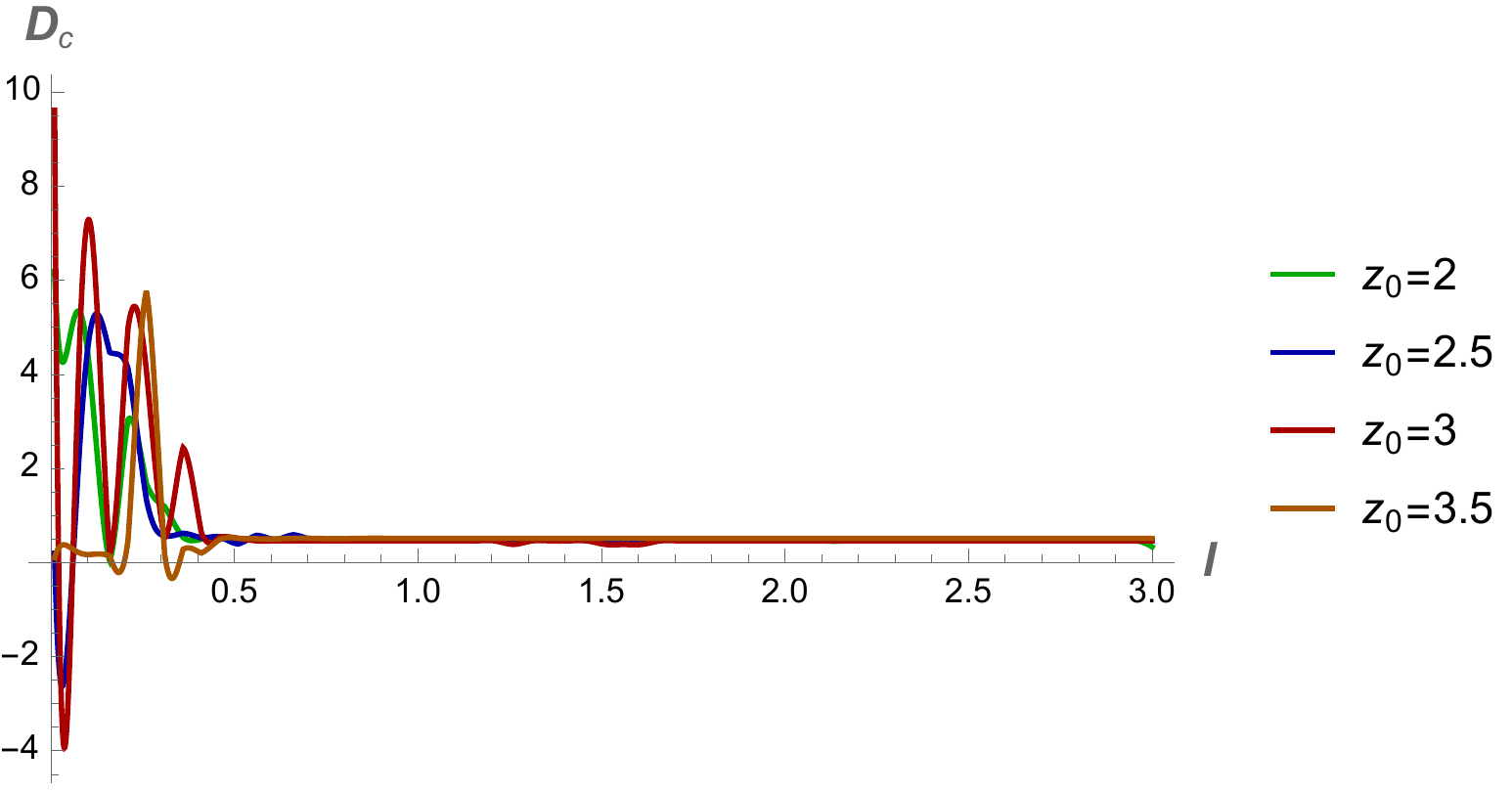} 
      \includegraphics[width=7.5 cm] {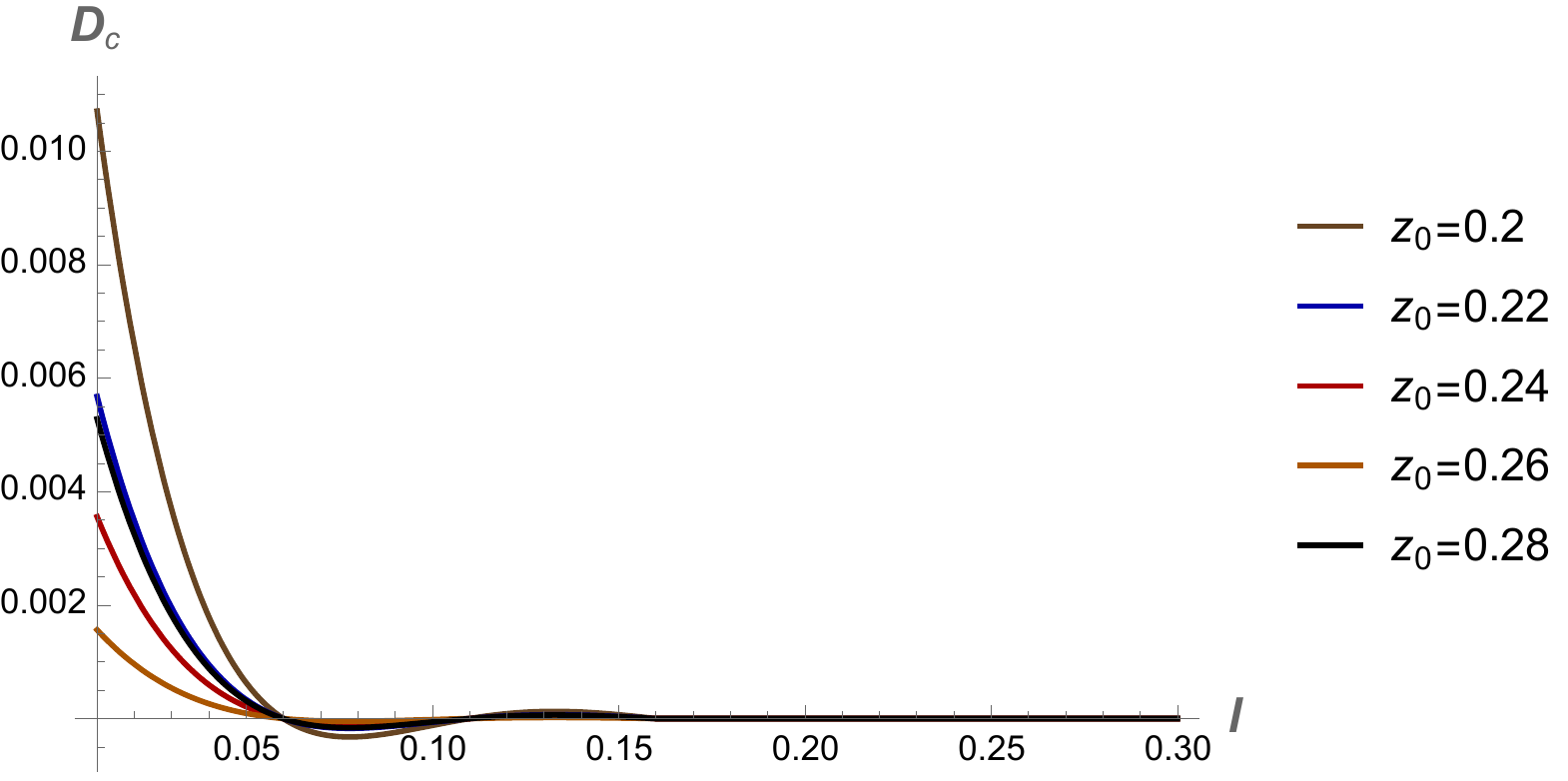} 
          \includegraphics[width=7.5 cm] {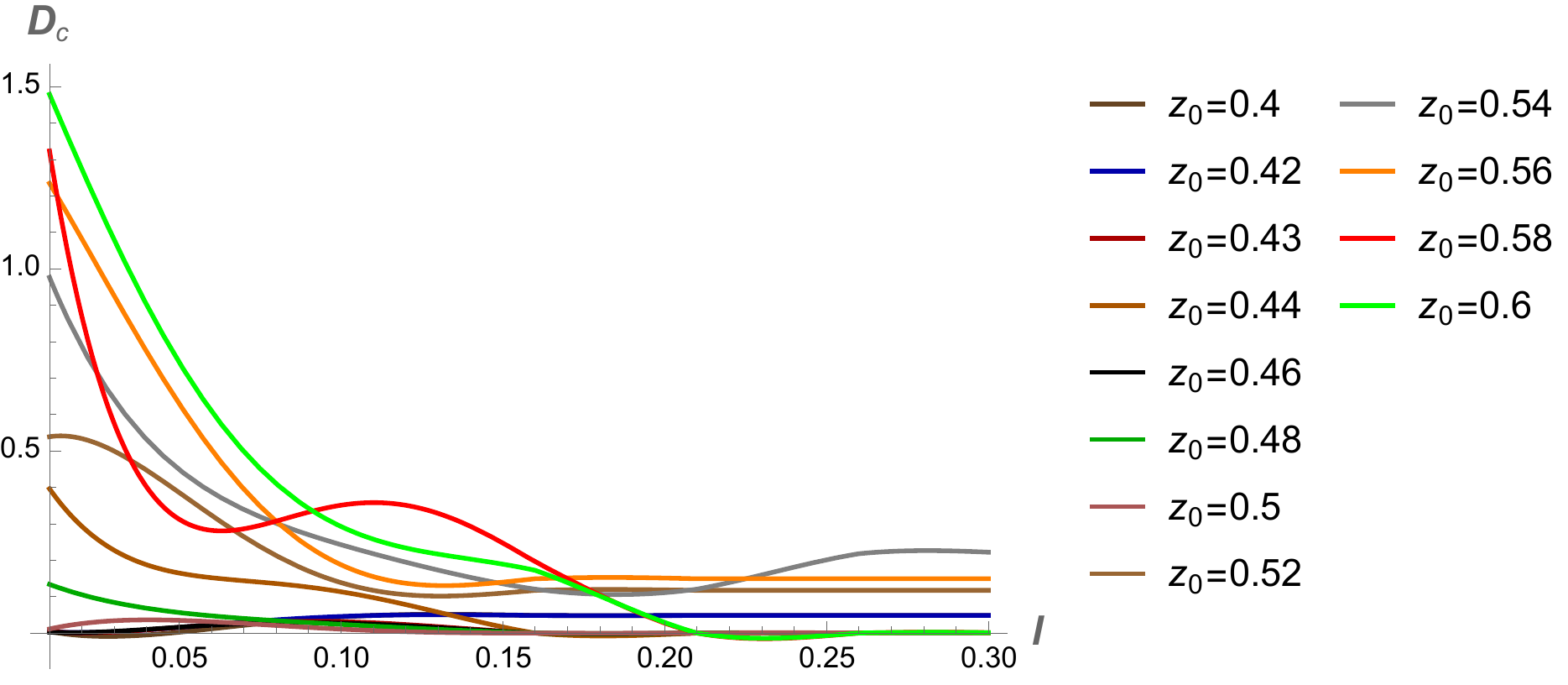}
  \caption{The four phases detected by the relationship between critical distance between two strips, $D_c$ and width of the strip in AdS soliton background, for the range of parameter $0.35<z_0<0.4$ in the up-left for the 1st phase and $2<z_0<3.5$ in the up-right for the 2nd phase, $0.2<z_0<0.3$ down-left in the 3rd phase and $0.4<z_0<0.6$ down-right in the 4th phase  which actually till $z_0=2$ would still be in the fourth phase.}
 \label{fig:Stripinsolitoncrit}
\end{figure}

Then, the position of the singularity could be changed, and the behavior of critical distance $D_c$  versus the width $l$ and therefore the phase structures of this theory could be found where the results are shown in figure \ref{fig:Stripinsolitoncrit}. By changing $z_0$ at least four different phases could be detected here where in each phase the behavior of $D_c$ is somehow distinct.

From plotting $D_c$ for various positions of the end of the world or singularity, i.e,  $z_0$ here  or $z_\Lambda$ in other geometries,  it could be noted that unlike the conformal case which has “two phases”, confining geometries have “four phases”. Also, the behavior of this critical distance in probing phases is similar to the behavior of logarithmic negativity in the setup of the end of the world brane as recently discussed in \cite{Dong:2021oad} by Dong, McBride, Weng. So, this shows that in confining cases, the measures of quantum mixed states such as mutual information and negativity show compatibility with each other.

One important feature of this theory is the capping or the gap where its effect could be detected in the correlation measures such as EWCS and MI. As the entangling surface would wrap around the compact dimension, all the UV scaling structures specifically for the entanglement entropy and also entanglement of purification would become different. For instance EWCS would be proportional to $\frac{L_\theta}{R} S_{UV}^{(d)}$ where $L_\theta$ is the fixed proper size of the compact circle, $R$ is the linear size of the entangling surface and $S_{UV}^{(d)}$ is the UV structure of entanglement entropy of $\text{AdS}_{d+1}$. These measures in specific regimes and phases show monotonic RG flows as in c-theorem.

Note that AdS-soliton case could be presented in the cylinder or disc topologies. As mentioned in figure 1 of  \cite{Ishihara:2012jg}, the cylinder topology corresponds to the trivial product state after RG flow in MERA, and the disc topology corresponds to the entangled states protected by the symmetry or the topological orders. The critical distance in each case could capture these quantum distinctions as well.

Our results would also be very useful in studying the dynamics of black holes and also chaotic systems. In principle, the structures of islands and replica worm holes could also be investigated for the black holes in such confining backgrounds using this new quantity. Then, the interplay between the UV and IR correlations could also be probed this way.

Note that the AdS soliton can also be written in other formats as well. One can write the general background which is asymptotically AdS as
\begin{gather}\label{eq:solitonform2}
ds^2= \frac{r^2}{\ell^2} \left \lbrack - \left( 1- \frac{r_0^{p+1} }{r^{p+1} } \right) dt^2 + (dx^i)^2 \right \rbrack + \left( 1-\frac{r_0^{p+1} }{r^{p+1}} \right ) ^{-1}  \frac{\ell^2}{r^2} dr^2,
\end{gather}
where $i=1, ...,p$, which for certain values of $p$ would arise in the near horizon geometry of p-branes. The energy here is $E_p= \frac{p V_p}{16 \pi G_{p+2} \ell^{p+2}  } r_0^{p+1}$.

The AdS soliton metric could then be derived by the analytical continuation $t \to i \tau$ and $x^p \to it$ and be written in the form of \cite{Horowitz:1998ha}
\begin{gather}
ds^2=\frac{r^2}{\ell^2} \left \lbrack \left(1- \frac{r_0^{p+1}}{r^{p+1}} \right) d\tau^2 +(dx^i)^2 -dt^2 \right\rbrack + \left (1-\frac{r_0^{p+1} }{r^{p+1} } \right)^{-1} \frac{\ell^2}{r^2} dr^2,
\end{gather}
where there are $p-1$ $x^i$'s. The energy then would be $E=- \frac{r_0^{p+1}\beta V_{p-1}   }{16 \pi G_{p+2} \ell^{p+2}} $, where $\beta=4\pi l^2 /(p+1)) r_0$ is the period.

\begin{figure}[ht!]
 \centering
  \includegraphics[width=4.5 cm] {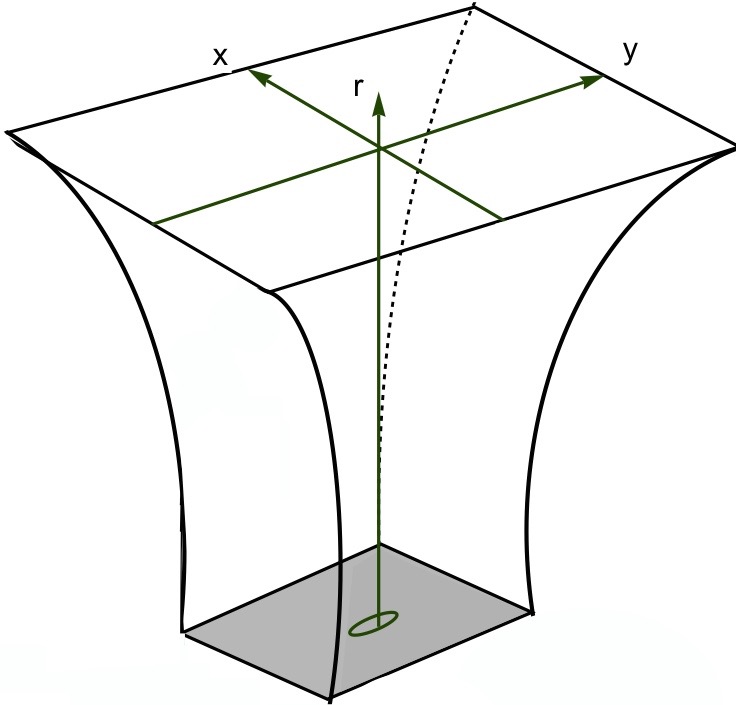}
    \includegraphics[width=4.5 cm] {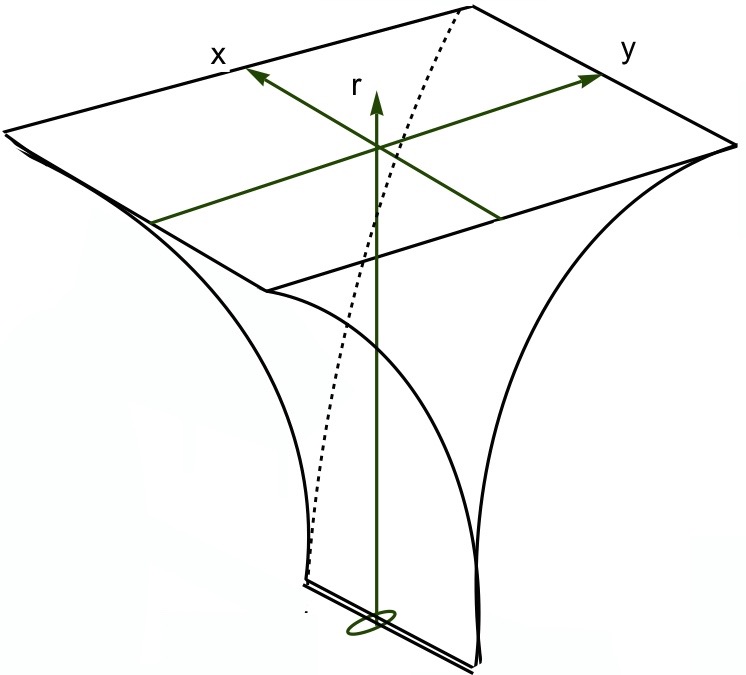} 
       \includegraphics[width=4.8 cm] {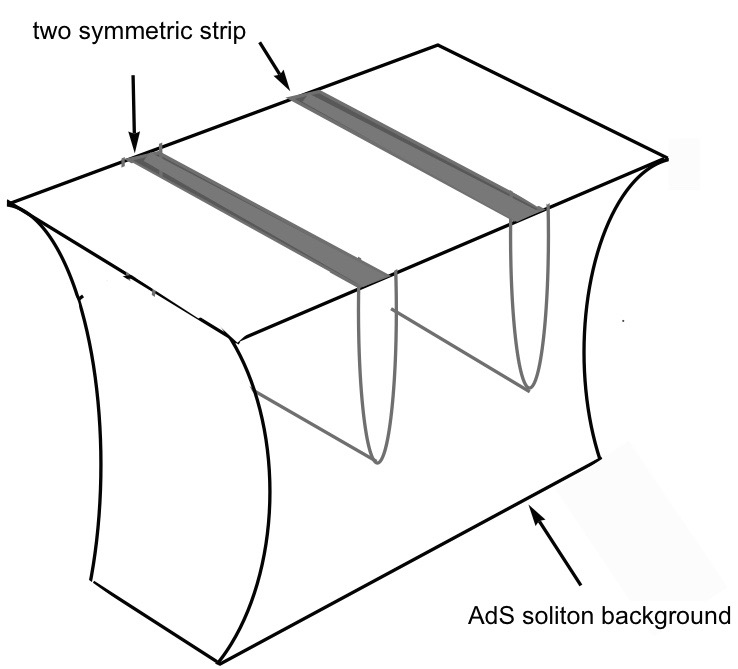}
  \caption{The picture in the left is the AdS-BH solution, the one in the middle is an AdS-soliton geometry and the one in the right shows our system of two symmetric strips in the AdS soliton background.}
 \label{fig:AdSBHsol}
\end{figure}

In order to understand this negative energy feature of the AdS soliton, in \cite{Shi:2016bxz}, the geodesic motions of massive and massless particles have been studied in this background. In the Poincare coordinate, the metrics have been written as
\begin{gather}
\text{AdS-BH} \ : \ \ ds^2=\frac{r^2}{\ell^2} (-h d\eta^2+ d \vec{x} . d \vec{x}+ dy^2 ) + \frac{\ell^2 dr^2}{r^2 h}, \nonumber\\
\text{AdS-soliton} \ : \ \ ds^2=\frac{r^2}{\ell^2} (+h dy^2+ d\vec{x} . d \vec{x} - d\eta^2) +\frac{\ell^2 dr^2}{r^2 h},
\end{gather}
where $h=1-\frac{r_0^n}{r^n}$. In that work the topologies of AdS-BH and AdS-soliton has been compared, as in figure \ref{fig:AdSBHsol}, where the horizon surface of AdS-BH has the topology of $R^{n-}$, but the $r=r_0$ ``line'' in AdS-soliton case has the topology of $R^{n-2,1}$. The two mixed systems in the form of two parallel strips with width $l$ and infinite lengths would be located on the boundary side.

Note that in AdS/CFT setup, and for the AdS soliton case, the cold and large black holes are dual to the deconfining phase and the hot and small black holes are dual to the confining phase.

\subsection{Witten-Sakai-Sugimoto model}
\begin{figure}[ht!]
 \centering
  \includegraphics[width=7.5 cm] {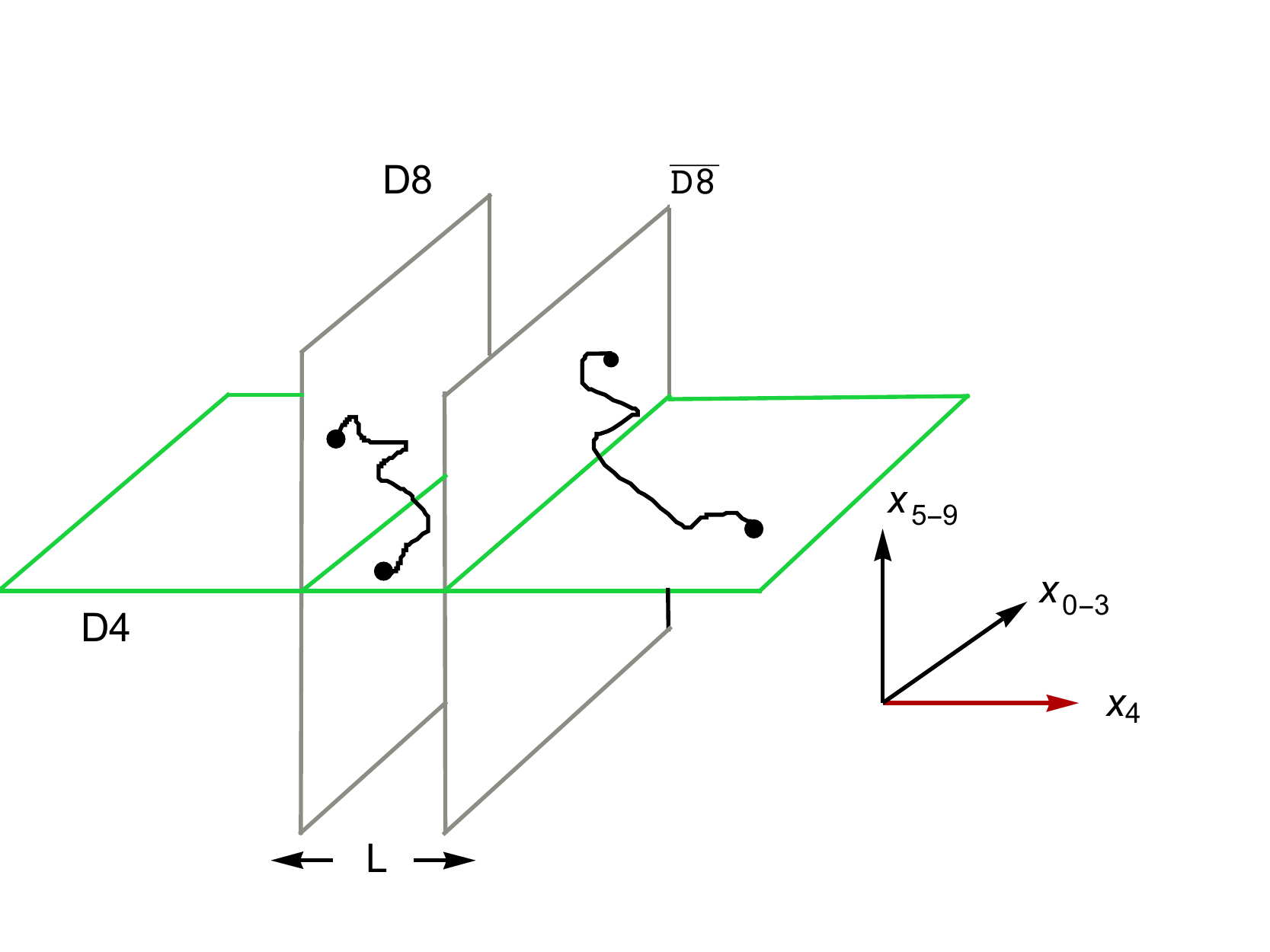}
    \includegraphics[width=7.5 cm] {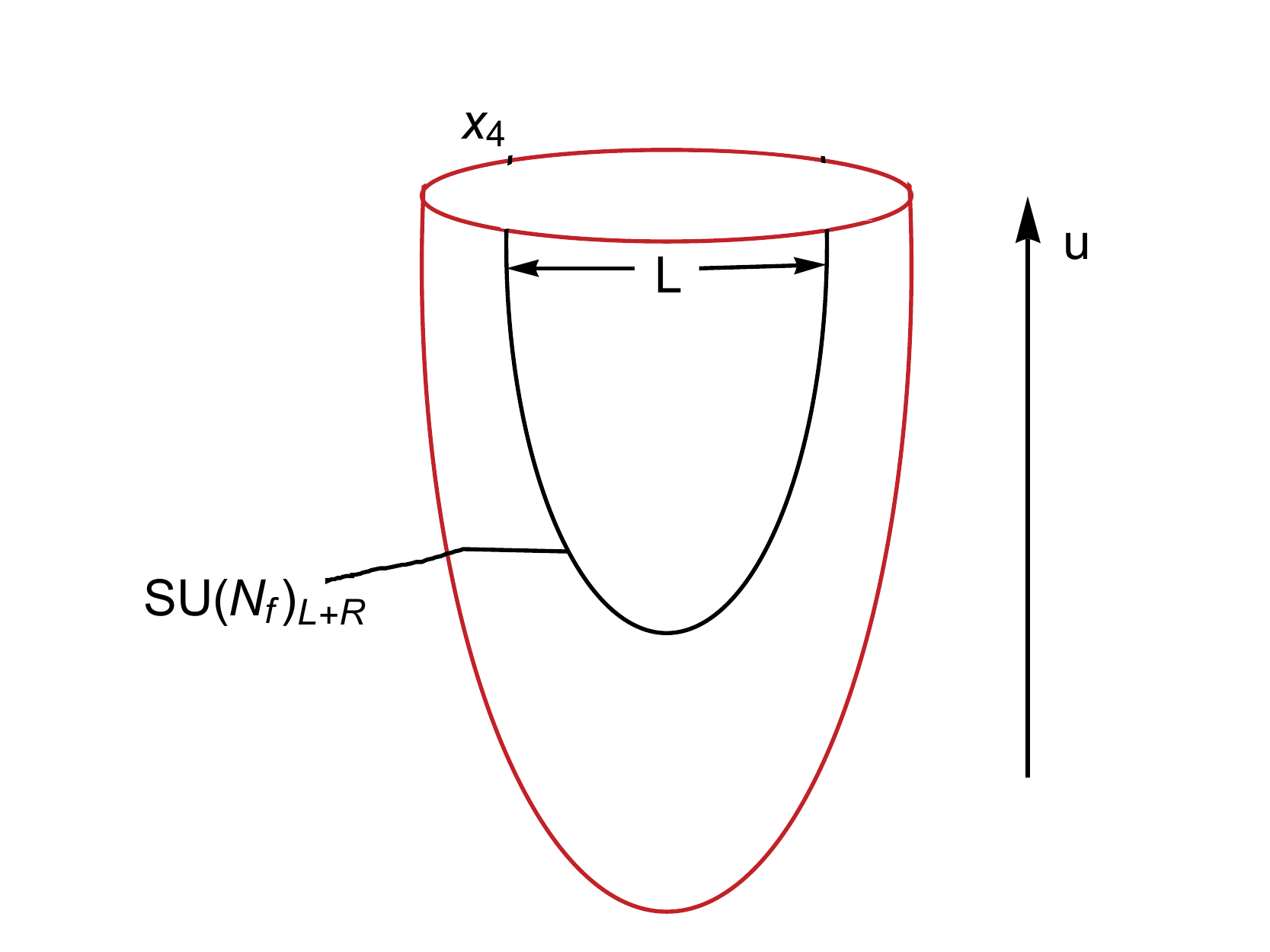}
  \caption{Intersecting D4-D8 brane model of Witten-Sakai-Sugimoto. Usually the D8-branes are treated as probe branes.}
 \label{fig:sakai}
\end{figure}

In this work we are more interested in the top-down holographic QCD models which are engineered by intersecting D-branes, like D4-D8 brane intersections known as Witten-Sakai-Sugimoto model shown in figure \ref{fig:sakai}. This theory in fact can nicely model the $SU(N_f)_L \times SU(N_f)_R$ chiral flavor symmetry breaking.

In this setup, $x^4$ is the spatial coordinate which is being compactified on $S^1$, which for fermions would have anti-periodic boundary conditions. The gauge theory is coupled to $N_f$ left-handed quarks and $N_f$ right-handed quarks which are localized at different points of the compact circle. The D8-branes are at $x^4=0$ and the anti-D8 branes are located in parallel and at $x^4=\pi R$, and $R$ is the radius of $S^1$.

In the above picture, the D4-branes correspond to color sector and the open strings which end on them produce the color/gluon gauge field.  On the other hand, the D8-branes are flavor branes and the strings ending on them give the flavor/meson gauge field. The strings which are stretched between D4 and D8 branes would then give rise to the quarks. The closed strings also correspond to the glue-balls. Using this picture,  the color/flavor duality would then mapped to open/closed string duality.

The entanglement and correlations structures among D8-D8, D4-D4 and D8-D4 branes would be different. The different correlation measures could then probe all these structures which we for the first step, we show that it can be captured first by the critical distance $D_c$.

The metric of D4-branes is
\begin{gather}
ds^2_{D4}= \left(\frac{u}{R_{D4}} \right)^{3/2} (-dt^2 +\delta_{ij} dx^i dx^j +f(u) (dx^4)^2 ) + \left(\frac{R_{D4} }{u}\right)^{3/2} \left( \frac{du^2}{f(u)} + u^2 d \Omega_4^2 \right ),
\end{gather}
where the dilaton, the Ramond-Ramond field, the function $f(u)$ and the AdS radius are defined as
\begin{gather}
e^\Phi= g_s  \left(\frac{u}{R_{D4}} \right)^{3/4}, \  \ \ F_4\equiv dC_3= \frac{2\pi N_c}{V_4} \epsilon_4, \ \ \ f(u) \equiv 1-\frac{u_{KK}^3 }{u^3}, \  \ \ R^3_{D4} \equiv \pi g_s N_c l_s^3.
\end{gather}
Here, $N_c$ is the number of colors in the gauge group, $g_s$ is the string coupling, $l_s$ is the string length which has the relation $l_s^2= \alpha^\prime$, $V_4$ is the volume of the unit four sphere $S^4$, and $s_4$ is the volume form of $S^4$. In addition, $u$ is the holographic radial direction, which is in the region of $u_{KK} \le u \le \infty$.

Then, assuming the position of D8 brane  at $x^4=0$ and anti-brane at $x^4=\pi R$, leading to relation $dx^4/du=0$, the induced metric of the D8-brane becomes
\begin{gather}\label{eq:metricd8}
ds_{D8}^2 = \left (\frac{u}{R_{D4} }  \right)^{3/2} (-dt^2+\delta_{ij} dx^i dx^j )+ \left (\frac{R_{D4} }{u}  \right)^{3/2} \left ( \frac{du^2}{f(u) } +u^2 d\Omega_4^2 \right). 
\end{gather}

As there are two metrics, one for D4-branes and one for D8 branes, here one might be confused that which one should be employed to calculate the holographic mixed quantum measures. In fact, the calculations should be done for the flavor D-branes metric which represent the quark sector, which here for the case of Sakai-Sugimoto, would correspond to the \textit{D8-branes}. In works such as \cite{Hashimoto:2014yya}, for calculating the pair production rates, also the imaginary part of D8-branes have been evaluated, which again is the main part for mixed correlations too. Therefore, for the calculation of Sakai-Sugimoto case, we consider the above metric \ref{eq:metricd8}.

We find $S_C(u_t)$, i.e, entanglement entropy at turning point $u_t$ of the strip with width $L(u_t)$ as
\begin{gather}\label{eq:SSakai}
S_C(u_t)=\frac{V_3 V_4 R_{D_4}^3 }{2 g_s^2 G_N^{(10)}}  \int_{u_t}^\infty du  \frac{u }{\sqrt{ \left(1-\frac{u_{KK}^3 }{u^3} \right)  \left(   1-  \frac{u_t^5}{u^5} \right )  } },
\end{gather}
and the width of the strip in terms of turning point as 
\begin{gather}\label{eq:LSakai}
L(u_t)= 2 R_{D_4}^{\frac{3}{2}} \int_{u_t}^ \infty du \frac{1}{\sqrt{u^3 \left(1- \frac{u_{KK}^3 }{u^3} \right)  \left( \frac{u^5}{u_t^5} -1 \right)   } }. 
\end{gather}

\begin{figure}[ht!]
 \centering
  \includegraphics[width=7.5 cm] {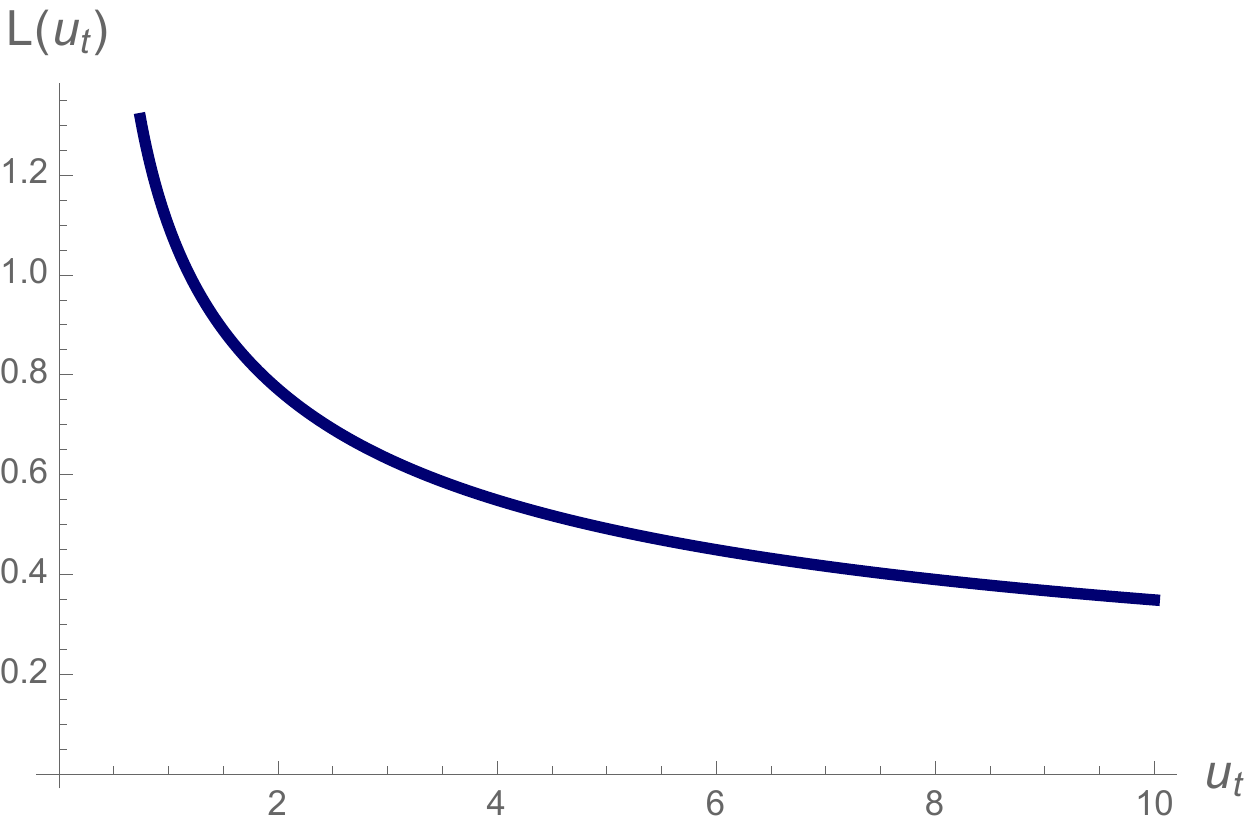} 
    \includegraphics[width=7.5 cm] {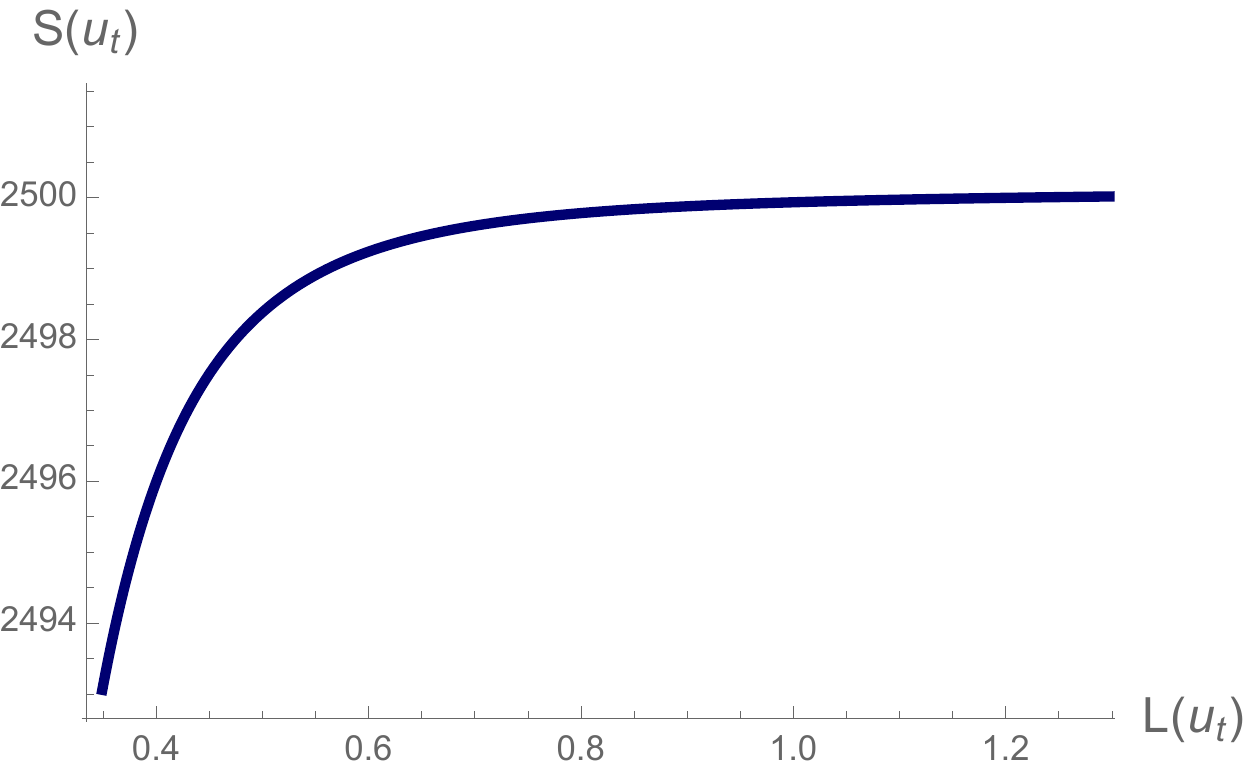} 
  \caption{The plot of $L(u_t)$ vs. turning point $u_t$ and $S(u_t)$ vs. $L(u_t)$ for the case of Sakai-Sugimoto, for $u_{KK}=0.5$.}
 \label{fig:SLsakaisugimoto}
\end{figure}

The plot of $L(u_t)$ versus turning point $u_t$ and entanglement entropy versus $L$ is shown in figure \ref{fig:SLsakaisugimoto}.  

Next, in this confining model, we would like to examine the behavior of critical distance between the two strips $D_c$ as a function of $u_{KK}$ and detect various phases in this specific background.

\begin{figure}[ht!]
 \centering
  \includegraphics[width=9 cm] {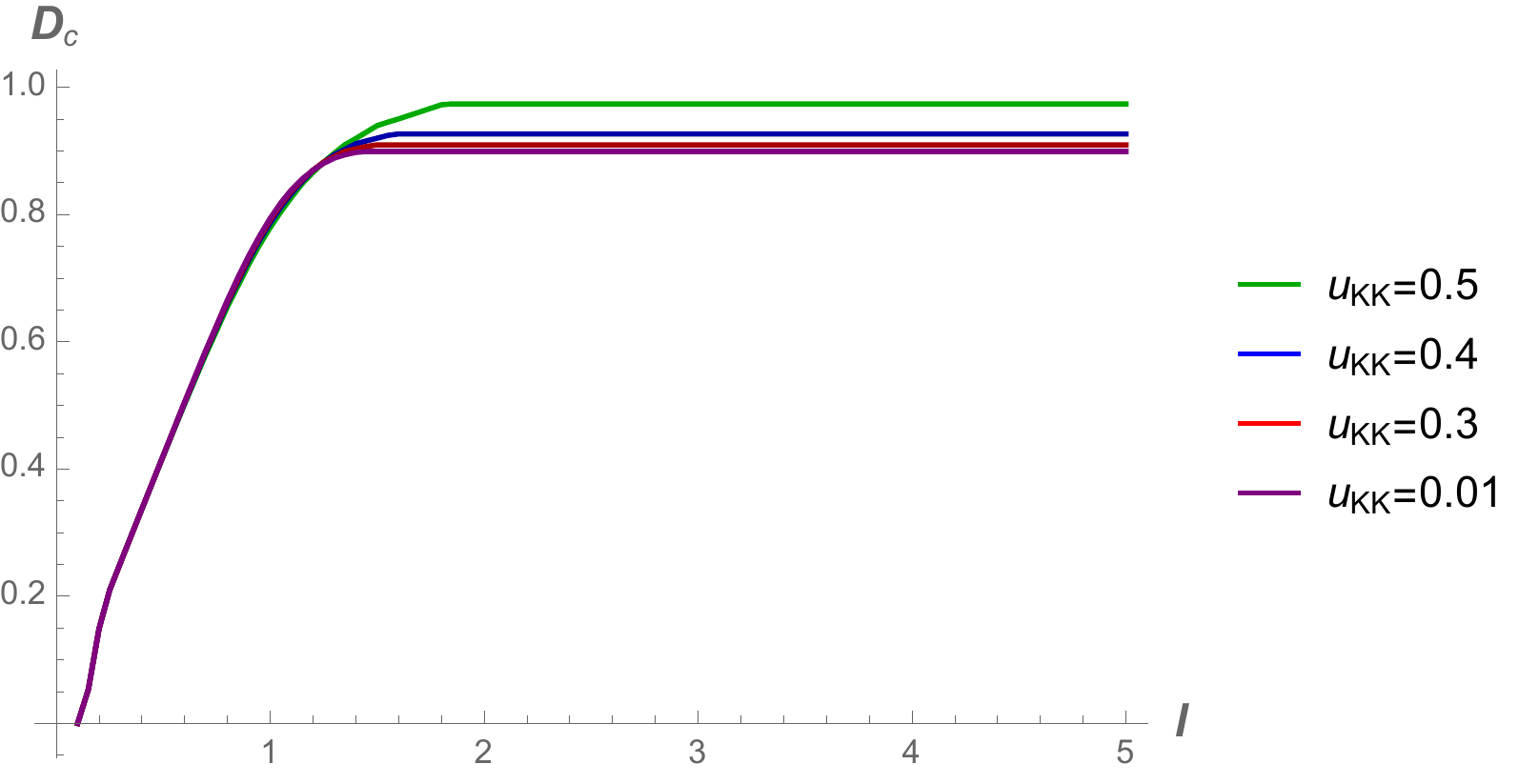} 
  \caption{The plot of $D_c$ vs. $l$ for different $u_{KK}$ in phase 1.}
 \label{fig:Dc1sakaisugimoto}
\end{figure}

Again, at least four distinct phases could be detected. In the first phase shown in figure \ref{fig:Dc1sakaisugimoto}, increasing $u_{KK}$ would increase $D_c$. When $u_{KK}$ becomes zero the critical distance would be around $D_c=0.899$. The upper bound for $u_{KK}$ in this phase is $u_{KK}=0.5$ corresponding to $D_c= 0.97$. After that increasing $u_{KK}$ decreases $D_c$ and we enter the next phase which by increasing $u_{KK}$, $D_c$ would decrease.

\begin{figure}[ht!]
 \centering
     \includegraphics[width=8.5 cm] {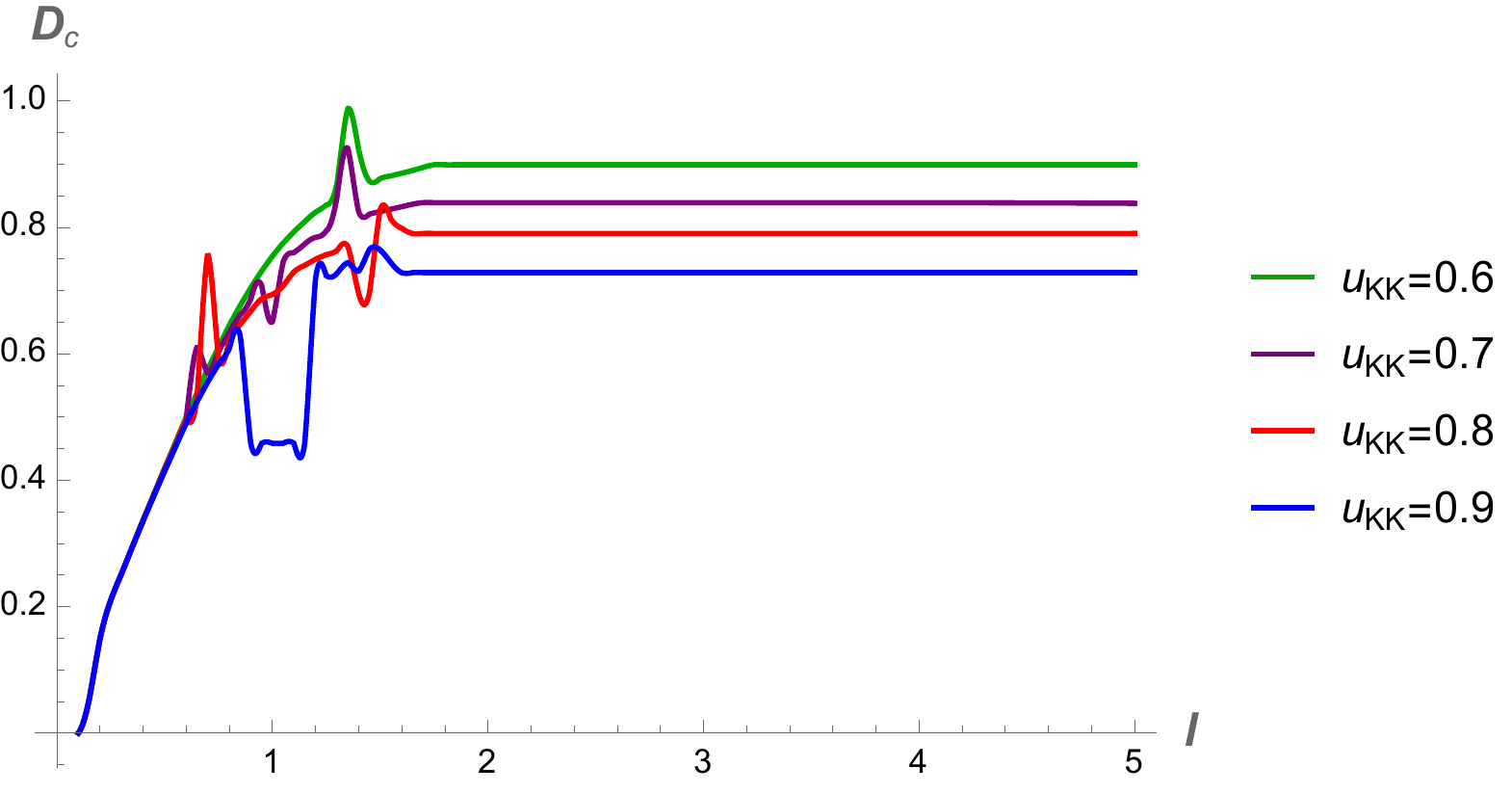} 
  \includegraphics[width=8.5 cm] {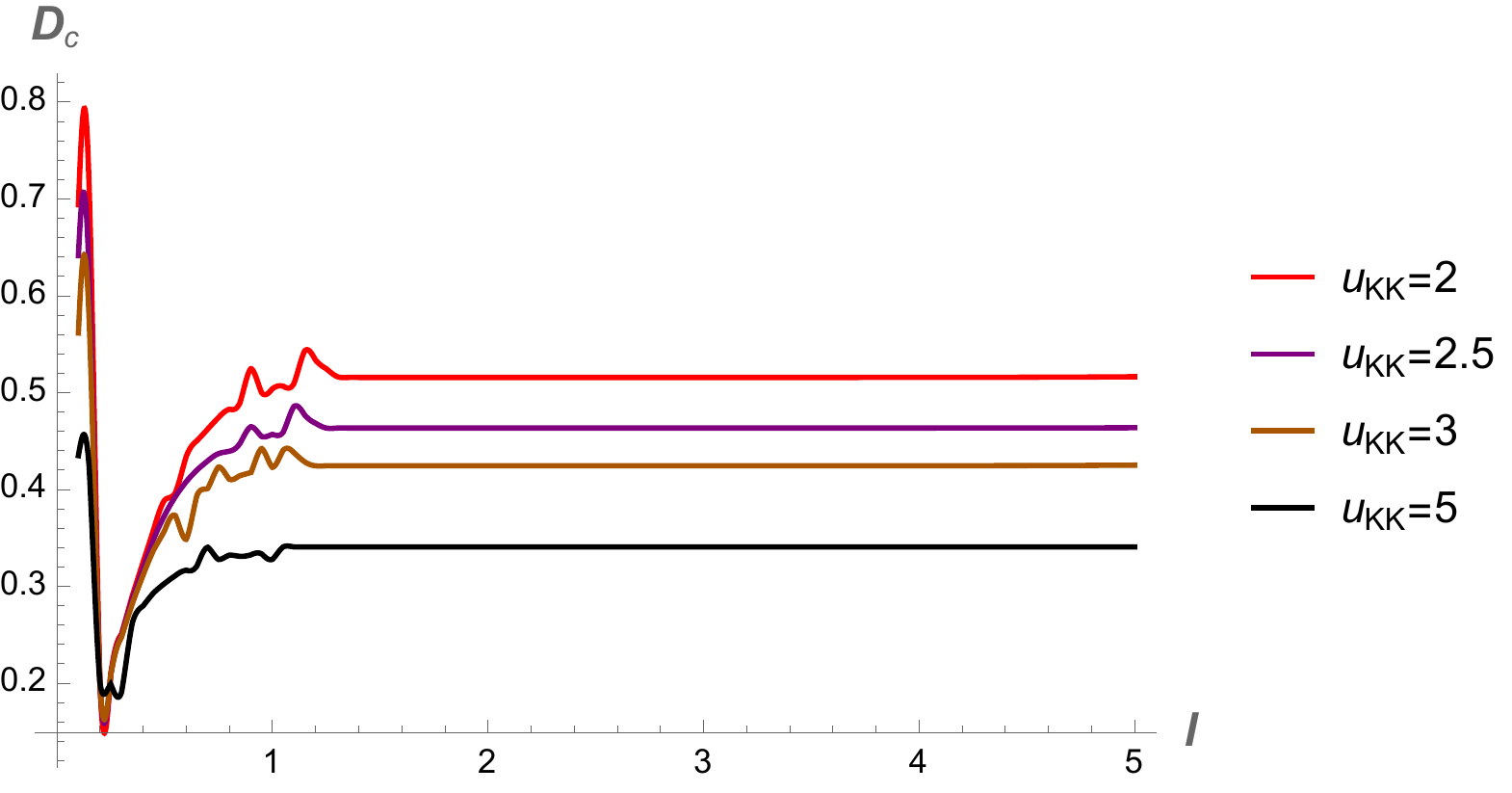} 
  \caption{The plot of $D_c$ vs. $l$ for different $u_{KK}$, in phase 2 in the left side and in phase 3 in the right side.}
 \label{fig:DPhase2Sakai}
\end{figure}

In the second phase, shown in the left side of figure \ref{fig:DPhase2Sakai}, one could note that increasing $u_{KK}$ would decrease critical distance $D_c$ unlike the previous phase of figure \ref{fig:Dc1sakaisugimoto}. Also, at smaller $I$ one could see that $D_c$ falls and then become saturated for the larger $l$. It could see that at the end of the second phase and the beginning of the third phase, i.e, for $u_{KK}=0.9$, the behavior is becoming more like the third phase in the right side. 

In the third phase shown in the right side of figure \ref{fig:DPhase2Sakai}, again one can see that after reaching its maximum at the previous phase, increasing $u_{KK}$ decreases $D_c$, and the behavior is different from the second phase.

\begin{figure}[ht!]
 \centering
  \includegraphics[width=9.5 cm] {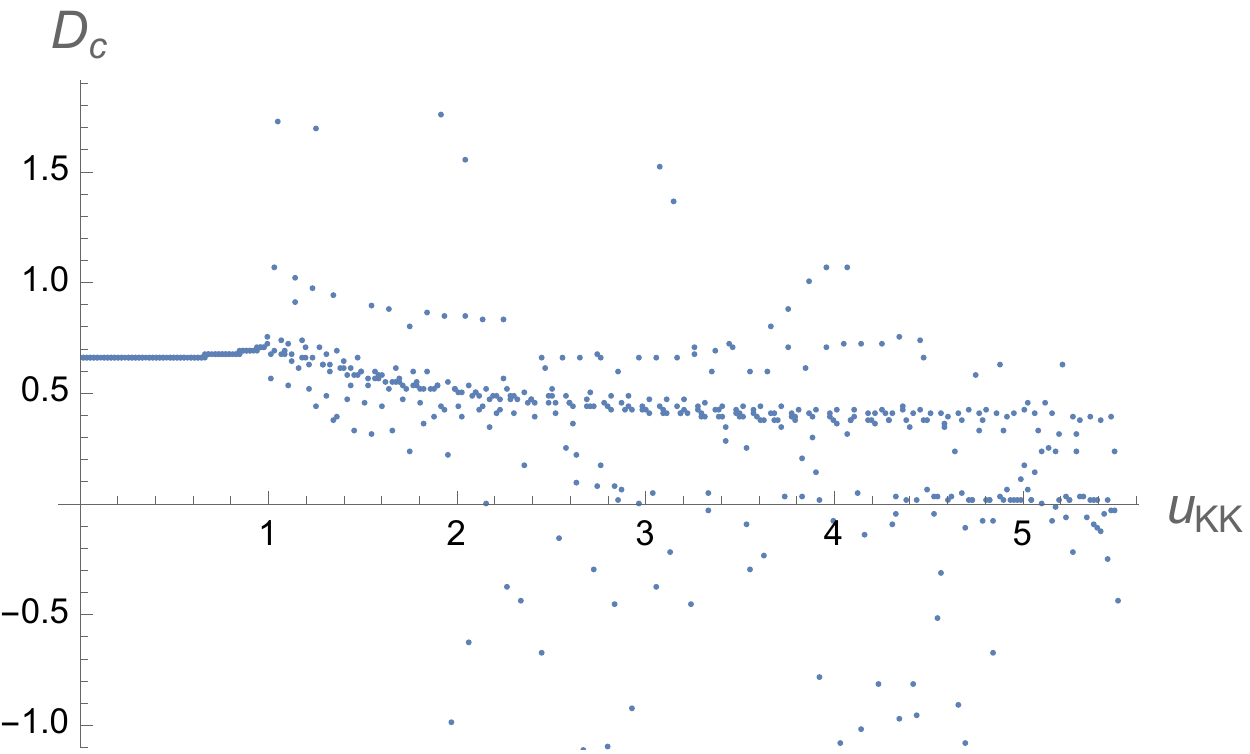} 
  \caption{The plots of $D_c$ vs. $u_{KK}$ in Witten-Sakai-Sugimoto background.}
  \label{fig:WittenSakaiDatapoint2}
\end{figure}

In figure \ref{fig:WittenSakaiDatapoint2}, we show in more details the behaviour of critical distance $D_c$ at various $u_{KK}$ for a constant $l$.  

So both the confinement and approximate spontaneous chiral symmetry breaking would affect the correlations between mixed systems in these confining backgrounds where mutual information and other mixed correlation measures such as entanglement of purification (and probably complexity of purification)  could detect their effects.

Note that in this background, the massless fields living on the D8-branes would lead to the low-spin mesons, while the strings which fall from the D8-branes down to the wall at $u=u_{KK}$ and then going back up again would lead to the higher spin mesons. This is another reason that changing the position of $u=u_{KK}$ would have such a strong effect on the entanglement and correlations between two mixed systems in this confining background as we have observed in figures \ref{fig:Dc1sakaisugimoto} and \ref{fig:DPhase2Sakai}.
Also, this change of position of $u_{KK}$ would change the behavior of the fluctuations of the gauge fields on the branes which are responsible for the creation of pseudo-vectors, scalar mesons and massless pions. These particles could also mediate information between the two mixed systems and therefore affect the phase structures of QCD probing by these quantum measures.

These phase transitions could also be examined from the point of view of traversable wormholes \cite{Maldacena:2018lmt, Numasawa:2020sty}. During these first order phase transitions,  the traversable wormholes would exchange their configurations. Also, it could be seen as Hawking-Page-like phase transition where the quark-anti-quark potential changes from the linear behavior in the low-temperature and finite values of $\sqrt{g_{tt} g_{xx}}$ and the free energy of the order $N_c^0$ in the confined phase to the decaying potential in higher temperatures and zero value of $\sqrt{g_{tt} g_{xx}}$, and free energy of the order $N_c^2$ in the deconfined phase.

In \cite{Ghodrati:2018hss}, we also established the connections between complexity and potential in the QCD models. The similarities between the behavior of potential and entanglement and complexity of purification could also be observed in these confining models,  specifically in Sakai-Sugimoto and Klebanov-Strassler case.

One should note that as mentioned in \cite{Aharony:2006da}, in these confining backgrounds there is another critical separation denoted by $L_c$ where when the quark separation is $L > L_c \simeq 0.97 * R$ the chiral symmetry would be restored at the temperature $T_d=1/2 \pi R$, and when $L < L_c \simeq 0.97 * R$ in an intermediate deconfined phase, the chiral symmetry is broken and again the chiral symmetry would be restored at $T_{\chi SB} \simeq 0.154/L$. Here $R$ is the radius of the circle where the $4+1$ supersymmetric gauge theory is compactified on.  Note also that here all the phase transitions are first order. 

The competitions between chiral symmetry breaking/restoration and the vanishing of mutual information due to increasing the distance between the two strips would in fact create the four phase transitions also observed in  \ref{fig:WittenSakaiDatapoint2}. In general the scale of chiral symmetry breaking and confinement would be independent, as the scale of chiral symmetry breaking would mostly depend on the distance between D8-branes or the fermions on the compactified circle, while the scale of confinement would depend on the radius where D4-brane is wrapped around, or the radius of the circle which the $SU(N_c)$ gauge theory is compactified on it.

So in the Sakai-Sugimoto model, the chiral symmetry breaking/restoration which is related to the mass scale of mesons, ($\frac{1}{L}$), is related to the distance between D8 branes. More precisely, the low spin mesons are described by strings attached between the two D8 branes while the high spin mesons fall from D8 branes on the wall at $u=u_{KK}$, being completely stretched on the wall and then go back to the D8 brane.  The confinement/deconfinement phase transition happens at the scale which is related to the mass of glueballs,$\frac{1}{R}$, which is related to the radius where D4 brane is wrapped around.

Also, note that in this model, the unstable modes have been observed and studied in the literature which specifically in high-energy regimes could significantly affect the quantum measures that we use here.

The relation for the minimal wedge cross section could be found using the relation
\begin{gather}
\Gamma_{WSS}= R_{D4}^3 \int_{u_D}^{u_{2l+D}} du \frac{  u ^5 }{1-\frac{u^3_{KK} }{u^3} }.
\end{gather}

Based on the AdS/CFT dictionary we have relations between the parameters $M_{KK}, g_{YM}, N_c$ in the boundary gauge side and $R_{D_4}, u_{KK}, g_s$ in the gravity side as
\begin{gather}
R_{D_4}^3=\frac{1}{2} \frac{\lambda l_s^2}{M_{KK}}, \ \ \ u_{KK}=\frac{2}{9} \lambda M_{KK} l_s^2, \ \ \ g_s=\frac{1}{2\pi} \frac{\lambda}{M_{KK} N_c l_s  }.
\end{gather}

The 't Hooft coupling $\lambda$ is $\lambda= g_{YM}^2 N_c $ and the gauge coupling $g_{YM}$ at the cutoff scale $M_{KK}$ can be written as $g_{YM}^2=(2\pi)^2 g_s l_s/ \delta x^4 $ and also we have the relation $M_{KK}\equiv \frac{2\pi}{\delta x^4}  $\cite{Hashimoto:2014yya}.

\begin{figure}[ht!]
 \centering
  \includegraphics[width=7 cm] {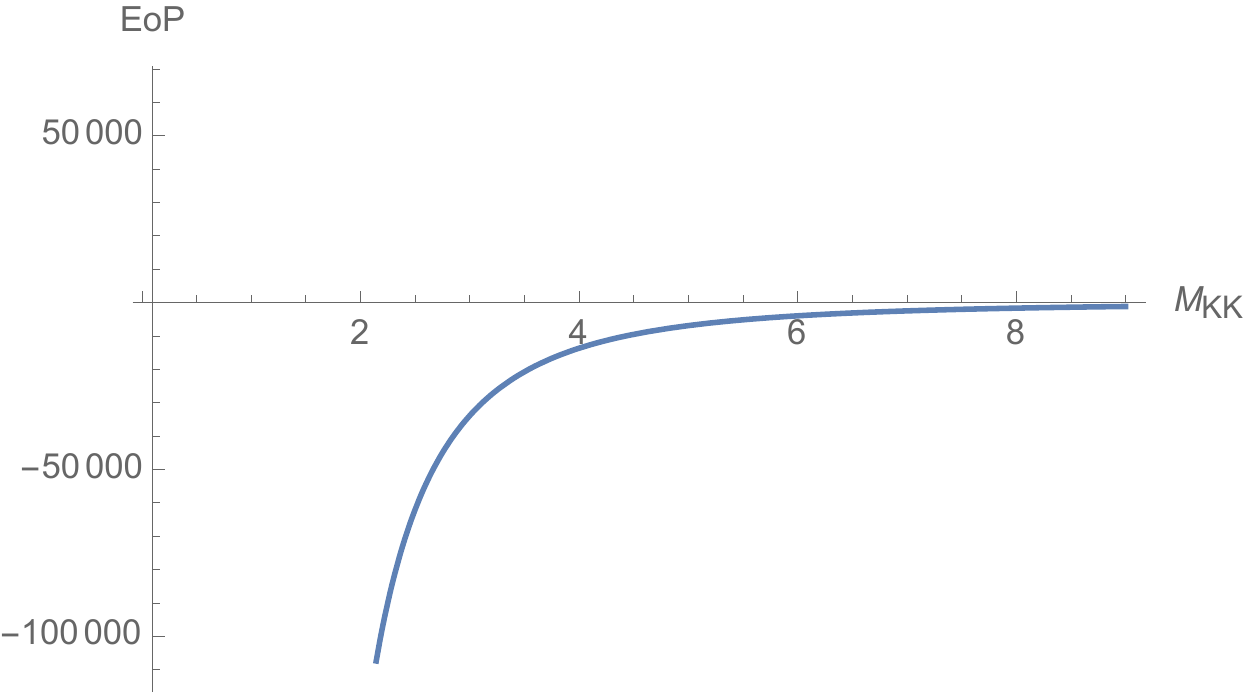} 
    \includegraphics[width=7 cm] {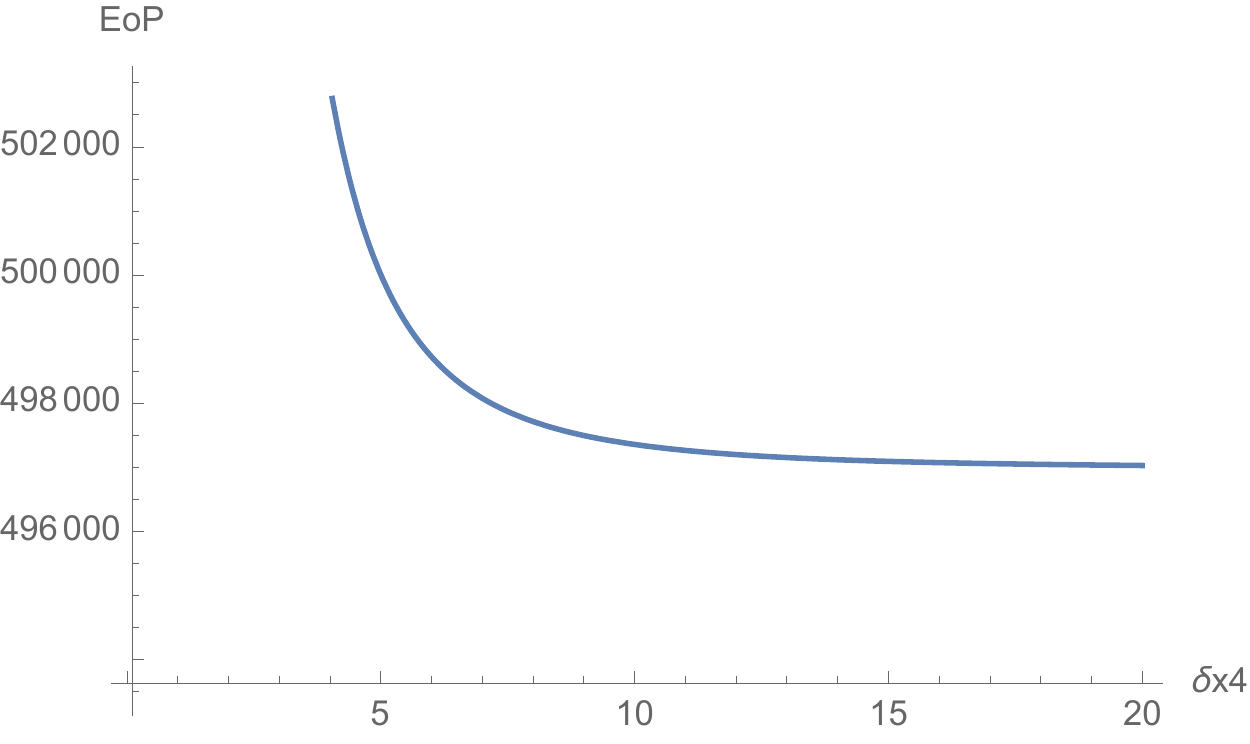} 
  \caption{The behavior of entanglement of purification versus $M_{KK}$ and $\delta x_4$ }
  \label{fig:dEoPMKK}
  \end{figure}

It could be seen that the string coupling $g_s$, the number of colors of gauge group $N_c$, the string length $l_s$ and the cutoff energy $M_{KK}$ increase the entanglement of purification while the periodicity of the boundary condition $\delta x^4$ would have an inverse effect and decreases the EWCS. The increasing behavior of EWCS versus $M_{KK}$ and its decrease versus $\delta x^4$ are shown in figure \ref{fig:dEoPMKK}. 

\subsubsection{Deformed Sakai-Sugimoto}
One can then consider the deformed Sakai-Sugimoto geometry \cite{Aharony:2006da,Hashimoto:2014yya}, where $x^4$ instead of being constant, depends on the coordinate $u$. 

\begin{figure}[ht!]
 \centering
  \includegraphics[width=7 cm] {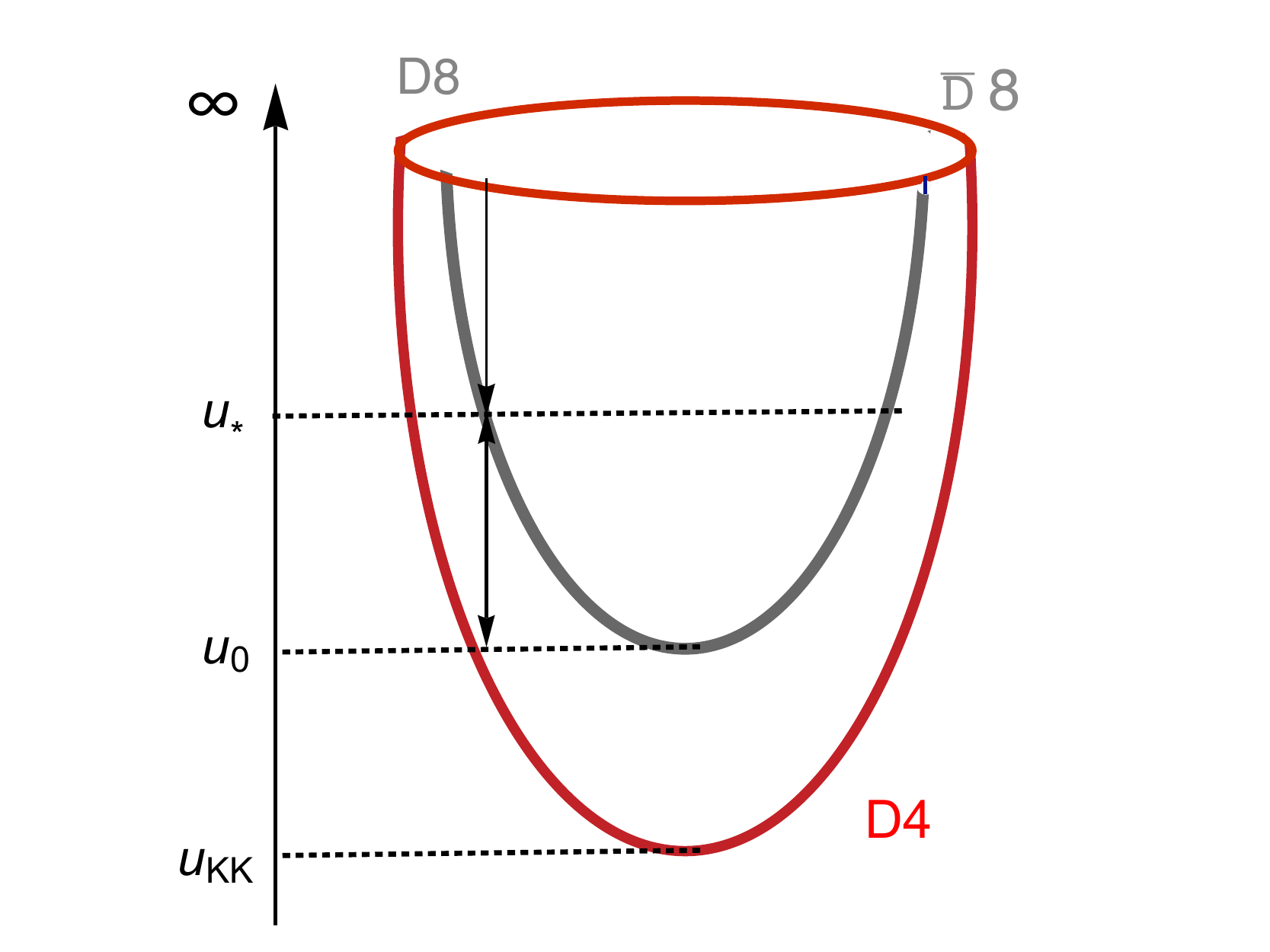} 
  \caption{The geometry of deformed Sakai-Sugimoto.}
  \label{fig:deformedSakai}
  \end{figure}

It has the configuration as shown in figure \ref{fig:deformedSakai} and its metric is as follows
\begin{gather}
ds^2_{D8}= \left ( \frac{u}{R_{D4}} \right)^{3/2} (-dt^2+\delta_{ij} dx^i dx^j ) +\left ( \frac{u}{R_{D4}} \right)^{3/2} \frac{du^2}{h(u)} +\left (\frac{R_{D4} }{u} \right)^{3/2} u^2 d \Omega_4^2,
\end{gather}
where 
\begin{gather}
h(u) \equiv \left \lbrack f (u) \left ( \frac{dx^4 (u) }{du} \right )^2 + \left ( \frac{R_{D4} }{u} \right)^3 \frac{1}{f(u)} \right \rbrack ^{-1}.
\end{gather}

The entanglement entropy of a strip in terms of the turning point is
\begin{gather}
S(u_t)=\frac{V_3 V_4 R_{D_4}^3 }{2 g_s^2 G_{N}^{(10) }  } \int_{u_t} ^\infty du \frac{u}{\sqrt { (1-\frac{u_{KK}^3 }{u^3} ) (1-\frac{u_t^5}{u^5} ) } } + \frac{V_3 V_4 R^{\frac{3}{2} }_{D_4} }{2 g_s^2 G_N^{(10) } } \int_{u_t}^ \infty du \left( \frac{dx^4 (u)}{du}  \right) \sqrt{ \frac{u^5 (1-\frac{u_{KK}^3 }{u^3} ) }{1-\frac{u_t^5}{u^5} }  }.  
\end{gather}

Note that the first term is similar to the previous usual Sakai-Sugimoto case, while in the deformed case the second term is added which increases the entanglement entropy compared to the previous case of relation \ref{eq:SSakai}. The gradient of the coordinate $x^4 (u)$ versus $u$ controls the difference.

Similarly, the width of the strip versus the turning point $u_t$ could be written as

\begin{gather}
L(u_t)= 2 R_{D_4}^{\frac{3}{2}} \int_{u_t}^ \infty du \frac{1}{\sqrt{u^3 \left(1- \frac{u_{KK}^3 }{u^3} \right)  \left( \frac{u^5}{u_t^5} -1 \right)   } }+2\int_{u_t}^\infty du \left( \frac{dx^4 (u) }{du} \right)  \sqrt{\frac{1-\frac{u_{KK}^3 }{u^3} }{\frac{u^5}{u_t^5} -1 } },
\end{gather}
which again the second term has been added to the previous case of relation \ref{eq:LSakai}.

For this case then, the relation for the minimal wedge cross section would be
\begin{gather}
\Gamma_{\text{Deformed SS}}=  R_{D4}^3 \int_{u_D}^{u_{2l+D}} du \frac{  u ^5 }{1-\frac{u^3_{KK} }{u^3} } +\int_{u_D}^{u_{2l+D}} du u^8 \left(1- \frac{u^3_{KK} }{u^3} \right)  \left(\frac{dx^4(u)}{du} \right)^2.
\end{gather}

So again in the deformed case the second term is being added. Therefore, the minimal wedge cross section of deformed Sakai-Sugimoto is bigger than the normal case, or rather the entanglement of purification of two mixed system is higher and the difference is being controlled by the derivative of $x^4(u)$ relative to $u$. 

By calculating the DBI action and then calculating its Hamiltonian which is conserved one would specifically get the relation for $u' =du/dx^4$ \cite{Aharony:2006da} as
\begin{gather}\label{eq:deformedrel}
u'^2= f^2 (u) \left ( \frac{u}{R_{D_4} } \right)^3 \left ( \frac{f(u)}{f(u_0)} \frac{u^8}{u_0^8}-1 \right ),
\end{gather}

Then, we can insert the inverse of this relation into all the previous results for the deformed Sakai-Sugimo and find the numerical solutions for this case as well. However, since the second integral contains a square term, the numerical solutions become much more complicated and noisy, and we leave its direct calculations for the future works.

\subsection{Witten-QCD}
In the string frame, the metric of Witten-QCD and its dilaton field is written as \cite{Witten:1998zw, Itzhaki:1998dd, Aharony:2002up,  Witten:1997ep},
\begin{gather}
ds^2=(\frac{u}{R})^{3/2}\left(\eta_{\mu\nu} dx^\mu dx^\nu+\frac{4R^3}{9u_t} f(u) d\theta^2\right)+\left(\frac{R}{u}\right)^{3/2} \frac{du^2}{f(u)}+R^{3/2} u^{1/2} d\Omega_4^2,\nonumber\\
f(u)=1-\frac{u_t^3}{u^3}, \ \ \ \ \ \ \ \ R=(\pi Ng_s)^{\frac{1}{3}} {\alpha^\prime}^{\frac{1}{2}},\ \ \ \ \ \
e^\Phi=g_s \frac{u^{3/4}}{R^{3/4}}.
\end{gather}

 \begin{figure}[ht!]
 \centering
  \includegraphics[width=7.5 cm] {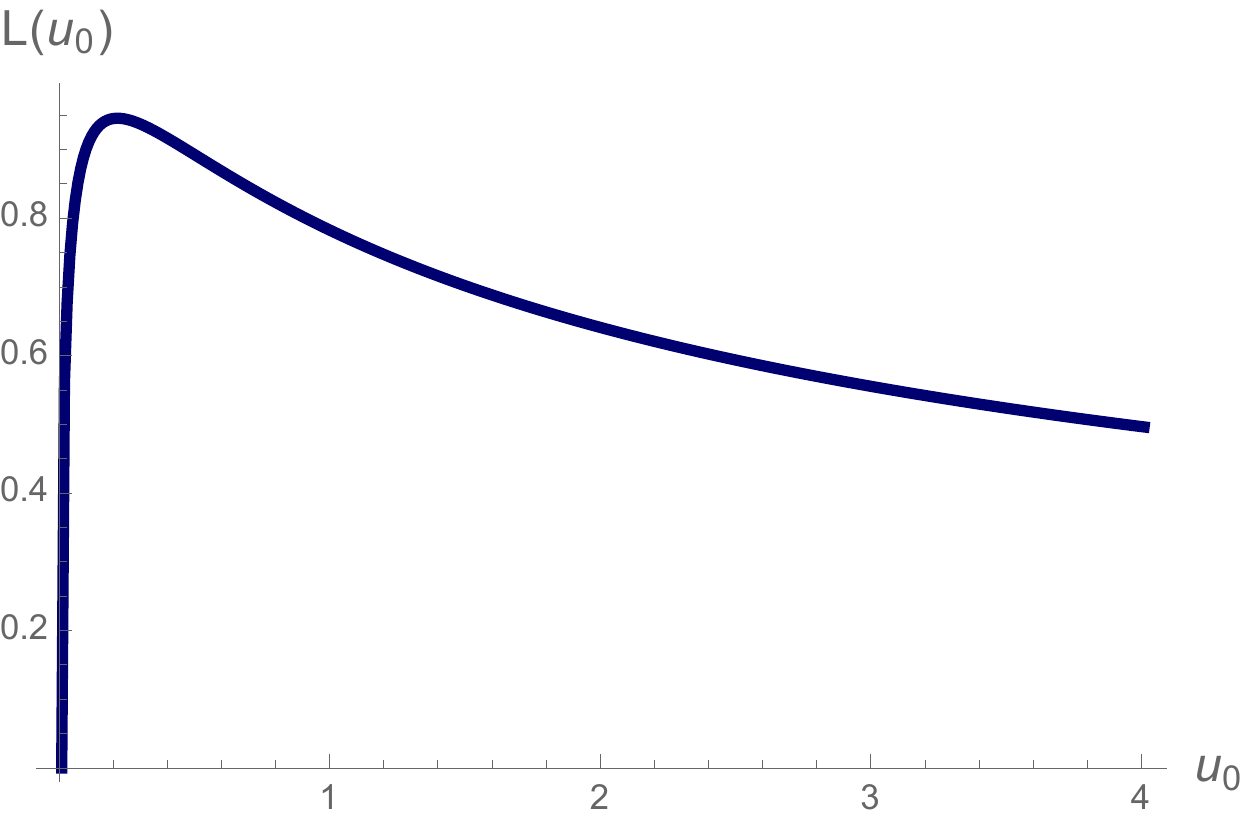} 
    \includegraphics[width=7.5 cm] {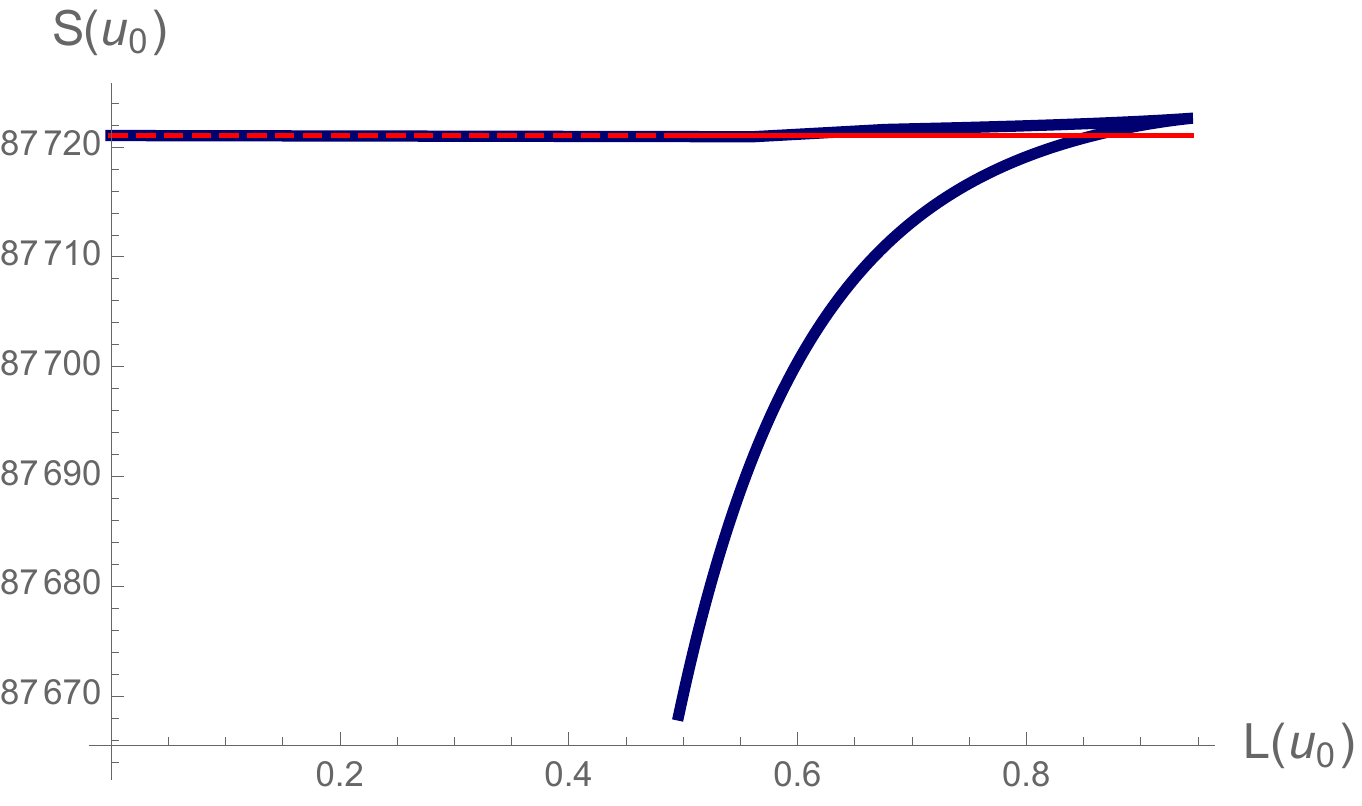} 
  \caption{The relationship between width of a strip versus turning point $\tau_0$ (in the left), and entanglement entropy versus $L$ (in the right) in the background on Witten-QCD.}
 \label{fig:StripinKS}
\end{figure}

 \begin{figure}[ht!]
 \centering
  \includegraphics[width=7.6 cm] {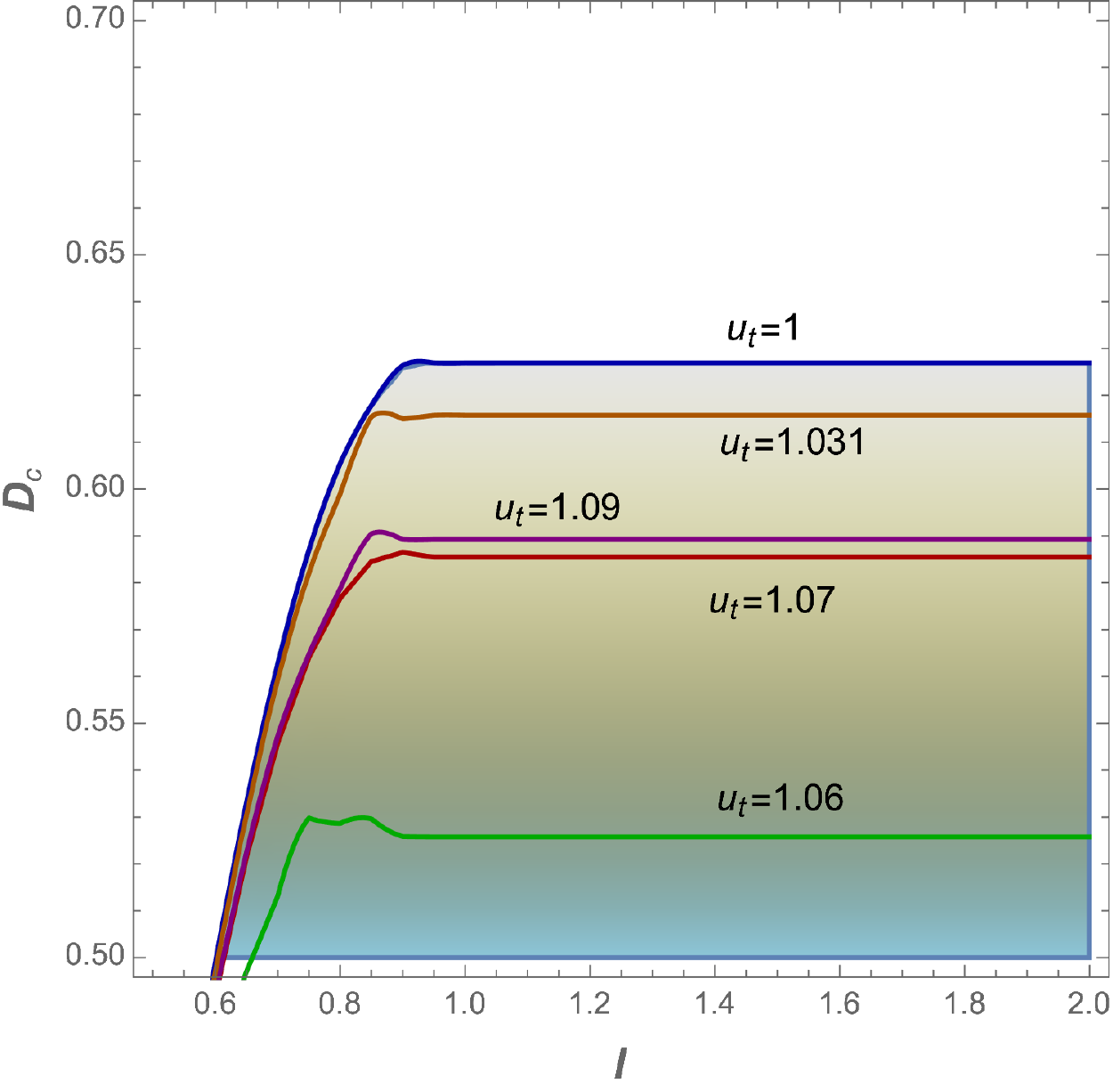} 
  \caption{The behavior of $D_c$ vs $l$ for various $u_t$ is shown where the region for non-zero mutual information for each parameter is the region covered below each curve.}
  \label{fig:WittenQCDcrit}
  \end{figure}

 \begin{figure}[ht!]
 \centering
  \includegraphics[width=10.5 cm] {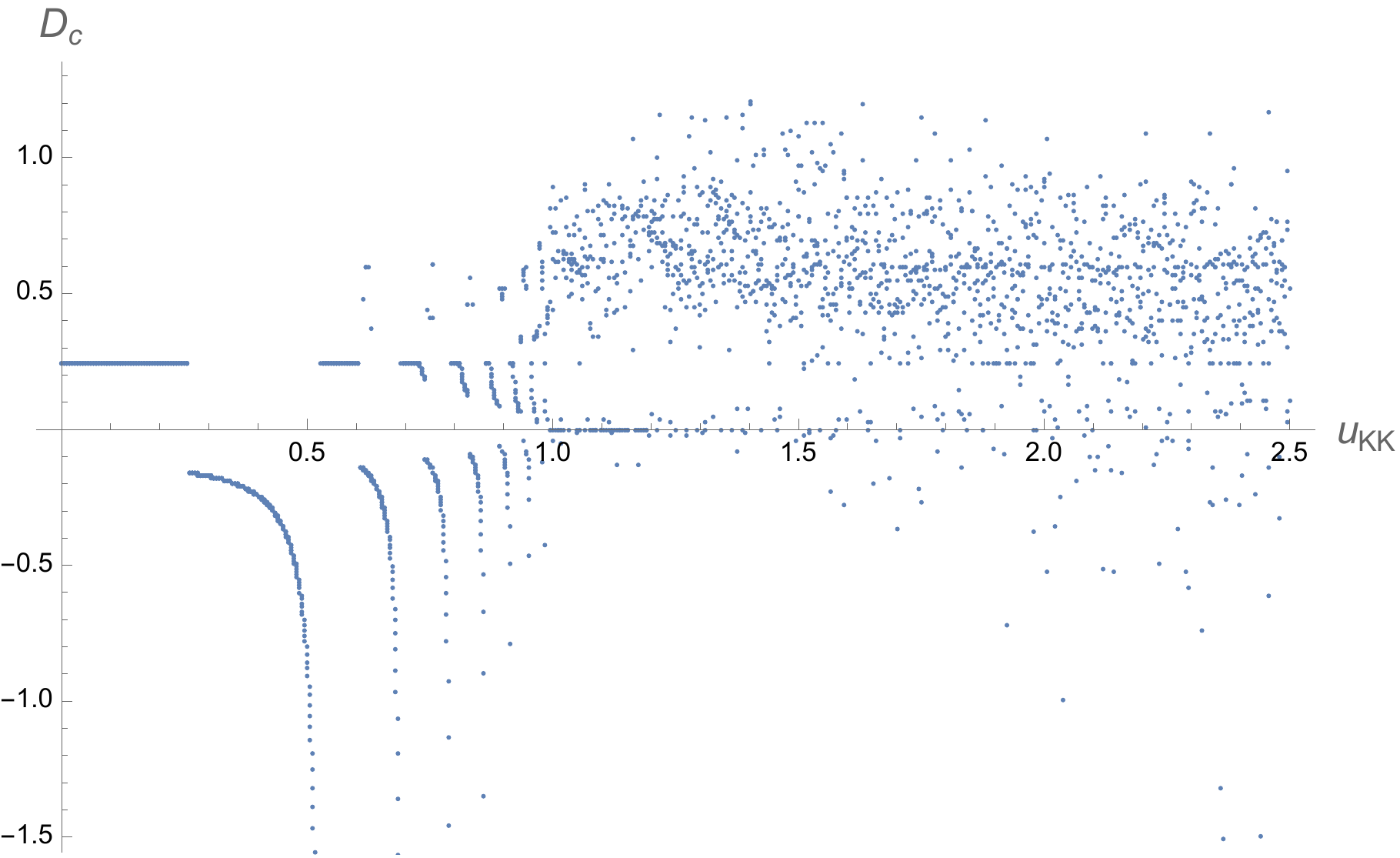} 
  \caption{The behavior of $D_c$ versus $u_{KK}$ in Witten-QCD background.}
  \label{fig:WittenQCDdatapoints}
  \end{figure}

The turning point of the RT surface depends on the width of the strip $L(u_0)$ and has been found in \cite{Kol:2014nqa, Ghodrati:2015rta} as
\begin{gather}
L(u_0)=2\int_{u_0}^\infty du \sqrt{ \frac{(\frac{R}{u} )^3 \frac{1}{f(u)}  } {\frac{f(u) u^5}{f(u_0) u_0^5}-1} },
\end{gather}
and the entropy of the connected solution $S_C$  would be \cite{Ghodrati:2015rta}
\begin{gather}
S_C(u_0)= \frac{V_2}{G_N^{(10)} }  \frac{8 \pi^2 R^{\frac{9}{2} } }{9g_s^2 \sqrt{u_t} } \int_{u_0}^\infty du \sqrt{\frac{u^2}{1- \frac{u_0^5}{u^5}. \frac{f(u_0)}{f(u)} } }. 
\end{gather}

Using the above relations, the phase structures can be numerically probed where the results are shown in figures \ref{fig:WittenQCDcrit} and \ref{fig:WittenQCDdatapoints}. 

The behavior of $D_c$ versus $u_{KK}$ in Witten-QCD background is shown in figure \ref{fig:WittenQCDdatapoints}. Note that specifically in figure \ref{fig:WittenQCDdatapoints}, the first phase is when $D_c$ versus $u_{KK}$ is constant, then, around $u_{KK}=1$ there is a drop and then a singularity or rather a jump which is the second phase, then another one around $u_{KK}=1.3$ and  the next around  $u_{KK}=1.7$, and then $D_c$ decreases by increasing $u_{KK}$ which is the last phase, a similar behavior to AdS-soliton case.

Then, if we use relation \ref{eq:generalizedeEW}, for this case, and then assuming that the coordinate $\theta(u)$ is the only coordinate as the function of $u$, then the minimal wedge cross section $\Gamma$ would be found as
\begin{gather}
\Gamma_{\text{WQCD}}=\frac{2 R^{\frac{9}{2} } }{3 \sqrt{u_t} g_s^2 }  \int_{u_D}^{u_{2l+D} } du u^{\frac{7}{4} } \sqrt{1+\frac{4}{9u_t} \left(\frac{d\theta}{du}\right )^2 f^2(u) },
\end{gather}
which its relation with various parameters of the theory could be seen.

\subsection{Klebanov-Strassler}

Another interesting geometry worths to study and check the behavior of quantum information measures and $D_c$ in it, would be the Klebanov-Strassler (KS) case \cite{Klebanov:2000hb}. This type-IIB supergravity solution has various fluxes and warp factors and are dual to the confining $\mathcal {N}=1$ SYM theories which describe the baryonic branch of the dual gauge theory.  

The KS metric is actually being constructed by the collection of N regular and M fractional D3-branes in the geometry of the deformed conifold. The fractional D3-branes could be thought as D5-branes that wrap the two-cycle of $T^{1,1}$ which collapse at the apex $\tau=0$ in a way that the world-volume would effectively becomes that of a 3-brane.

There are two free parameters in the KS model which the dual interpretations of them are baryonic VEV and the gauge coupling constant. The Maldacena-Nunez solution which we study later, could also be considered as the end point of a flow from KS by changing these two parameters.

The gauge group on the stack of N D3-branes and M fractional D3-branes is $SU(N+M) \times SU(N)$ \cite{Gubser:1998fp}. 
The limit where $M <<N$ corresponds to the Klebanov-Tseytlin solution which we also check later. 

The fluxes follow the relations
\begin{gather}
\int_{T^{1,1}} F_5 \propto N, \ \ \ \int_{S^3} F_3 \propto M,
\end{gather}
and the NSNS-flux would be
\begin{gather}
H_3 \propto \frac{g_s M}{\tau} d\tau \wedge \omega_2, \ \ \ \ \  \text{therefore:}  \ \ \int_{S^2} B_2 = g_s M \log (\tau/\tau_0),
\end{gather}
where $\omega_2$ is the volume form of the two-sphere of $T^{1,1}$ and $\tau_0$ is just an integration constant.

The form of the metric is
\begin{gather}
ds_{\text{def} }^2= \frac{1}{2} \epsilon^{\frac{4}{3} } K \Big \lbrack \frac{1}{3K^3} (d\tau^2+g_5^2) + \sinh^2 (\frac{\tau}{2} ) (g_1^2 +g_2^2) +\cosh^2 (\frac{\tau}{2} ) (g_3^2+g_4^2) \Big \rbrack, 
\end{gather}
and $K$ is a decreasing function of the radial coordinate $\tau$, as
\begin{gather}
K(\tau)= \frac{(\sinh 2\tau - 2 \tau )^{\frac{1}{3}}  }{2^{\frac{1}{3} } \sinh \tau}. 
\end{gather}

\begin{figure}[ht!]
 \centering
  \includegraphics[width=7 cm] {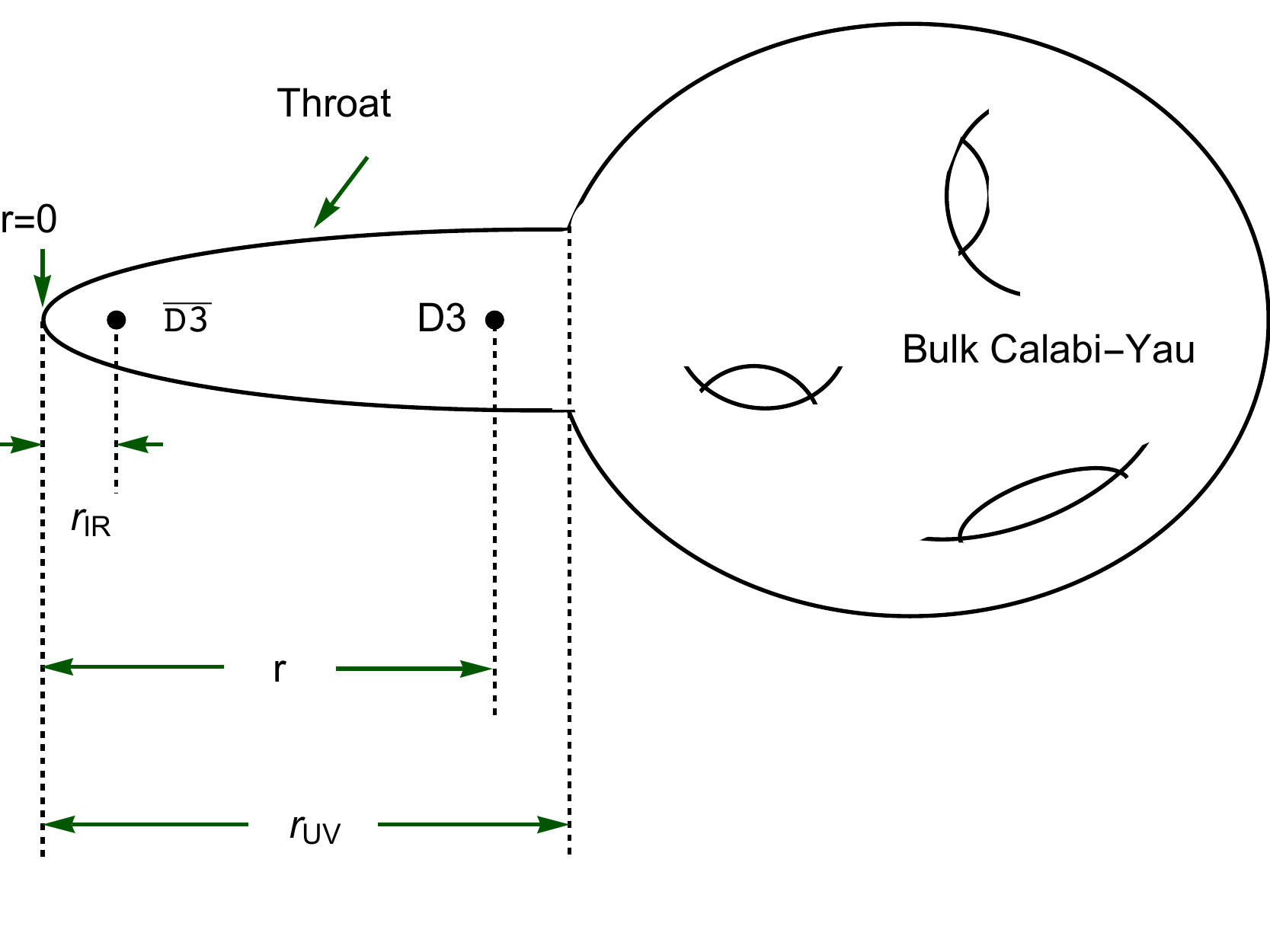} 
  \caption{Klebanov-Strassler is the part of the warped throat in this geometry.}
 \label{fig:Klebanov}
\end{figure}

For these confining geometries which has an exotic throat as shown in figure \ref{fig:Klebanov}, the tools of bulk reconstruction such as Kinematic space, Crofton form, tensor networks or even the form of bit threads or ``quantum bit threads'' \cite{Rolph:2021hgz,Agon:2021tia}, entanglement contours and Page curves could give further information about the interplay between geometry and topology in the bulk, and the information in the boundary. The complicated relation for the Crofton form of KS metric is presented in equation \ref{deformedcroft}. 

The Klebanov-Strassler (KS) metric which is known also as warped deformed conifold is obtained by a collection of $N$ regular and $M$ fractional D3-branes  \cite{Klebanov:2000hb}. 

The metric is
\begin{gather}
ds_{10}^2=h^{-\frac{1}{2}}(\tau) dx_\mu dx^\mu+h^{\frac{1}{2}}(\tau) {ds_6}^2,
\end{gather}
where again the $ds_6^2$ is the metric of the deformed conifold which is
\begin{gather}
{ds_6}^2=\frac{1}{2} \epsilon^{\frac{4}{3}} K(\tau) \Big[ \frac{1}{3 K^3(\tau)} (d\tau^2+ (g^5)^2 )+\cosh^2 (\frac{\tau}{2}) [(g^3)^2+(g^4)^2]+\sinh^2 (\frac{\tau}{2}) [(g^1)^2+(g^2)^2] \Big].
\end{gather}
The parameters of the metric are
\begin{gather}
K(\tau)=\frac{(\sinh(2\tau) -2\tau)^{\frac{1}{3}}} {2^{\frac{1}{3}} \sinh \tau },\ \ \ \ \ \ \ \ \ h(\tau)=(g_s M \alpha')^2 2^{2/3} \epsilon^{-8/3} I(\tau),\nonumber\\  I(\tau)=\int_\tau^\infty dx \frac{x \coth x-1}{\sinh^2 x} (\sinh(2x)-2x)^{\frac{1}{3}}, 
\end{gather}
and
\begin{gather}
g^1=\frac{1}{\sqrt{2}} [ -\sin \theta_1 d\phi_1-\cos \psi \sin\theta_2 d\phi_2+\sin \psi d\theta_2],\ \ \ \ \
g^2=\frac{1}{\sqrt{2}}[d\theta_1 -\sin \psi \sin\theta_2 d\phi_2-\cos\psi d\theta_2],\nonumber\\
g^3=\frac{1}{\sqrt{2}}[-\sin\theta_1 d\phi_1+\cos \psi \sin \theta_2 d\phi_2-\sin\psi d\theta_2],\ \ \ \ \ 
g^4=\frac{1}{\sqrt{2}}[d\theta_1 +\sin \psi\sin\theta_2 d\phi_2+\cos \psi d\theta_2],\nonumber\\
g^5=d\psi+\cos \theta_1 d\phi_1+\cos\theta_2 d\phi_2.
\end{gather}

Note that $h(\tau)$ is  the warp factor here. Also, as mentioned, $ds_6^2$ is the deformed conifold which is a cone over $S$ pace that is topologically $S^3 \times S^2$. The $(g_1, g_2)$ would build the $S^2$ part which shrinks to zero size at the apex of the deformed conifold while $(g_3, g_4, g_5)$ build the $S^3$ part which have a finite size.

In \cite{Basu:2012ae}, it has been mentioned that the IR behavior of the function $I$ acts as $I(\tau \to 0) \to a_0-a_2 \tau^2 + \mathcal{O} (\tau^4)$, and since there is no linear term in $\tau$, the expansion is really around the end of space and the Wilson loop would find it more favorable to arrange themselves there. This point then should show its effects on the behaviours of other quantum information measures such as entanglement of purification, mutual information and negativity for the two strips in the mixed setup as we examine here.

The length of the strip versus turning point of the RT surface can be written as \cite{Ghodrati:2019hnn}
\begin{gather}
L(\tau_0)=\frac{2^{\frac{5}{6} } \epsilon^{\frac{2}{3} } }{\sqrt{3}} \int_{\tau_0}^\infty d\tau \frac{\sinh(\tau)  }{ (\sinh (2\tau) -2\tau )^{\frac{1}{3} } }  \sqrt{\frac {h(\tau) }{\frac{\sinh^2(\tau) }{\sinh^2(\tau_0)} \left(\frac{\sinh(2\tau)-2\tau}{\sinh(2\tau_0)-2\tau_0}\right )^{\frac{2}{3}} -1} },
\end{gather}
and the entanglement entropy for the connected solution is
\begin{gather}
S_C(\tau_0)=\frac{V_2 \pi^3 \epsilon^4}{3 G_N^{(10)} } \int_{\tau_0}^\infty d\tau \frac{h(\tau) \sinh^2 (\tau) }{\sqrt{1- \left(\frac{\sinh \tau_0}{\sinh \tau} \right)^2  \left(\frac{ \sinh (2\tau_0)-2\tau_0 }{ \sinh(2\tau)-2\tau } \right )^{\frac{2}{3}}}}.
\end{gather}

 \begin{figure}[ht!]
 \centering
  \includegraphics[width=7.5 cm] {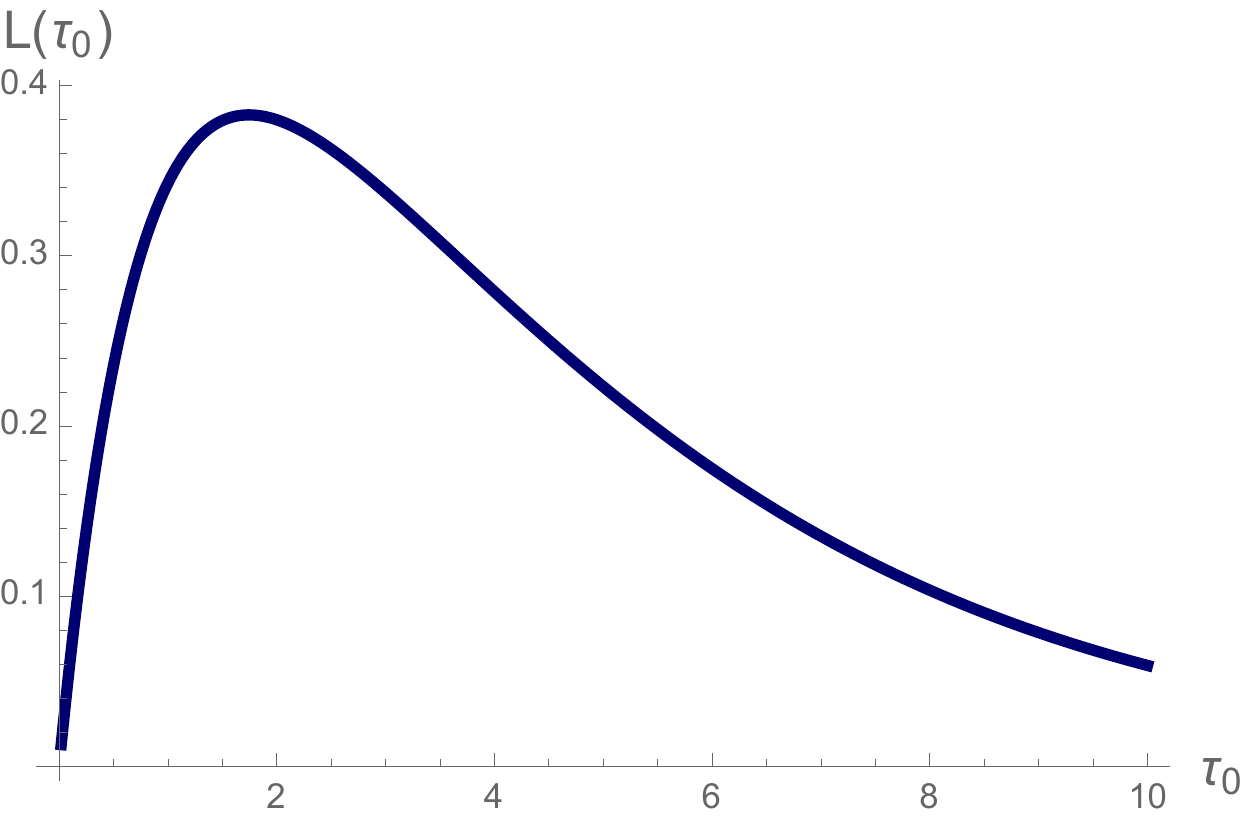} 
    \includegraphics[width=7.5 cm] {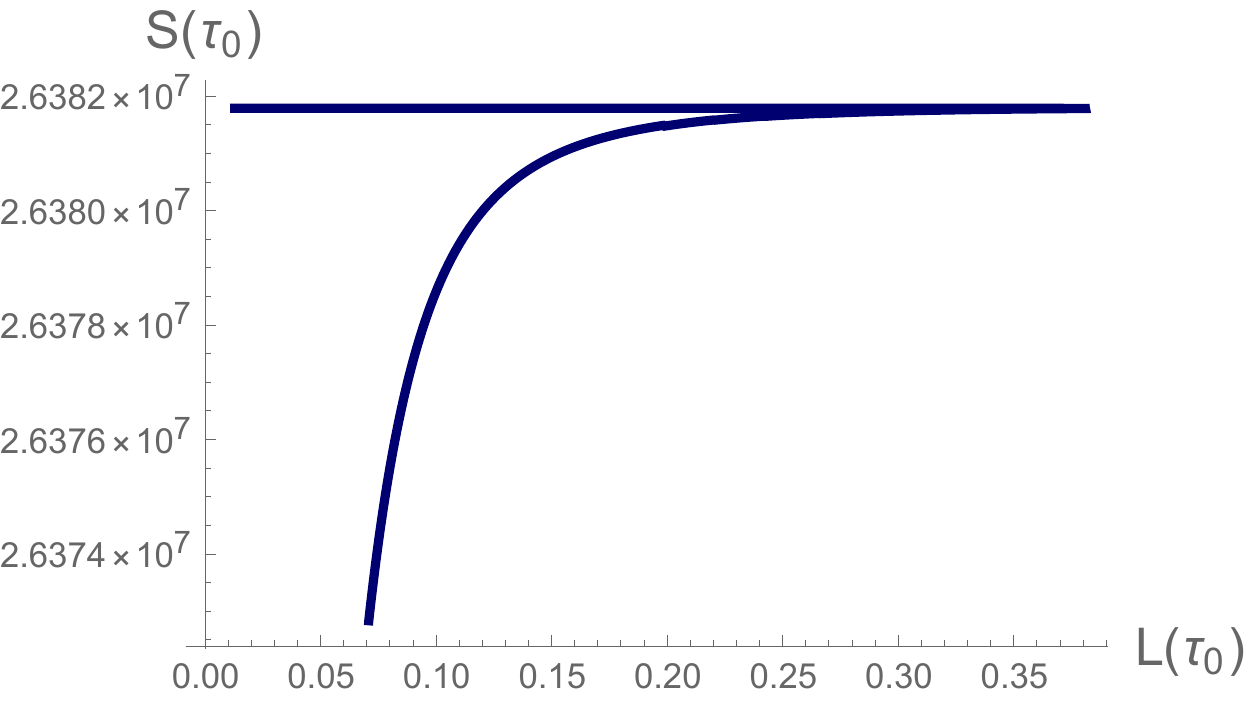} 
  \caption{The relationship between width of a strip versus turning point $\tau_0$ (in the left), and entanglement entropy versus $L$ (in the right) in Klebanov-Strassler background.}
 \label{fig:StripinKS}
\end{figure}

The determinant of the eight-dimensional matrix which is needed would be found where only one of the coordinate $\theta_1$ would be considered as a function of the radial coordinate $\tau$.

So we get
\begin{align*} 
\Gamma_{\text{KS}}&= \int_{\tau_D}^{\tau_{2l+D} } d\tau \frac{ h   \epsilon ^4 \sin \left(\theta _1\right)}{192 \sqrt{6} K^{3/2} } \left ( 3 K^3 \left(\frac{\text{d$\theta $}_1}{\text{d$\tau $}}\right)^2 \cosh (\tau )+2\right )^{\frac{1}{2}}\times \nonumber\\ & \left (  \cosh (\tau ) \left(4 \cos \left(2 \theta _2\right) \sinh ^2(\tau )-1\right)+6 K^3 \sin ^2\left(\theta _2\right) \sin ^2(\psi ) (\cosh (2 \tau )+\cos (2 \psi ))+\cosh (3 \tau )\right )^{\frac{1}{2}}
\end{align*}

The parameter $M$, $\alpha^\prime$ or $g_s$ could be changed which correspondingly change $h(\tau)$, and therefore the behavior of the critical $D$ in this background can be studied which probes the phase structures.

\begin{figure}[ht!]
 \centering
  \includegraphics[width=7 cm] {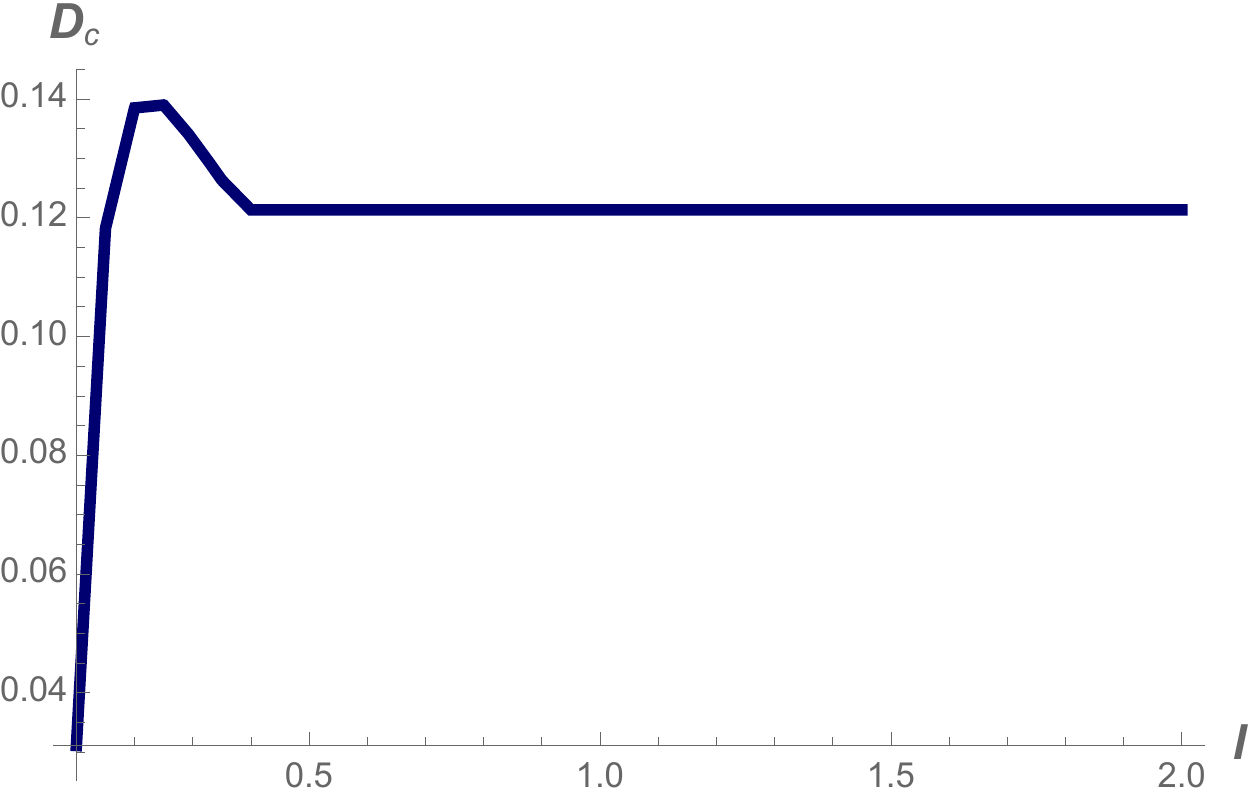} 
  \caption{The behavior of critical distance, $D_c$, between the two strips in mixed systems, in the Klebanov-Strassler background.}
 \label{fig:StripinMN}
\end{figure}

Generally, the behavior of $D_c$ versus the width of the two symmetric strips $l$ in the Klebanov-Strassler background would be as shown in figure \ref{fig:StripinMN}. 

It could be seen that for small width of the strips the critical distance would increase by increasing $l$, but after reaching to a maximum it would decrease at then becomes constant.

\subsection{Klebanov-Tseytlin}
The Klebanov-Tseytin (KT) metric is a singular solution which is dual to the chirally symmetric phase of the Klebanov-Strassler model which has D3-brane charges that dissolve in the flux \cite{Bena:2012ek}.

The metric is
\begin{gather}
ds_{10}^2=h(r)^{-1/2} \left[ -dt^2+d \vec{x}^2 \right]+h(r)^{1/2} \left[ dr^2+r^2 ds_{T^{1,1} }^2 \right].
\end{gather}

Here $ds_{T^{1,1} }^2$ is a base of a cone with the definition of
\begin{gather}
ds_{T^{1,1} }^2=\frac{1}{9} (g^5)^2+\frac{1}{6} \sum_{i=1}^4 (g^i)^2.
\end{gather}
It is actually the metric on the coset space $T^{1,1}=(SU(2) \times SU(2))/U(1)$.

Also, $g^i$ are some functions of the angles $\theta_1, \theta_2, \phi_1, \phi_2, \psi$ as
\begin{gather}
g^1=(-\sin \theta_1 d\phi_1-\cos \psi \sin \theta_2 d\phi_2 +\sin \psi d\theta_2)/ \sqrt{2}, \ \ \ \ 
g^2=(d\theta_1-\sin \psi \sin \theta_2 d\phi_2-\cos \psi d\theta_2) / \sqrt{2},\nonumber\\
g^3=(-\sin \theta_1 d\phi_1+\cos \psi \sin \theta_2 d \phi_2-\sin\psi d \theta_2) \sqrt{2}, \ \ \ \ \ 
g^4=(d\theta_1+ \sin \psi \sin \theta_2 d\phi_2+\cos \psi d\theta_2) /\sqrt{2},\nonumber\\
g^5=d\psi+\cos \theta_1 d\phi_1+\cos \theta_2 d\phi_2,
\end{gather}
and also we have the relations
\begin{gather}
h(r)=\frac{L^4}{r^4} \ln{\frac{r}{r_s}}, \ \ \ \ \ \ \ L^4=\frac{81}{2} g_s M^2 \epsilon^4.
\end{gather}

 \begin{figure}[ht!]
 \centering
  \includegraphics[width=7.5 cm] {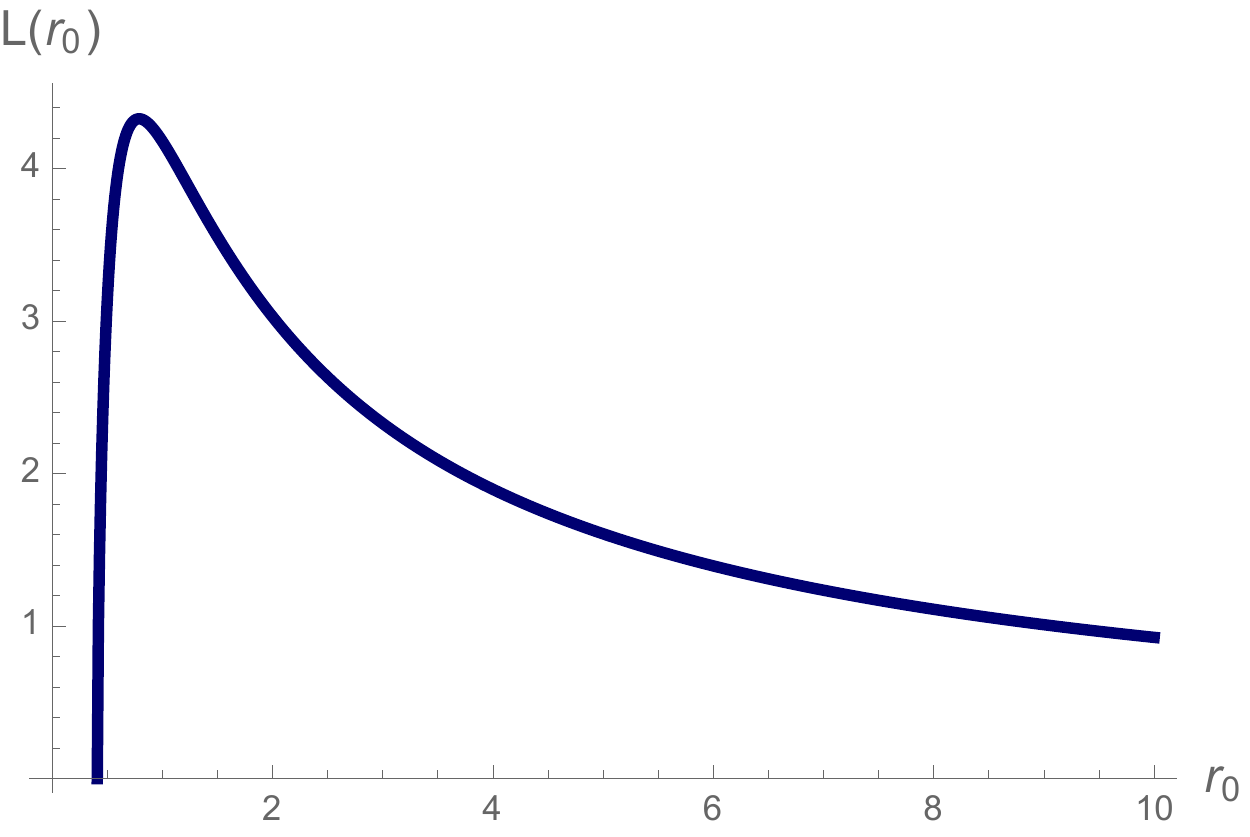} 
    \includegraphics[width=7.5 cm] {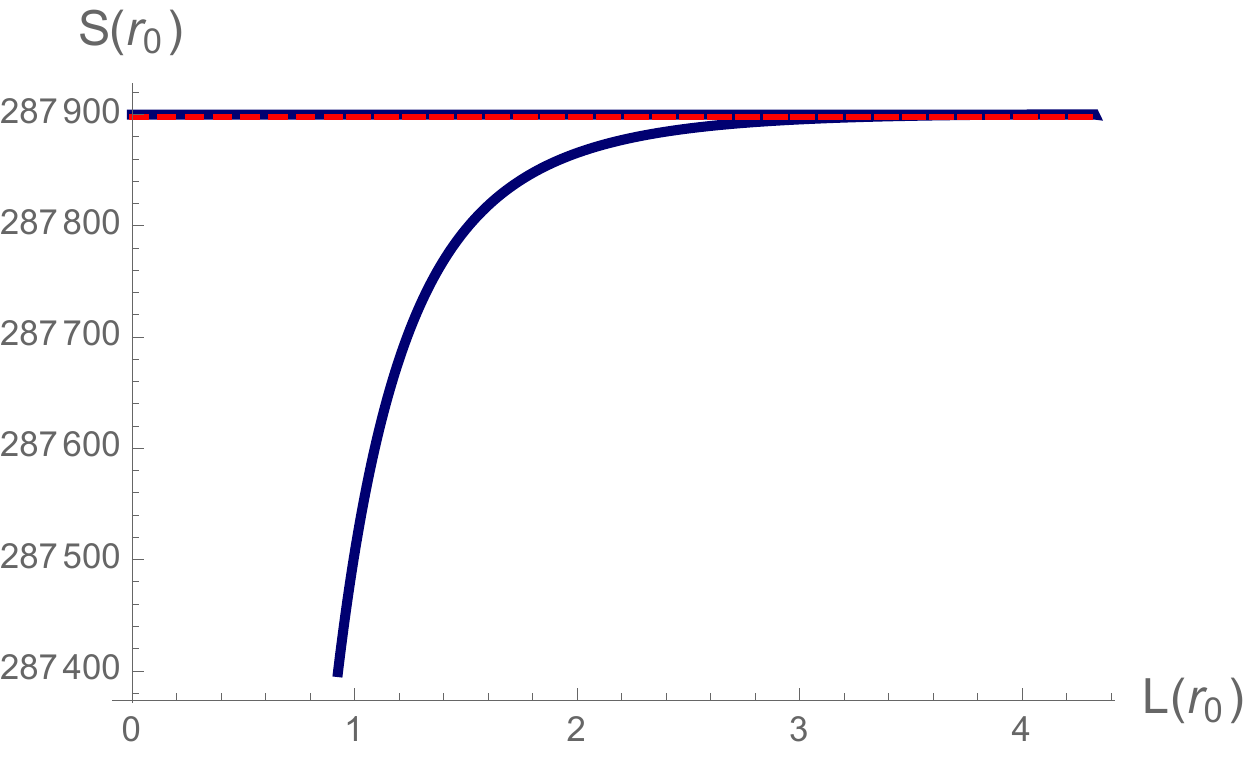} 
  \caption{The relationship between width of a strip $L(r_0)$ versus turning point $r_0$ (in the left), and entanglement entropy $S(r_0)$ versus $L$ (in the right) in the Klebanov-Tseytlin background.}
 \label{fig:regionDKT1}
\end{figure}

 \begin{figure}[ht!]
 \centering
  \includegraphics[width=8.5 cm] {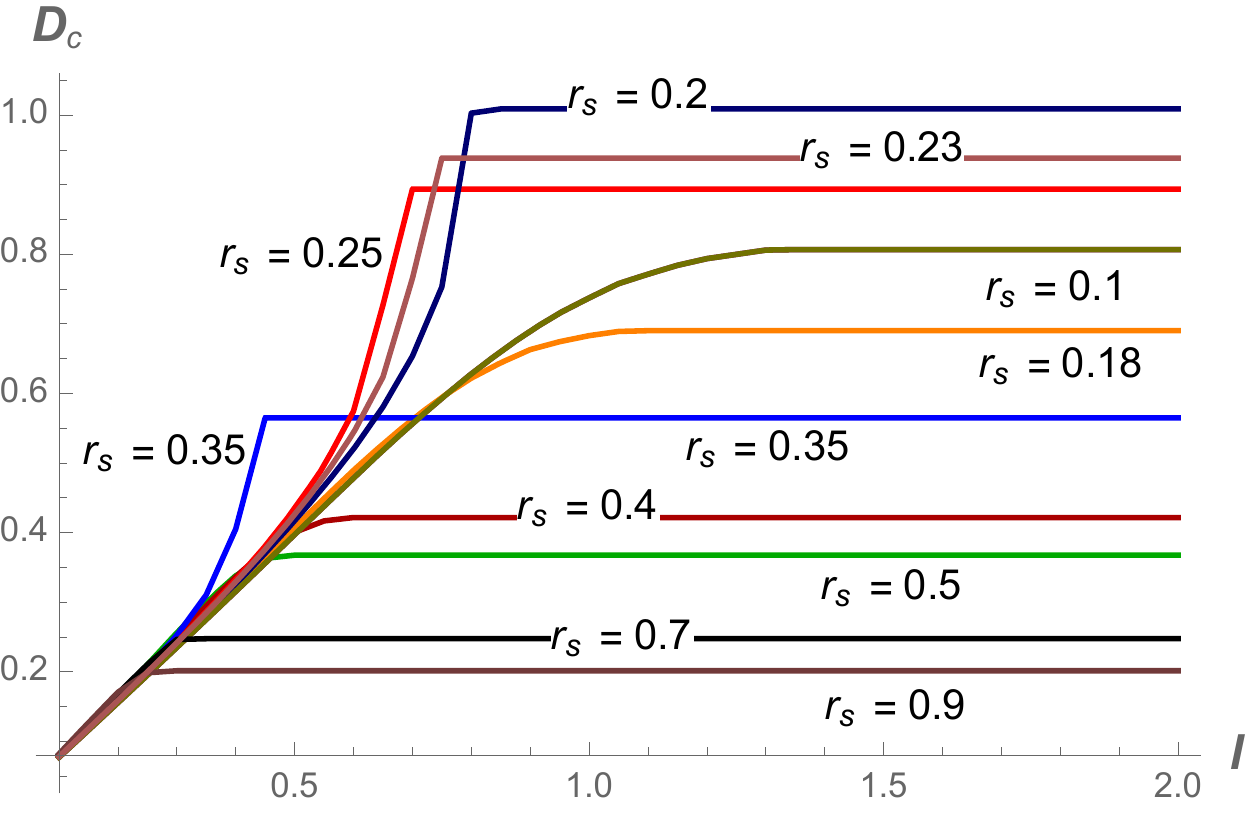} 
  \caption{The relationship between the critical distance $D_c$ between the two strips versus $l$ in the Klebanov-Tseytlin geometry. The regions below each curve for each $r_s$ correspond to non-zero mutual information and $r_s$ corresponds to the position of the singularity.}
 \label{fig:regionDKT2}
\end{figure}

 \begin{figure}[ht!]
 \centering
  \includegraphics[width=11 cm] {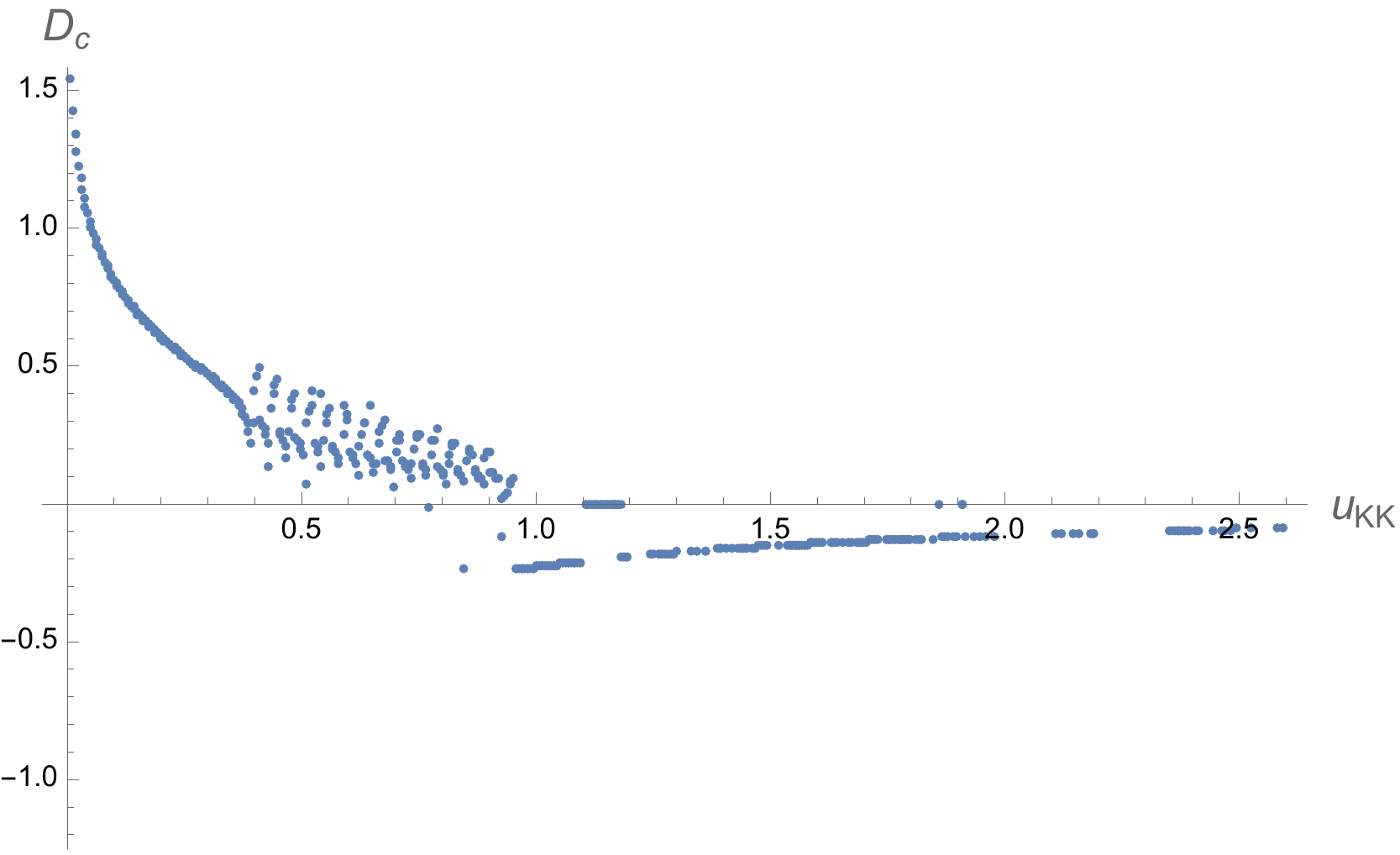} 
  \caption{The relationship between the critical distance $D_c$ between the two strips versus $u_{KK}$ in the Klebanov-Tseytlin geometry. }
 \label{fig:KTdatapoint3}
\end{figure}

In this frame, the asymptotic flat region has been eliminated. Also, $r=r_s$ is where the naked singularity is located.

The width of one strip versus turning point of RT surface in KT background can be written as \cite{Ghodrati:2015rta,Kol:2014nqa}
\begin{gather}
L(r_0)=9\sqrt{2} M \sqrt{g_s} \epsilon^2 \int_{r_0}^\infty dr \frac{\sqrt{\ln \frac{r}{r_s} } }{r^2 \sqrt {\frac{r^6}{r_0^6} \frac{\ln \frac{r}{r_s} }{\ln \frac{r_0}{r_s} }-1 } }, 
\end{gather}
and the entanglement entropy of the connected solution is
\begin{gather}
S_C(r_0)= \frac{12 V_2 \pi^3 M^2 g_s \epsilon^4}{G_N^{(10)} } \int_{r_0}^\infty dr \frac{r \ln \frac{r}{r_s} }{\sqrt{1-\frac{r_0^6}{r^6} \frac{\ln \frac{r_0}{r_s} }{\ln \frac{r}{r_s} } }},
\end{gather}
where their behavior is shown in figure \ref{fig:regionDKT1}.

The behavior of $D_c$ versus $L$ is shown in figure \ref{fig:regionDKT2} and the behavior of $D_c$ versus $u_{KK}$ is shown in figure \ref{fig:KTdatapoint3}.

Then, the determinant of the induced metric for calculating $\Gamma$ could be found and then we get
\begin{gather}
\Gamma_{\text{KT}}= \frac{1}{648} h^{3/4} r^4 \sin \left(\theta _1\right)  \left ( r^2 \left(\frac{\text{d$\theta $}_1}{\text{dr}}\right){}^2+6\right )^{\frac{1}{2}} \left (   \left(2 \sqrt{h} r^2+1\right) \cos \left(2 \theta _2\right)+2 \sqrt{h} r^2-5  \right )^{\frac{1}{2}}.
\end{gather}

The phase structure coming from changing $r_s$ is presented in figure \ref{fig:regionDKT2}.  From figure \ref{fig:regionDKT2}, it could be seen that the relation of critical distance between the two strip versus $l$ is not a monotonic function in all the phase space, and again, by studying the behavior of $D_c$ and therefore EWCS, at least four different phases could be detected in these top-down QCD models, and again this is consistent with the result of \cite{Jain:2020rbb}.

Note that again the numerics are very sensitive to the parameter $r_s$ which is the position of naked singularity here, specially for smaller value of $r_s$.

\subsection{Klebanov-Witten}
The Klebanov-Witten solution is similar to the KT throat geometry but with no logarithmic warping \cite{Kaviani:2015rxa}. Unlike the other four mentioned metrics, it is just a conformal geometry. We study this background to compare our previous results  in confining cases with the conformal one.  

The metric is
\begin{gather}
ds^2=h^{-\frac{1}{2}} g_{\mu\nu} dx^\mu dx^\nu+h^{\frac{1}{2}} (dr^2+r^2 ds_{T^{1,1}} ^2),
\end{gather}
where
\begin{gather}
h=\frac{L^4}{r^4}, \ \ \ \ \ \  \text{and} \ \ \ \ L^4=\frac{27\pi}{4} g_s N (\alpha')^2.
\end{gather}

Since the Klebanov-Witten solution is very similar to Klebanov-Tseytlin, the functional for calculating $\Gamma$ would be similar, and only the relation for $h$ is different, as it has no Logaritmic term.

\begin{gather}
\Gamma_{\text{KW}}=\frac{1}{108 \sqrt{6}} \int_{r_D}^{r_{2l+D}} dr h^2(r) r^5   \int_0^{2\pi}  \int_0^{2\pi} d\theta_1 d\theta_2 \sqrt{-3+\cos 2 \theta_1+2 \cos 2 \theta_2} \sin \theta_1.
\end{gather}

\begin{figure}[ht!]
 \centering
  \includegraphics[width=7 cm] {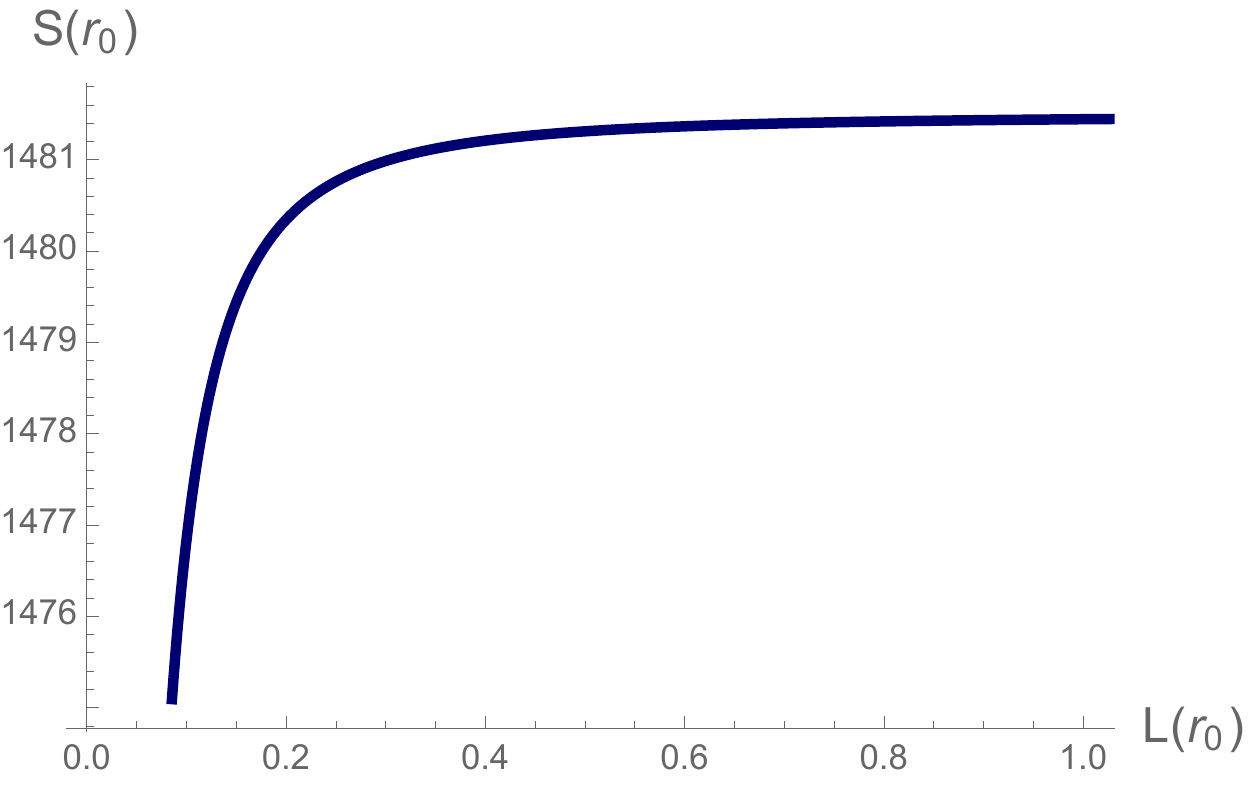} 
    \includegraphics[width=7 cm] {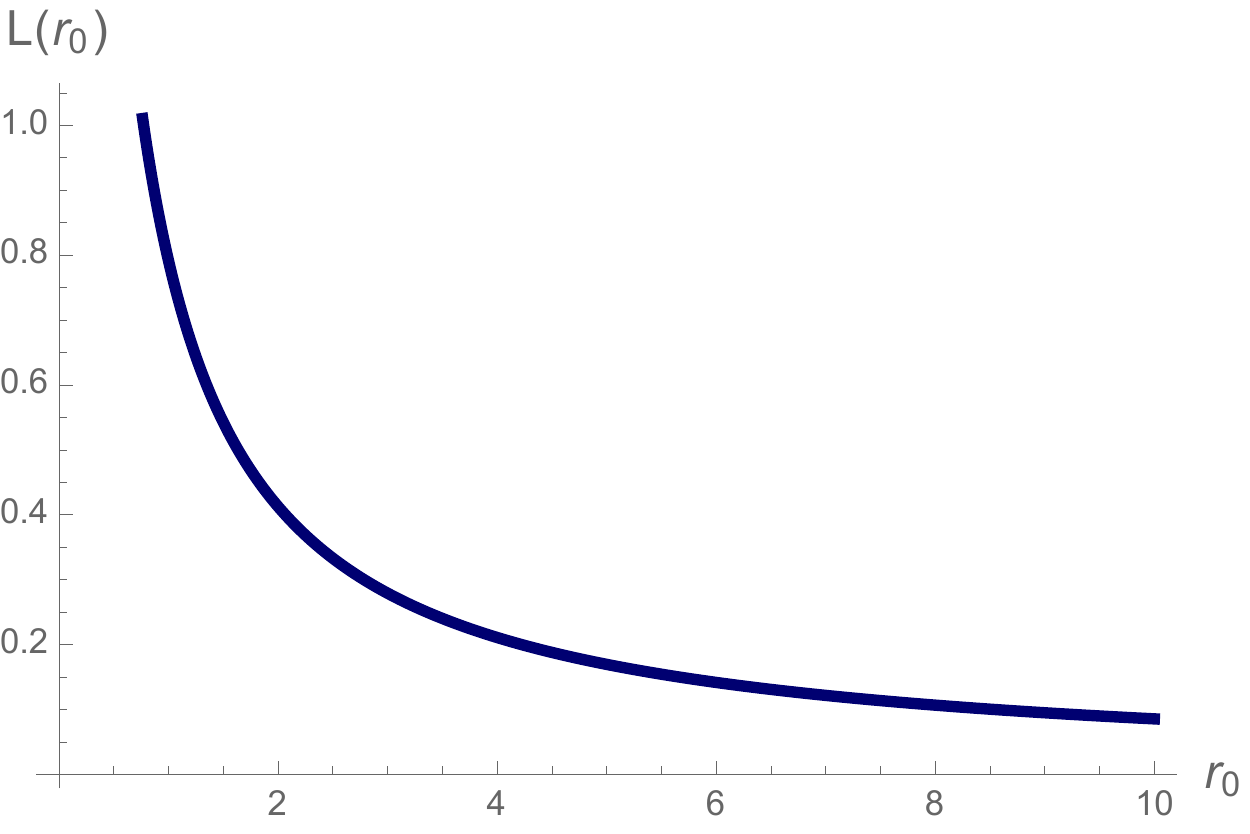} 
  \caption{The relationship between width of a strip versus turning point $r_0$ (in the left), and entanglement entropy versus $L$ (in the right) in the Klebanov-Witten (KW) background which shows similar behavior to AdS case.}
 \label{fig:SLKT}
\end{figure}

The width of the strip in terms of the turning point and the entanglement entropy could be written as
\begin{gather}
L(r_0)=2 L^2 \int_{r_0}^\infty dr \frac{1}{r^2 \sqrt{\frac{r^6}{r_0^6}-1} },
\end{gather}
and 
\begin{gather}
S(r_0)= \frac{8V_{d-1} \pi^4 L }{27 G_N^{(10)} } \int_{r_0}^\infty  dr \frac{r}{\sqrt{1-\frac{r_0^6}{r^6} } }.
\end{gather}
where their behavior is shown in figure \ref{fig:SLKT}.

\begin{figure}[ht!]
 \centering
  \includegraphics[width=6 cm] {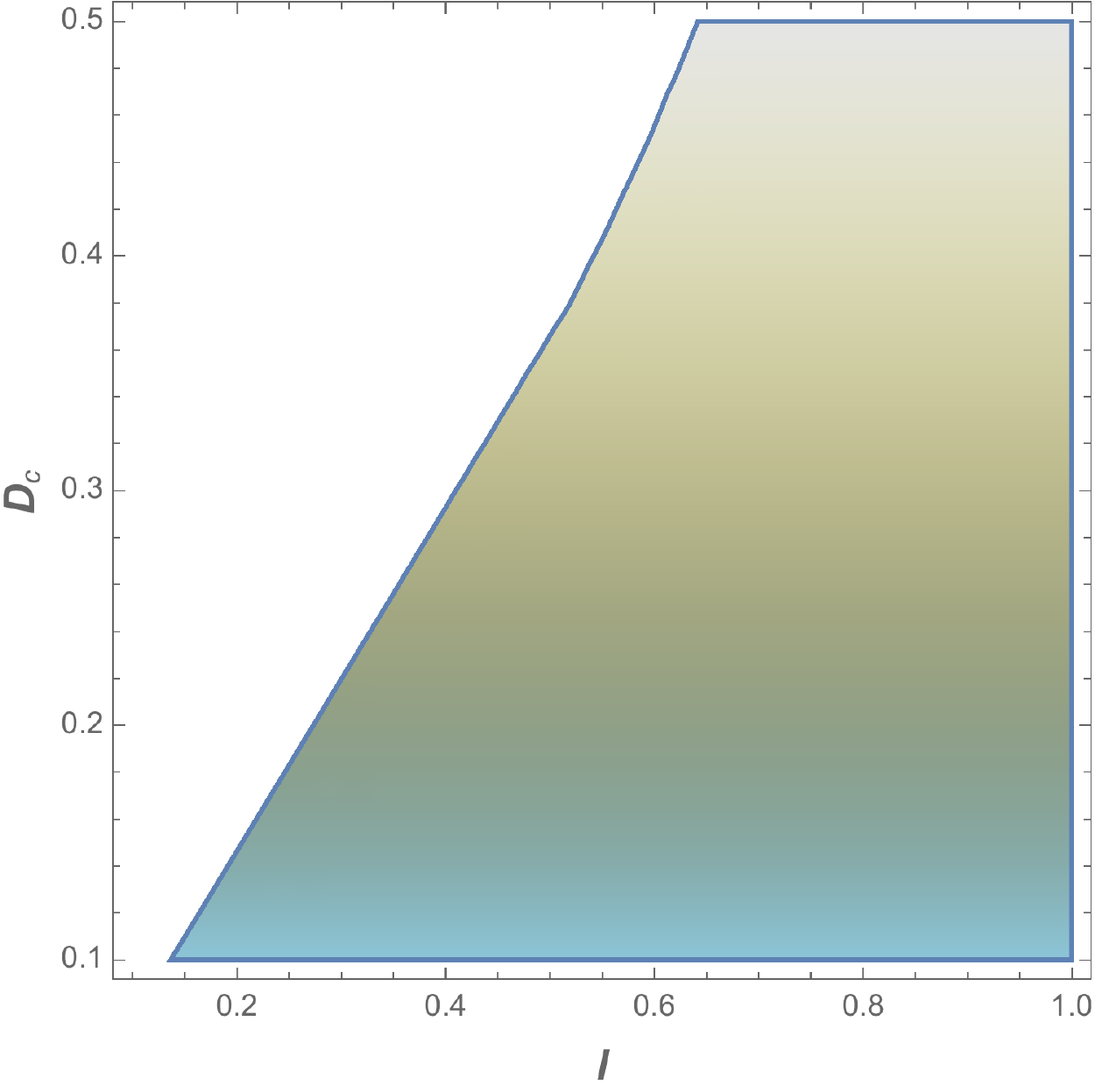} 
    \includegraphics[width=8 cm] {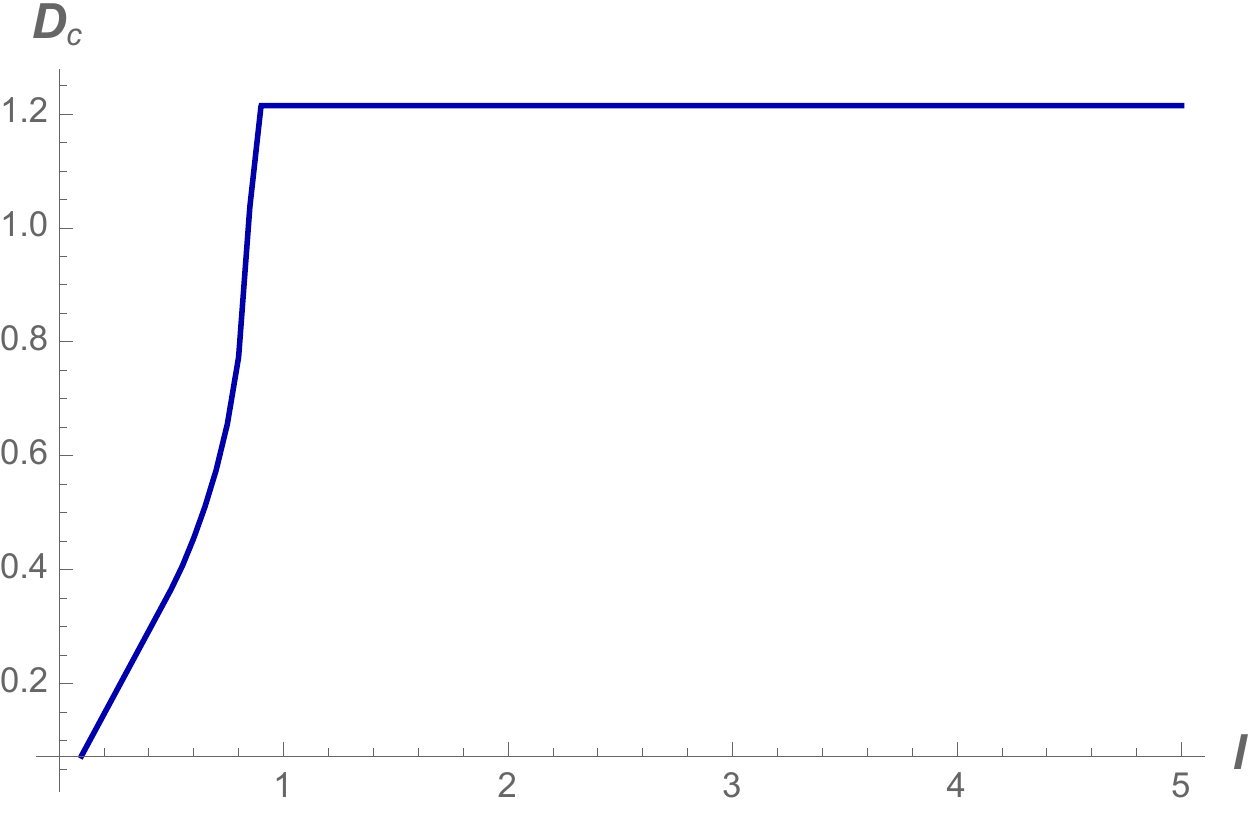} 
  \caption{The behavior of $D_c$ versus $l$ in KW background in the right part and non-zero region of mutual information in the left.}
 \label{fig:regionDKW2}
\end{figure}

The behavior of $D_c$ versus $l$ is shown in figure \ref{fig:regionDKW2}. Since this geometry is conformal and there is no position of the wall here, as in $r_s$, no rich phase structure could be detected, but the behavior of $D_c$ is similar to the first phase of Klebanov-Tseytlin shown in figure \ref{fig:regionDKT2}, as one expects.

\subsection{Maldacena-Nunez}
The Maldacena-Nunez (MN) metric is obtained by a large number of D5-branes wrapping on $S^2$ \cite{Maldacena:2000yy}. In the string frame the metric and the fields are written as \cite{Basu:2012ae}
\begin{gather}
ds^2_{10}=e^\Phi [ -dt^2+{dx_1}^2+{dx_2}^2+{dx_3}^2+e^{2h(r)}({d\theta_1}^2+\sin^2 \theta_1 {d\phi_1}^2)+dr^2+\frac{1}{4}(w^i-A^i)^2 ],
\end{gather}
where
\begin{gather}
A^1=-a(r)d\theta_1, \ \ \ \ \ \ \ \ A^2=a(r)\sin \theta_1 d\phi_1, \ \ \ \ \ \ \ \ 
A^3= -\cos \theta_1d\phi_1,
\end{gather}
and the $\omega^i$s parametrize the 3-sphere compactification, which are
\begin{gather}
w^1 =\cos \psi d\theta_2+\sin \psi \sin\theta_2 d \phi_2,\ \ \ \ \ \ \ \ \ 
w^2 =-\sin \psi d \theta_2+\cos \psi \sin \theta_2 d\phi_2,\nonumber\\
w^3=d\psi+\cos \theta_2 d\phi_2,
\end{gather}
and also the other parameters of the metric would be
\begin{gather}
a(r)=\frac{2r}{\sinh 2r}, \ \ \ \ 
e^{2h}= r\coth {2r}-\frac{r^2}{{\sinh 2r}^2}-\frac{1}{4},\ \ \ \ 
e^{-2\Phi} =e^{-2\Phi_0} \frac{2 e^h}{\sinh 2r}.
\end{gather}

For this geometry we get
\begin{gather}
\alpha(r)= e^\Phi,  \ \ \ \ \ \beta(r)=1, \ \ \ \ \ H(r)=64 e^{4h+4\Phi}=  64 \left ( r \coth 2r - \frac{r^2}{\sinh 2 r^2} - \frac{1}{4} \right )^2.
\end{gather}

So using the relation \ref{eq:L} the width of the strip versus turning point can be written as
\begin{gather}
L(r_0)= 2\int_{r_0}^\infty dr \frac{1}{\sqrt{\frac{r \coth 2r- \frac{r^2}{\sinh 2r^2} - \frac{1}{4} }{r_0 \coth 2 r_0 - \frac{r_0^2}{\sinh 2 r_0^2} - \frac{1}{4} }-1} },
\end{gather}
and the entanglement entropy of the connected solution of the strip from the relation \ref{eq:connected} would be found as
\begin{gather}
S_C (r_0) = \frac{4 V_2}{ G_N^{(10)} } \int_{r_0}^\infty \frac{r \coth 2r - \frac{r^2}{\sinh 2 r^2} - \frac{1}{4}  }{\sqrt{1- \frac{r_0 \coth 2 r_0- \frac{r_0^2}{\sinh 2 r_0^2} - \frac{1}{4} }{r \coth 2r - \frac{r^2}{\sinh 2r^2} - \frac{1}{4} } } }.
\end{gather}

\begin{figure}[ht!]
 \centering
  \includegraphics[width=7.5 cm] {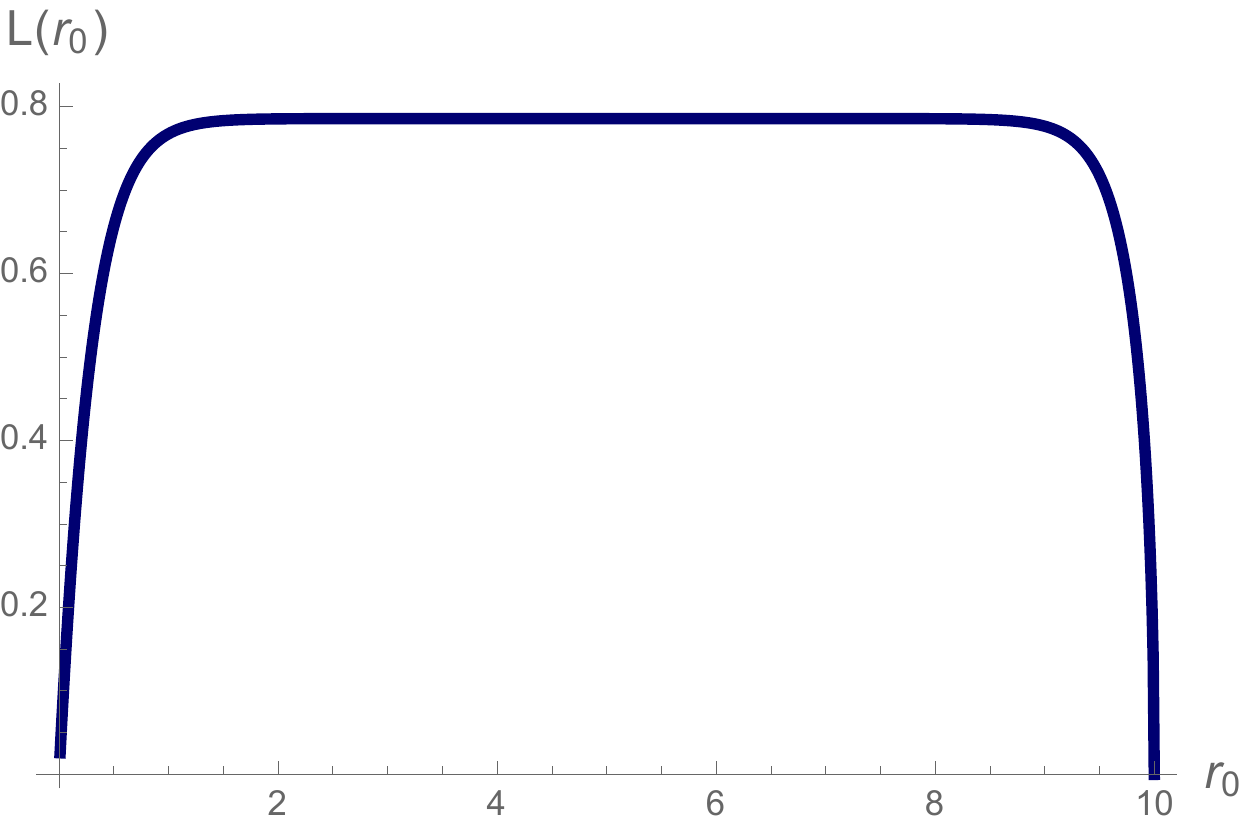} 
    \includegraphics[width=7.5 cm] {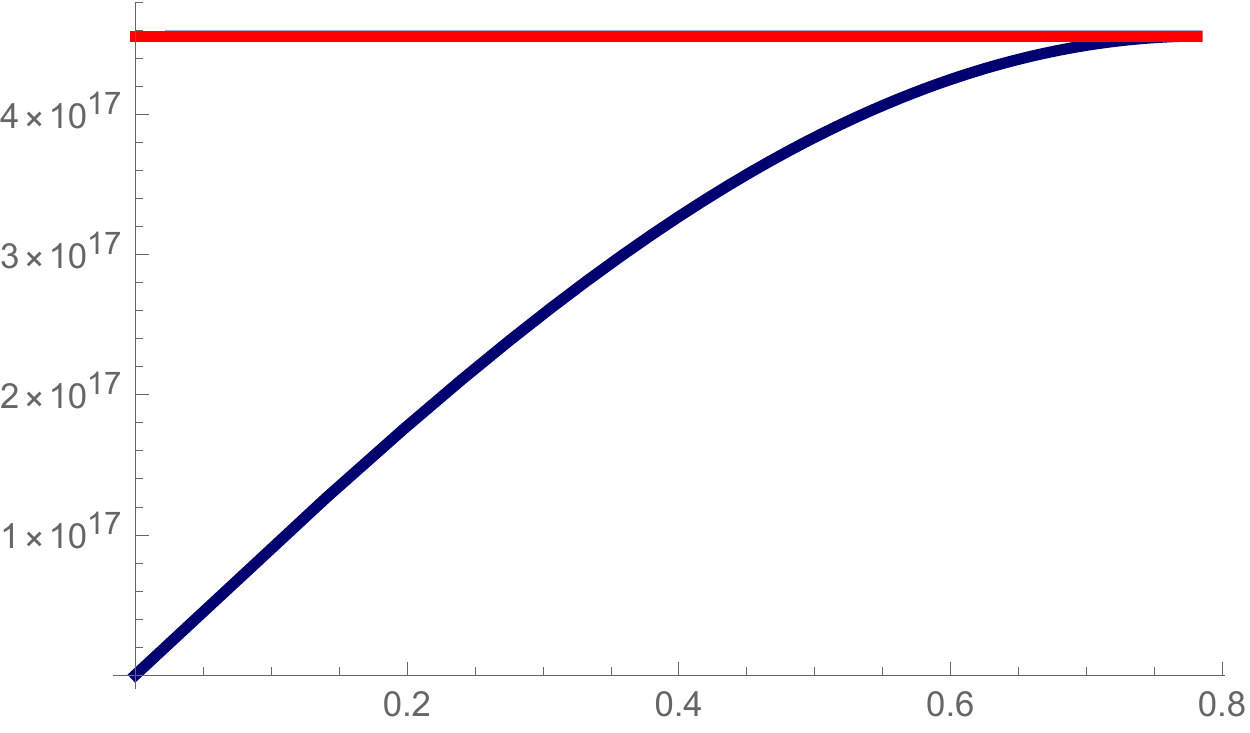} 
  \caption{The relationship between width of a strip versus turning point $r_0$ (in the left), and entanglement entropy versus $L$ (in the right) in the Maldacena-Nunez background.}
 \label{fig:SLinMN}
\end{figure}

The behavior of $L(r_0)$ and $S$ are shown in figure \ref{fig:SLinMN}.

\begin{figure}[ht!]
 \centering
  \includegraphics[width=8.3 cm] {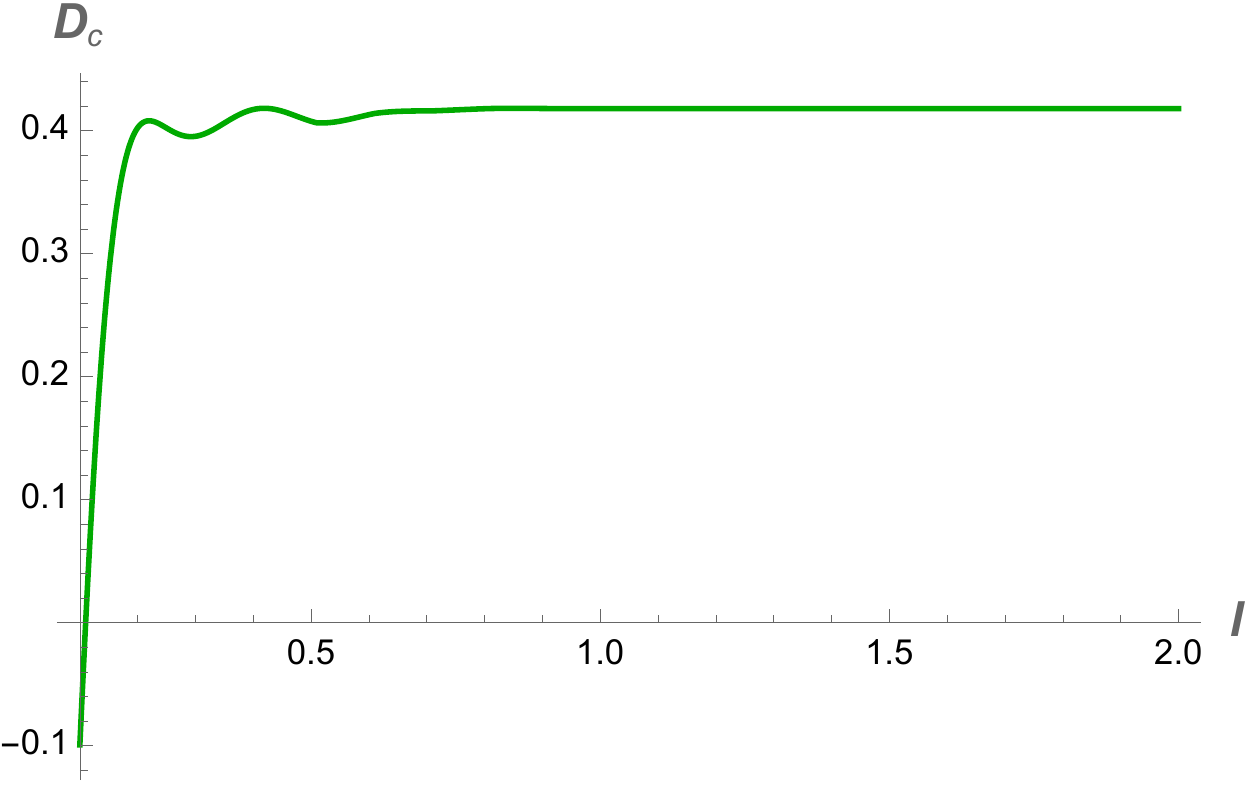} 
  \caption{The relationship between width of a strip $L(r_0)$ versus turning point $r_0$ (in the left), and the entanglement entropy $S(L) $ versus $L$ (in the right) in the Maldacena-Nunez background.}
 \label{fig:regionDMN1}
\end{figure}

The behavior of the critical distance versus $L$ (one strip) is then shown in figure \ref{fig:regionDMN1}.  Note that unlike other confining geometries, we don't have many free parameters in the MN case and therefore similar to the conformal case of KW, no phase structure could also be detected. The only difference here would be the early oscillations and noises in the behavior of $D_c$ for small values of width, $l$, (for the case of two strips) which we have explained their origins in the confining backgrounds before.

Then, for this geometry, the determinant of the eight-dimensional induced metric  $\Gamma$ could be calculated and then after simplifications we get the following result for the minimal wedge cross section as
\begin{align*}
\Gamma_{\text{MN}}& = \int_{r_D}^{r_{2l+D}} dr \frac{e^{4 \Phi }}{64} \left ( \left(a^2+4 e^{2 h}\right) \left(\frac{d\theta_1}{dr}\right)^2+4   \right )^{\frac{1}{2}} \nonumber\\ &
\Bigg(-2 e^{2 h} (-a^2 \sin ^2 (\theta _1) (\cos  (2 \theta _2 ) (6 \cos (2 \psi )-4)+3 \cos (2 \psi )+1 ) \nonumber\\ & -8 a \sin (\theta _1) \sin (\theta _2) \cos (\theta _2) \cos (\theta _1) \cos (\psi )+2 \cos (2 \theta _2) \cos ^2 (\theta _1 ) )-a^2 \sin ^2(\psi ) (a (a \sin ^2 (\theta _1) (9 \cos ^2 (\theta _2 )
\nonumber\\    & -\sin ^2 (\theta _2 ) (\cos (2 \psi )-2))+\sin (2 \theta _1) \sin (2 \theta _2) \cos (\psi ))+\cos ^2(\theta _1))+16 e^{4 h} \sin ^2 (\theta _1) (2 \cos (2 \theta _2)+1) \Bigg)^{\frac{1}{2}},
\end{align*}
where again the only free parameters are $\Phi_0$ or $r_0$ and no rich phase structure could be found here.

\subsection{Domain Wall AdS/QCD }\label{sec:DomainWall}

In the recent work \cite{Evans:2021zzm,Janik:2021jbq}, using the domain wall structure, a new holographic model for QCD has been constructed which consists of probe D7 branes in a D5 brane geometry. When one dimension of the D5 geometry is being compactified, the confinement will come into play and affects the gauge degrees of freedom in the theory.

The D5 geometry is 
\begin{gather}\label{eq:D5metric}
ds^2=\frac{u}{R} \big( \eta_{\mu \nu} dx^\mu dx^\nu+dx_4^2+dx_5^2 f(u; u_\Lambda) \big)+\frac{R}{u} \frac{du^2}{f(u; u_\Lambda)} +R \ u\  d\Omega_3^2,
\end{gather}
where
\begin{gather}
f(u;u_\Lambda) = \left ( 1- \frac{u_\Lambda^2 }{u^2} \right),
\end{gather}
and the non trivial dilaton field and 3-form flux are
\begin{gather}
e^\Phi=g_s \frac{u}{R}, \ \ \ \ \ F_3=\frac{2R^2}{g_s} \Omega_3,
\end{gather}
where $\Omega_3$ is the volume form of the unit 3-sphere and the parameter $R$ is related to the string parameters as $R^2= g_s N_c \alpha^\prime$.

The entanglement entropy as the function of turning point $u_t$  could be found as
\begin{gather}
S_C(u_t)=\frac{(2 \pi)^2 V_2 V_3}{2 g_s^2 G_N^{(10)} } \int_{u_t}^\infty du \frac{u}{\sqrt {1- \frac{u_t^4}{u^4} \left(\frac{1-\frac{u_\Lambda^2}{u_t^2} }{1-\frac{u_\Lambda^2}{u^2} }  \right ) } },
\end{gather}
and the width of the strip versus the turning point is
\begin{gather}
L(u_t)=2R \int_{u_t}^\infty \frac{du}{u}  \frac{1}{ \sqrt{ \left(1- \frac{u_\Lambda^2}{u^2} \right) \left(\frac{u^4}{u_t^4} \frac{1-\frac{u_\Lambda^2}{u^2}  }{1-\frac{u_\Lambda^2}{u_t^2} } -1\right)}}.
\end{gather}

\begin{figure}[ht!]
 \centering
  \includegraphics[width=7 cm] {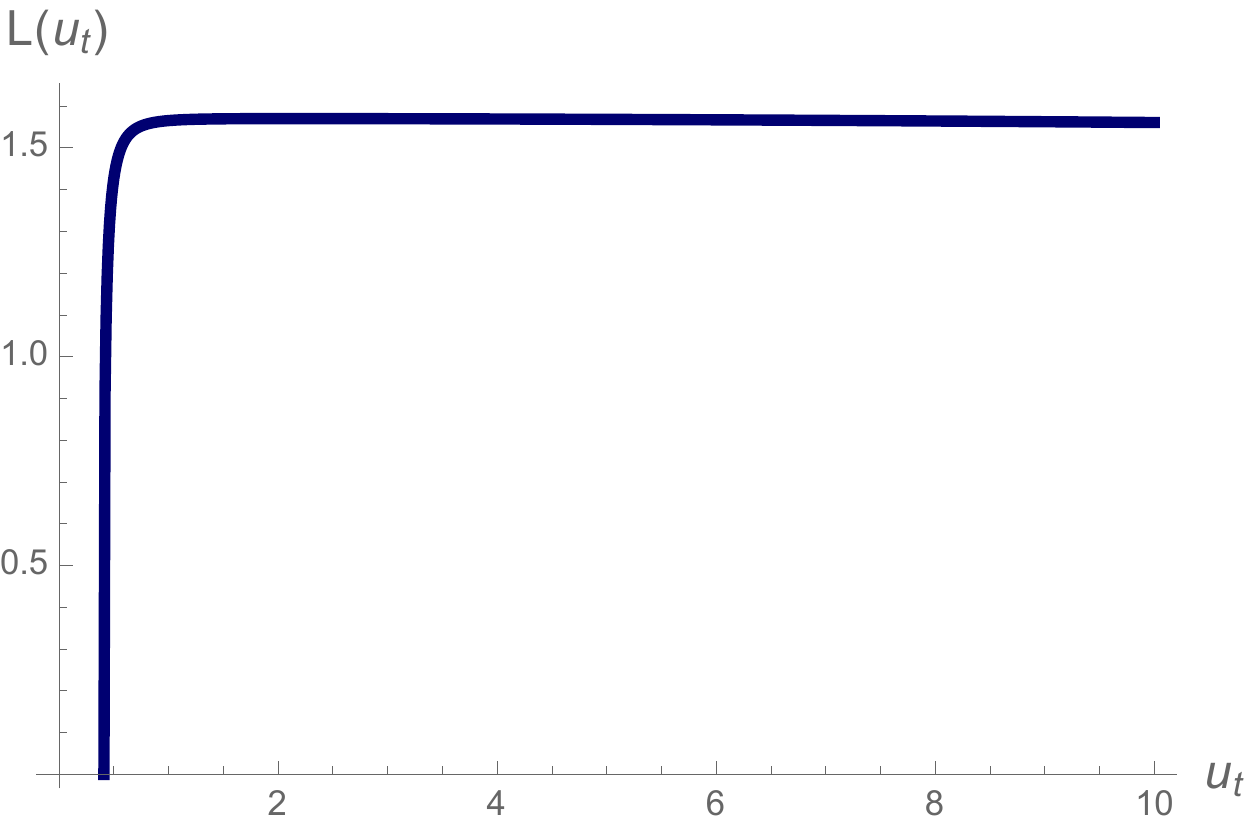} 
    \includegraphics[width=7 cm] {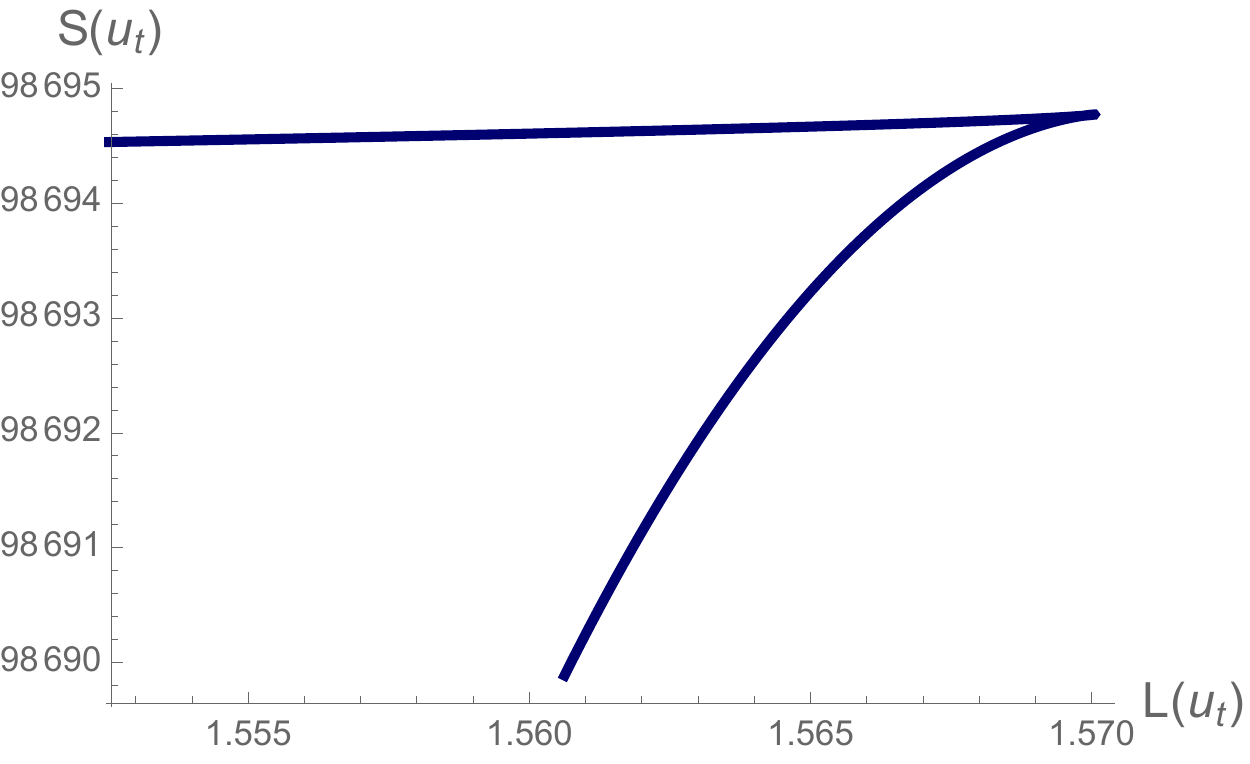} 
  \caption{The behavior of $L(u_t)$ versus $u_t$ and $S(u_t)$ versus $L(u_t)$ in the domain wall QCD.}
 \label{fig:domainwallLS}
\end{figure}

Their behavior is shown in figure \ref{fig:domainwallLS}, where again the typical and universal behavior of the QCD models, as the butterfly geometry for $S$, can be detected.

\begin{figure}[ht!]
 \centering
  \includegraphics[width=6.5 cm] {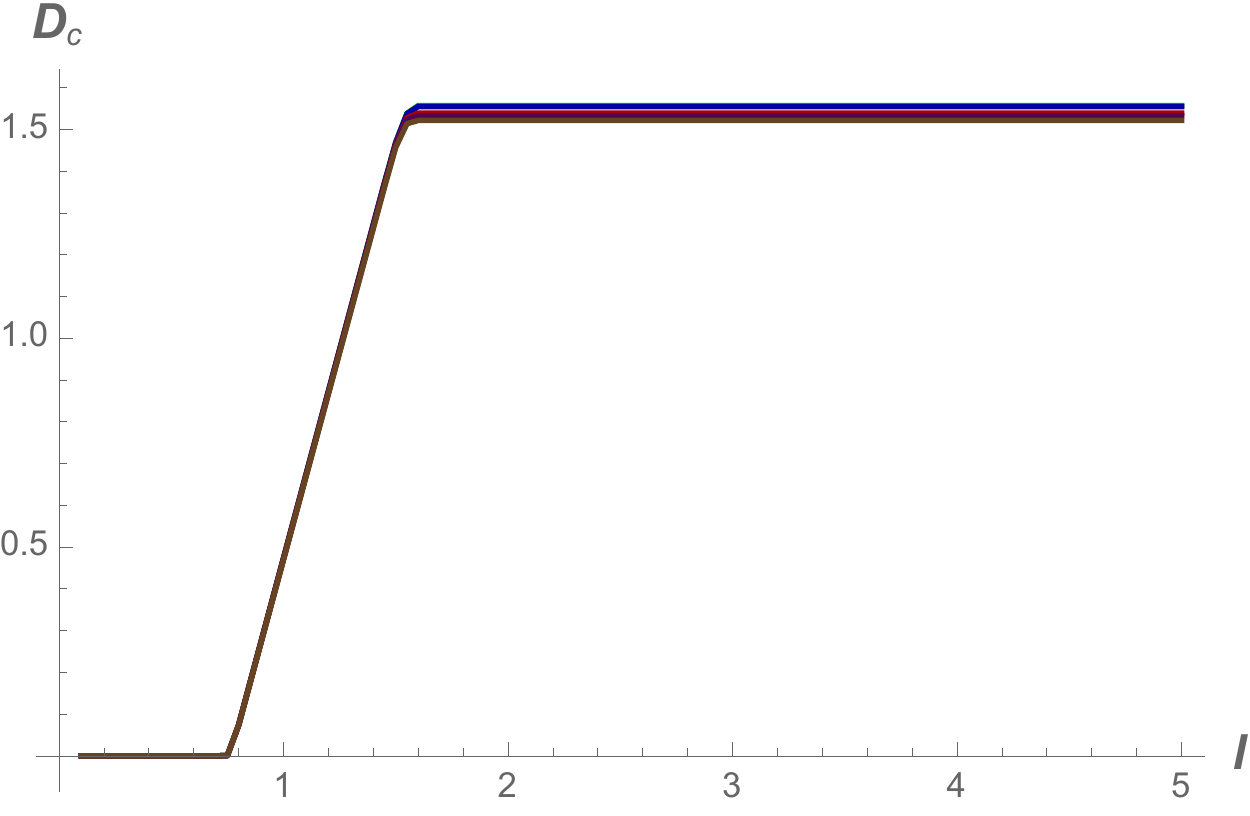} 
    \includegraphics[width=8.5 cm] {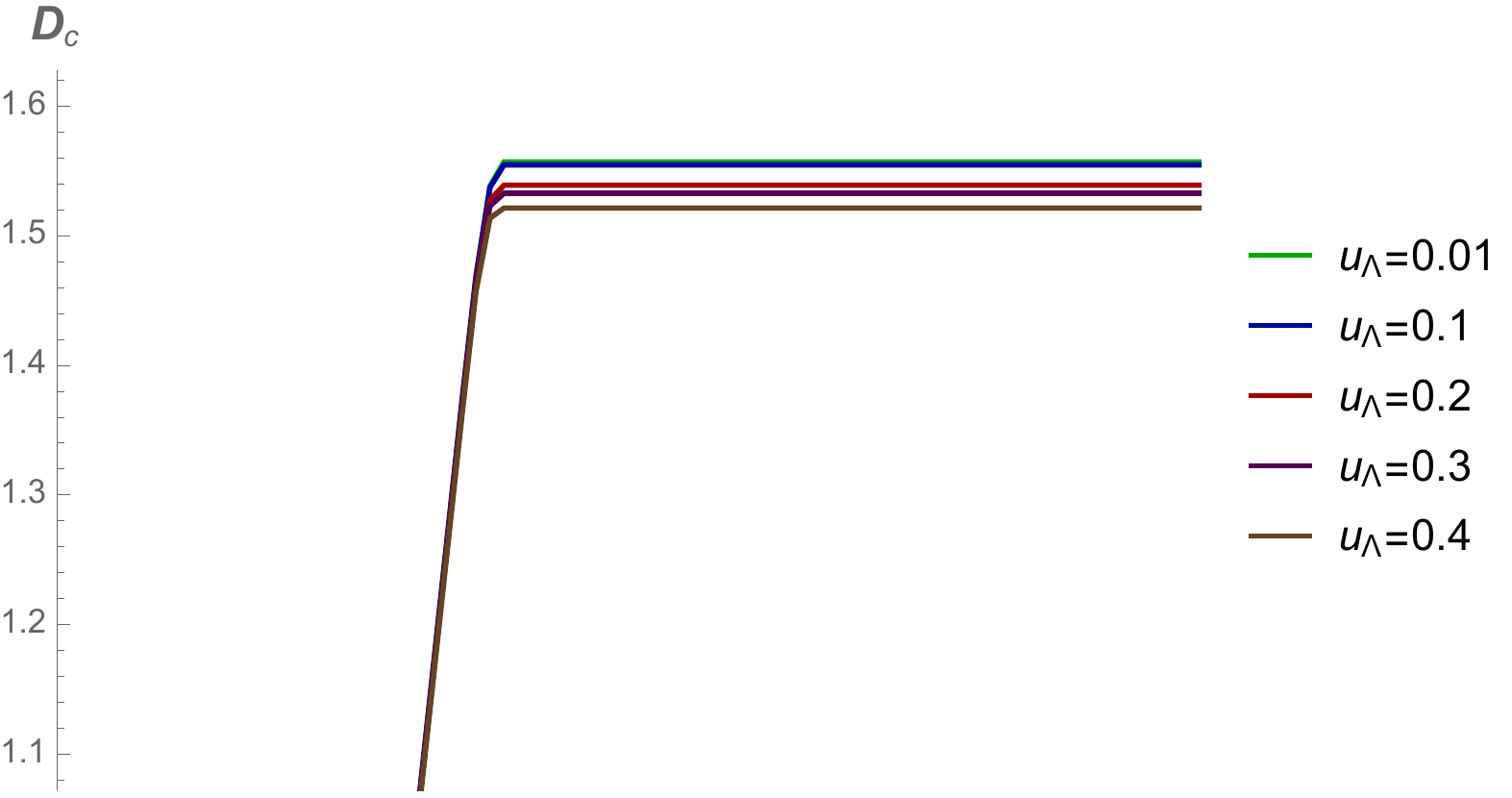} 
  \caption{Phase one in D5 background. }
 \label{fig:Dcwallp1}
\end{figure}

\begin{figure}[ht!]
 \centering
  \includegraphics[width=6.5 cm] {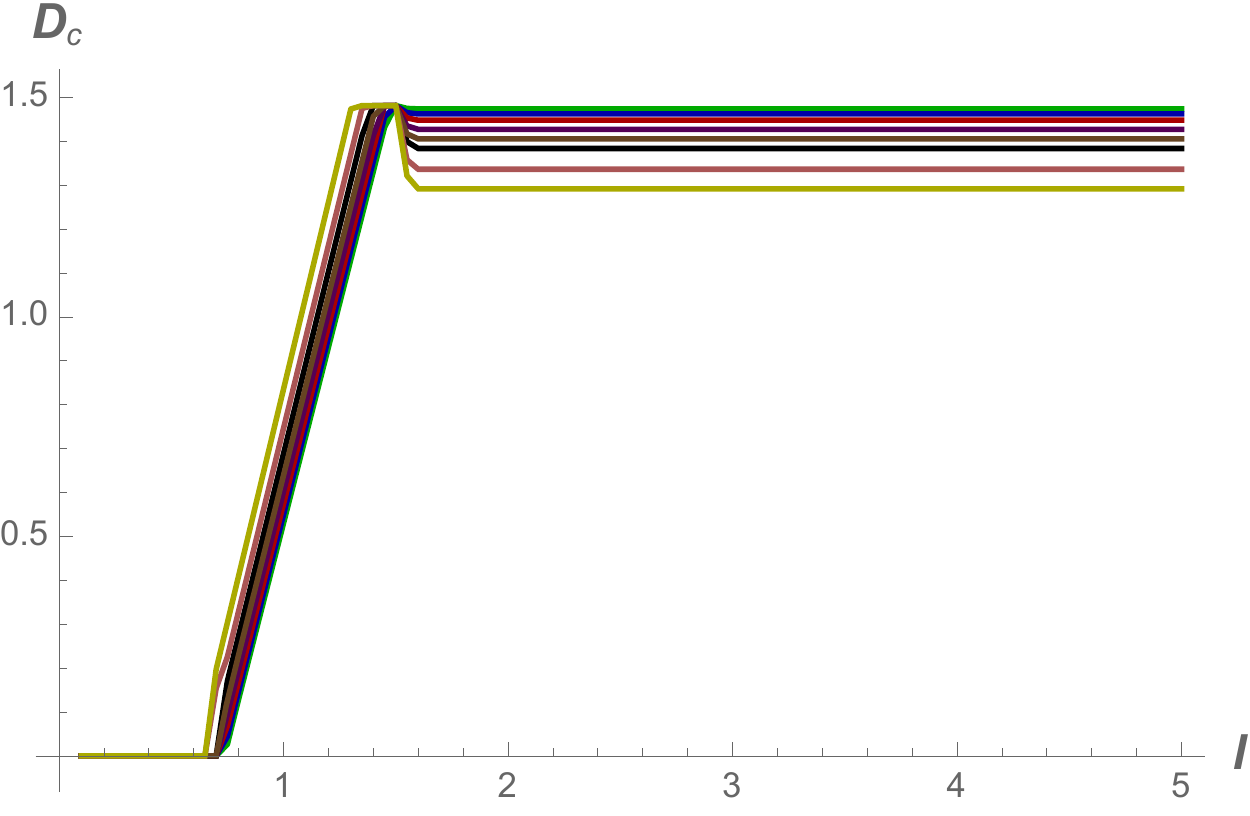} 
    \includegraphics[width=8.5 cm] {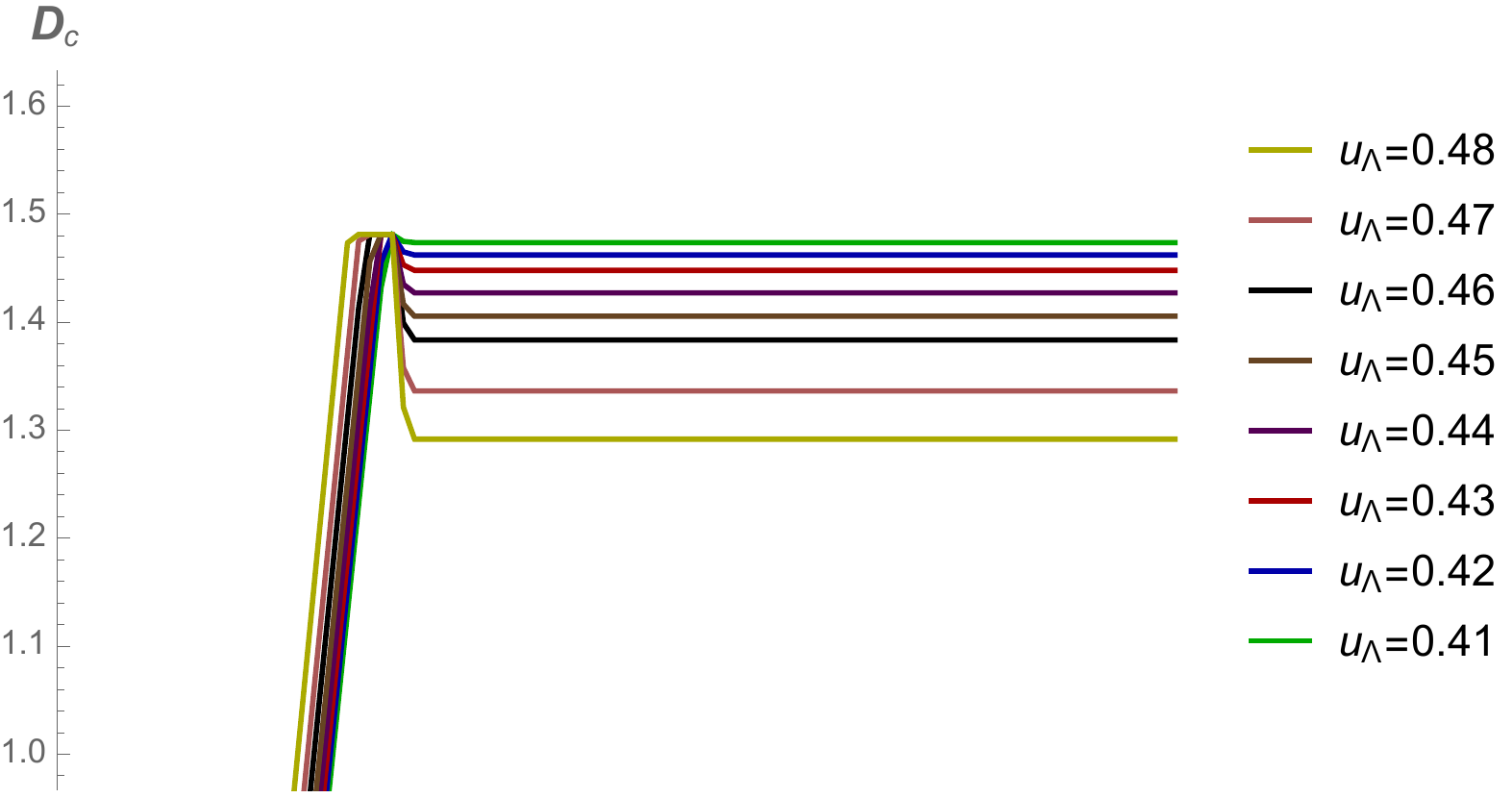} 
  \caption{Phase two in D5 background. }
 \label{fig:Dcwallp2}
\end{figure}

\begin{figure}[ht!]
 \centering
  \includegraphics[width=8.5 cm] {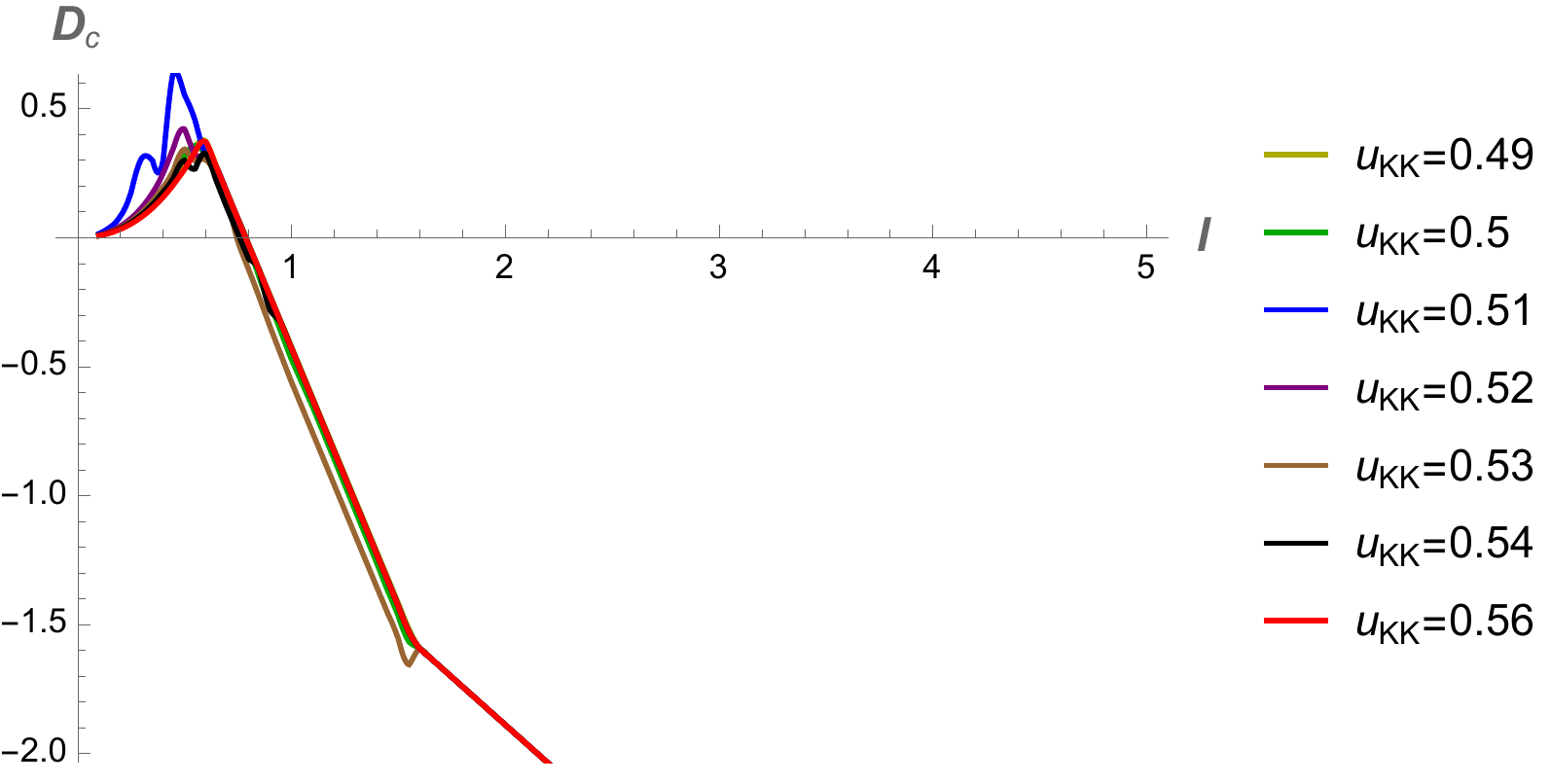}  
    \includegraphics[width=8.5 cm] {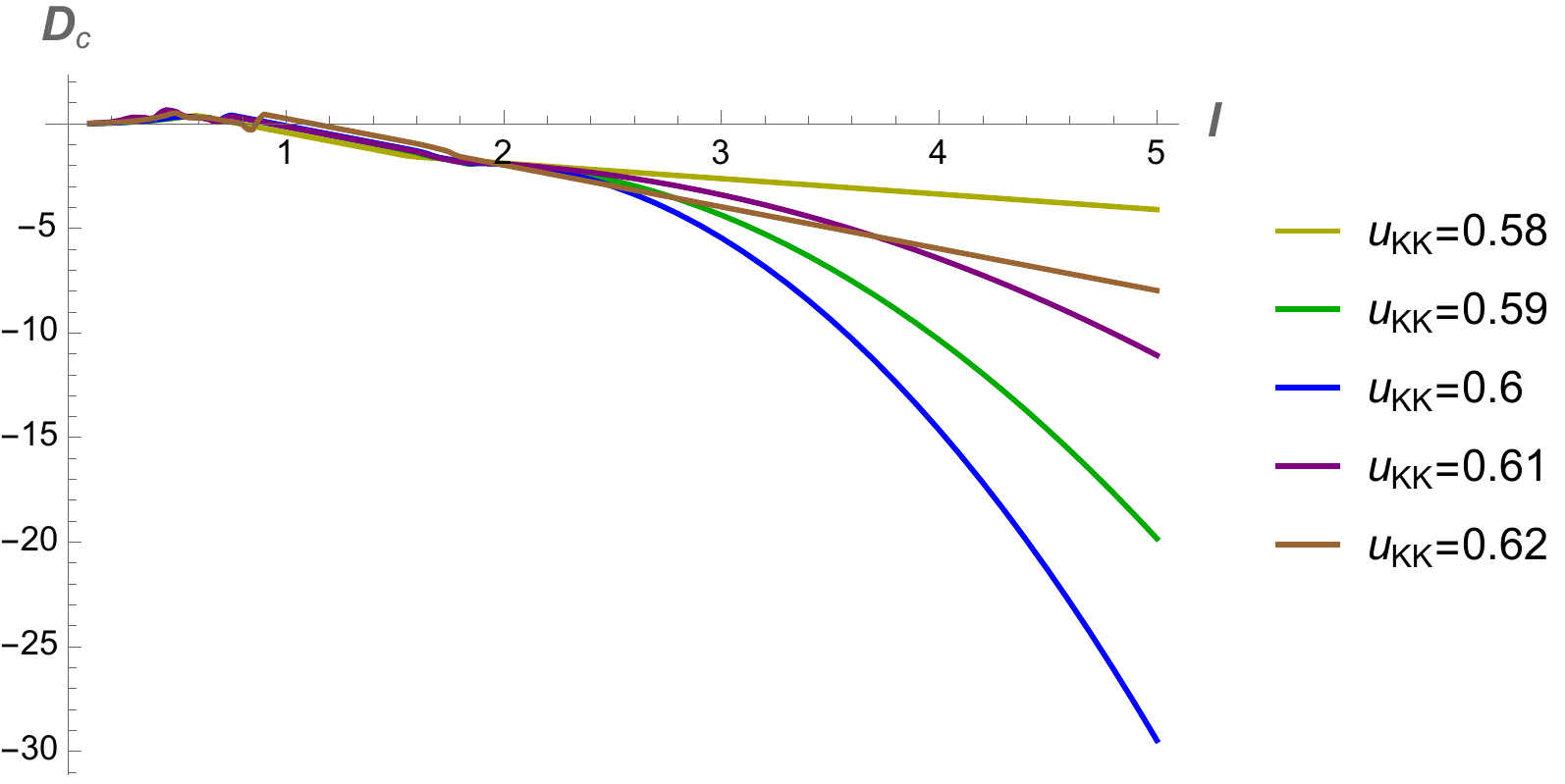}  
  \caption{ Phase three or the range of $u_\Lambda$ which $D_c$ is negative is shown in the left part, and the phase four is shown in the right part. }
 \label{fig:Dcwpzero}
\end{figure}

Then, the critical distance $D_c$ between the two strips versus their width $l$ could be studied, where again different values of $u_\Lambda$ can capture many interesting phase structures in this background, where the results are shown in figures \ref{fig:Dcwallp1}, \ref{fig:Dcwallp2}, and \ref{fig:Dcwpzero}.

Increasing $u_\Lambda$ after 0.49 would make $D_c$ negative which is not an acceptable range of parameter for this quantity, but still the behavior is shown in figure \ref{fig:Dcwpzero}.

More phases can be detected by increasing $u_\Lambda$, but it would become more noisy.

For this case, the results are less noisy but again some sharp jumps can be observed. The behavior of the quasi-normal modes would have significant effects on the quantum information measures in this case.

\begin{figure}[ht!]
 \centering
  \includegraphics[width=9.5 cm] {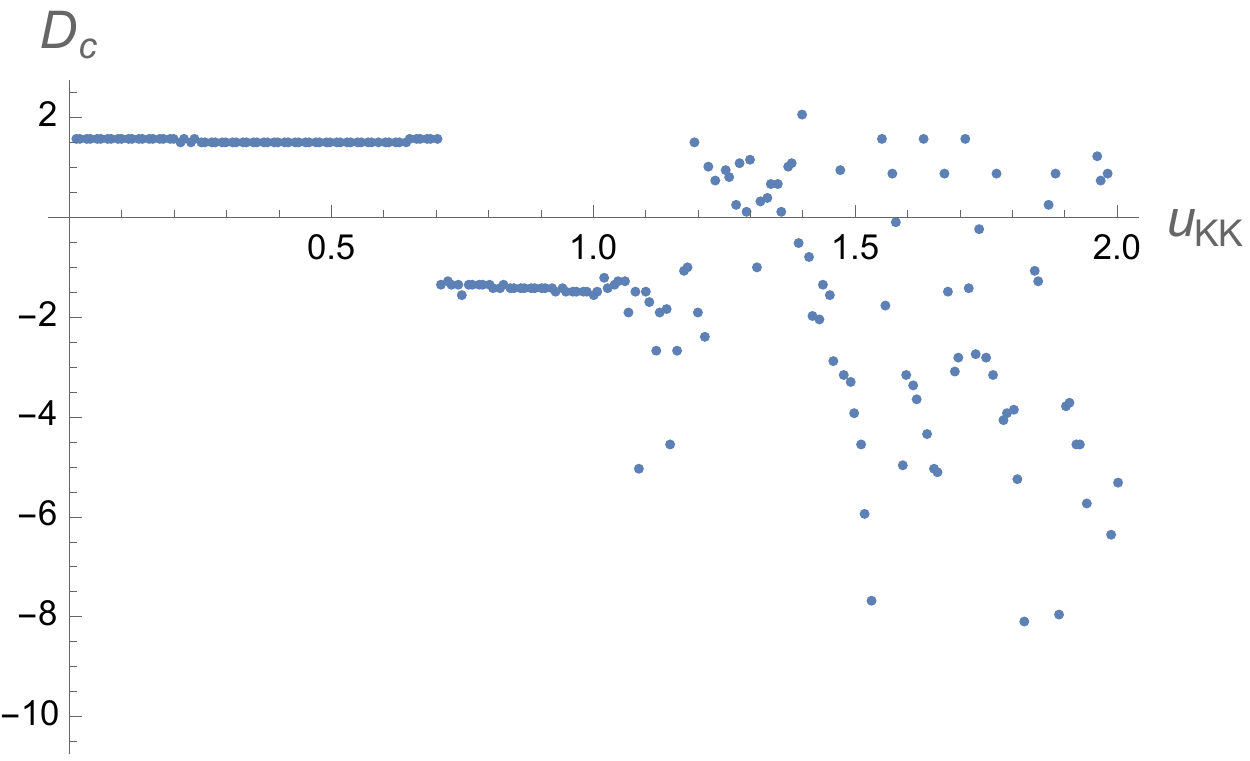} 
  \caption{The behavior of $D_c$ versus $u_{KK}$ (or $u_{\Lambda}$) in the Domain-Wall metric. }
 \label{fig:DomainWallDataPoint}
\end{figure}

The behavior of $D_c$ versus $u_{KK}$ (or $u_{\Lambda}$) in Domain-Wall metric is shown in figure \ref{fig:DomainWallDataPoint}.

It would then be interesting to compare the behavior of quasi-normal modes (QNM) of D-branes, and the geodesic motions in each of these models as in \cite{Bianchi:2021mft} and check the connections between the quantum Sieberg-Witten (SW) curves and the gravitational perturbations dubbed SW-QNM correspondence. Specially, the behavior of QNMs around the phase transitions and the connections with the critical distance, mutual information, EoP and negativity could be studied.

\subsection{Nunez-Legramandi metric}
Another interesting confining geometry that can be studied, is the new geometry constructed in \cite{Legramandi:2021aqv}, written in the form of
\begin{gather}
ds_{10, st}^2= f_1 \Big \lbrack ds^2 (\text{AdS}_6) +f_2 ds^2 (S^2) + f_3 ( d \sigma^2 + d \eta^2 ) \Big \rbrack, \nonumber\\
f_1= \frac{2}{3} \sqrt {\sigma^2 + \frac{3\sigma \partial_\sigma V}{\partial^2_\eta V} }, \ \ \ f_2=\frac{\partial_\sigma V \partial^2_\eta V}{3 \Lambda}, \ \ \
f_3=\frac{\partial^2_\eta V}{3\sigma \partial_\sigma V}, \nonumber\\ 
 \Lambda=3 \partial_\eta^2 V \partial_\sigma V + \sigma \left \lbrack (\partial^2_{\eta \sigma} V)^2+(\partial^2_\eta V)^2      \right \rbrack, \nonumber\\
f_4= \frac{2}{9} \left ( \eta -  \frac{ (\sigma \partial_\sigma V)(\partial_\sigma \partial_\eta V)  } {\Lambda} \right), \ \ \ f_5= 4 \left ( V -\frac{\sigma \partial_\sigma V}{\Lambda} (\partial_\eta V ( \partial_\sigma \partial_\eta V)-3(\partial^2_\eta V ) ( \partial_\sigma V )) \right ), \nonumber\\ f_6=18^2 \frac{3 \sigma^2 \partial_\sigma V \partial^2_\eta V}{(3 \partial_\sigma V + \sigma \partial^2_\eta V)^2} \Lambda, \ \ \ f_7= 18 \left ( \partial_\eta V + \frac{(3\sigma \partial_\sigma V)(\partial_\sigma \partial_\eta V)}{3 \partial_\sigma V + \sigma \partial^2_\eta V} \right),
\end{gather}
or , it could be written as
\begin{gather}
ds^2_{\text{st}}= f_1 \left \lbrack \frac{2 \tilde{g}^2 }{9}  \left ( e^{2 \rho} ( - d \tau^2 + d \vec{x}_3^2 )+ \frac{d\rho^2}{g(\rho)}+e^{2 \rho} g(\rho) d \psi^2 \right)+f_2 ds^2 (S^2) + f_3 ( d\sigma^2 + d \eta^2 ) \right \rbrack,\nonumber\\
B_2=f_4 \text{Vol} (S^2), \ \ \ \ C_2=f_5 \text{Vol}(S^2), \ \ \ \ C_0=f_7, \ \ \ \ \ e^{-2 \Phi}=f_6,
\end{gather}
where in order to avoid conical singularities, the function $g$ has been chosen as $g(\rho)= 1- e^{5 (\rho_* - \rho ) }$ with the constant $e^{2 \rho_* }= \frac{4}{125}$.

Here $V(\sigma, \eta)$ are potential function which solves the Laplace partial differential equation 
\begin{gather}
\partial_\sigma ( \sigma^2 \partial_\sigma V ) + \sigma^2 \partial^2_\eta V=0,
\end{gather}
and each $V$ which solves this equation, is a solution in Type IIB which contains a circle that asymptotes to $AdS_6 \times S^2 \times \Sigma_2 (\sigma, \eta)$ and a four dimensional Minkowski space-time.

In this metric the flavor branes are not probe branes and actually their back-reactions have been included in the construction of the geometry unlike the case of Sakai-Sugimoto or domain wall, which makes it very interesting in our discussion of using quantum information quantities to probe confinement in the bulk geometry \footnote{We thank Carlos Nunez for this comment.}.

For this geometry we find
\begin{gather}
\alpha=f_1 \frac{2 \tilde{g}^2 }{9} e^{2\rho}, \ \ \ \ \beta(\rho)=\frac{1}{e^{2\rho} g(\rho) }, \nonumber\\
H(\rho)= \frac{2^4}{9^4} \tilde{g}^8 V_5^2 g(\rho) e^{8\rho} f_1^8 f_2^2 f_3^2 f_6^2,
\end{gather}

\begin{figure}[ht!]
 \centering
  \includegraphics[width=6.5 cm] {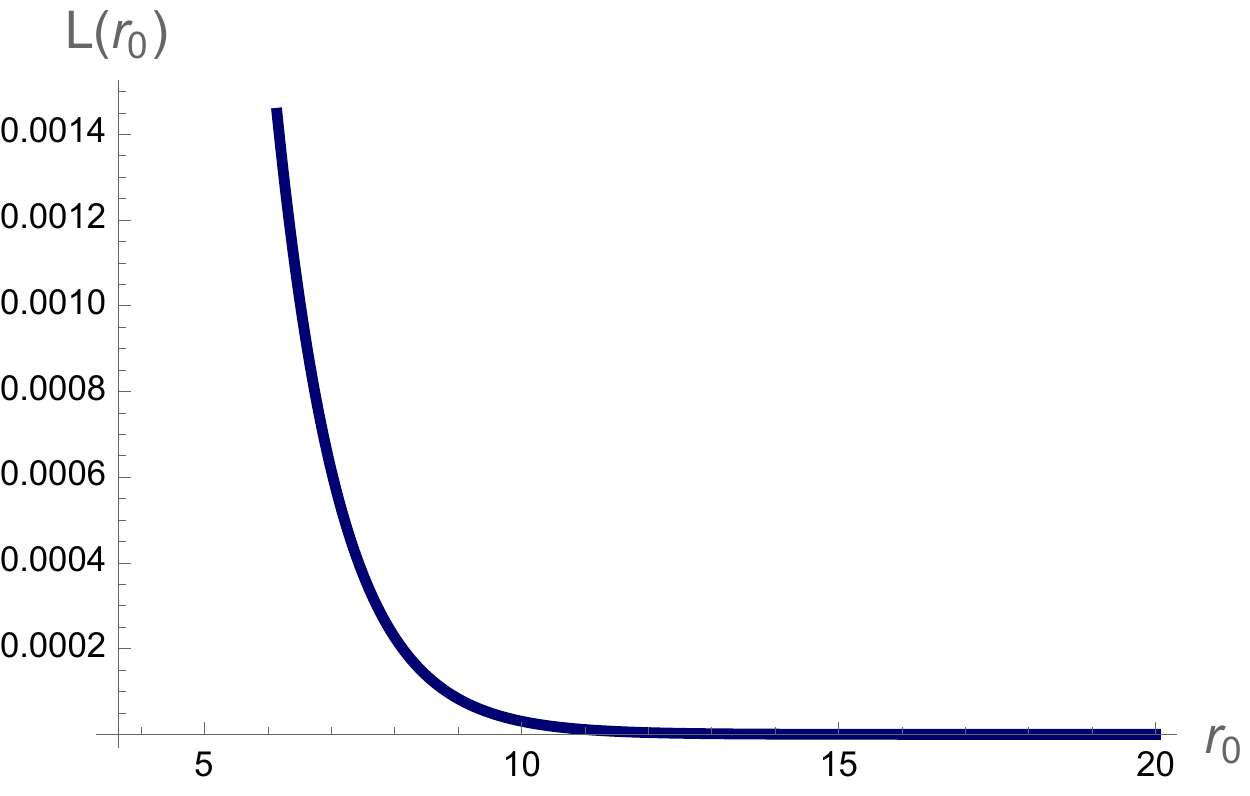} 
    \includegraphics[width=8.5 cm] {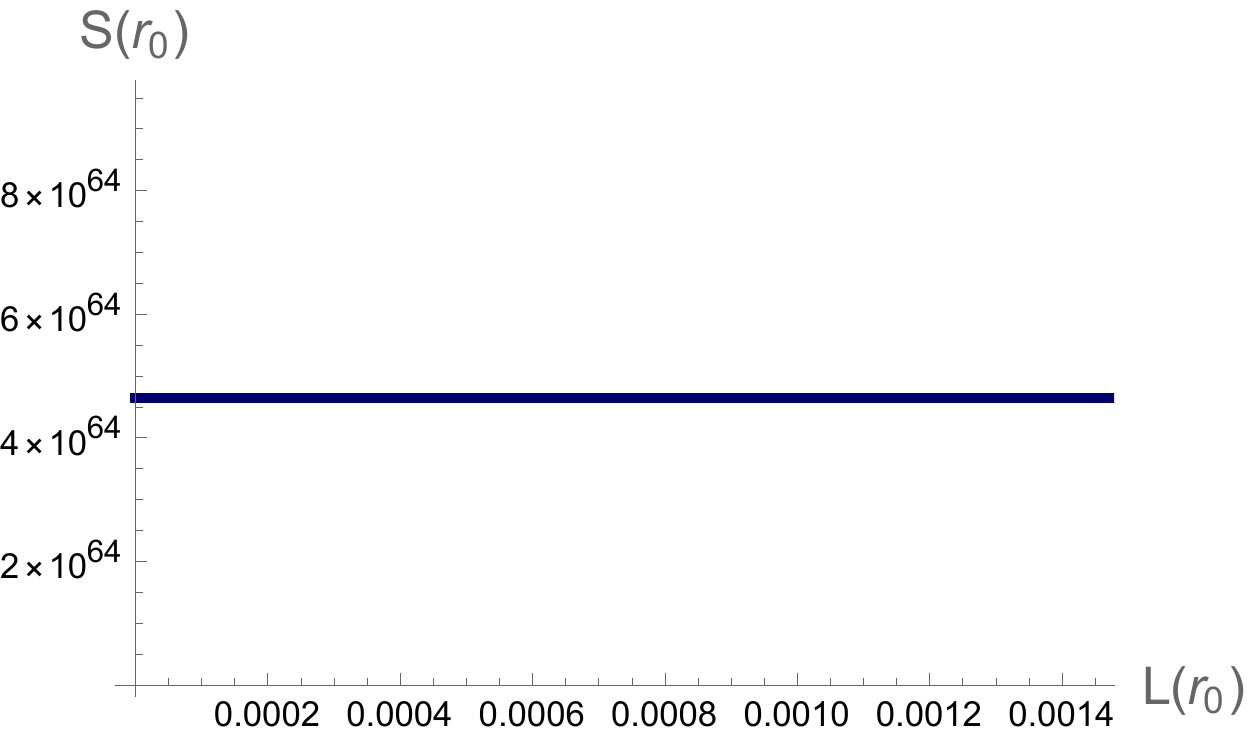} 
  \caption{$S$ and $L$ in Nunez-Legramandi metric. }
 \label{fig:LSNunez1}
\end{figure}

Note that $f_i (\sigma, \eta)$s are not a function of the coordinate $\rho$. So for width and entanglement entropy we get
\begin{gather}
L(\rho_0)= 2 \int_{\rho_0}^\infty d\rho  \frac{e^{-\rho}}{\sqrt{\frac{\left(1- \frac{2^5}{ (5 \sqrt{5} )^5 }  e^{-5\rho} \right)^2}{1- \frac{2^5}{ (5 \sqrt{5} )^5 }  e^{-5\rho_0} } e^{8(\rho-\rho_0)}  -1}}, \nonumber\\
S_C(\rho_0)= \frac{2}{9^2 G_N^{(10)} } \tilde{g}^4 \int dV_{\text{int}} f_1^4 f_2 f_3 f_6 \int_{\rho_0}^\infty d\rho \frac{e^{3\rho} }{ \sqrt{1- \frac{e^{3\rho_0} }{e^{3\rho} }\frac{e^{5\rho_0}- \frac{2^5}{ (5 \sqrt{5} )^5 } }{e^{5\rho}-\frac{2^5}{ (5 \sqrt{5} )^5 } } } },
\end{gather}
which their behaviors are shown in \ref{fig:LSNunez1}.

\begin{figure}[ht!]
 \centering
  \includegraphics[width=7.5 cm] {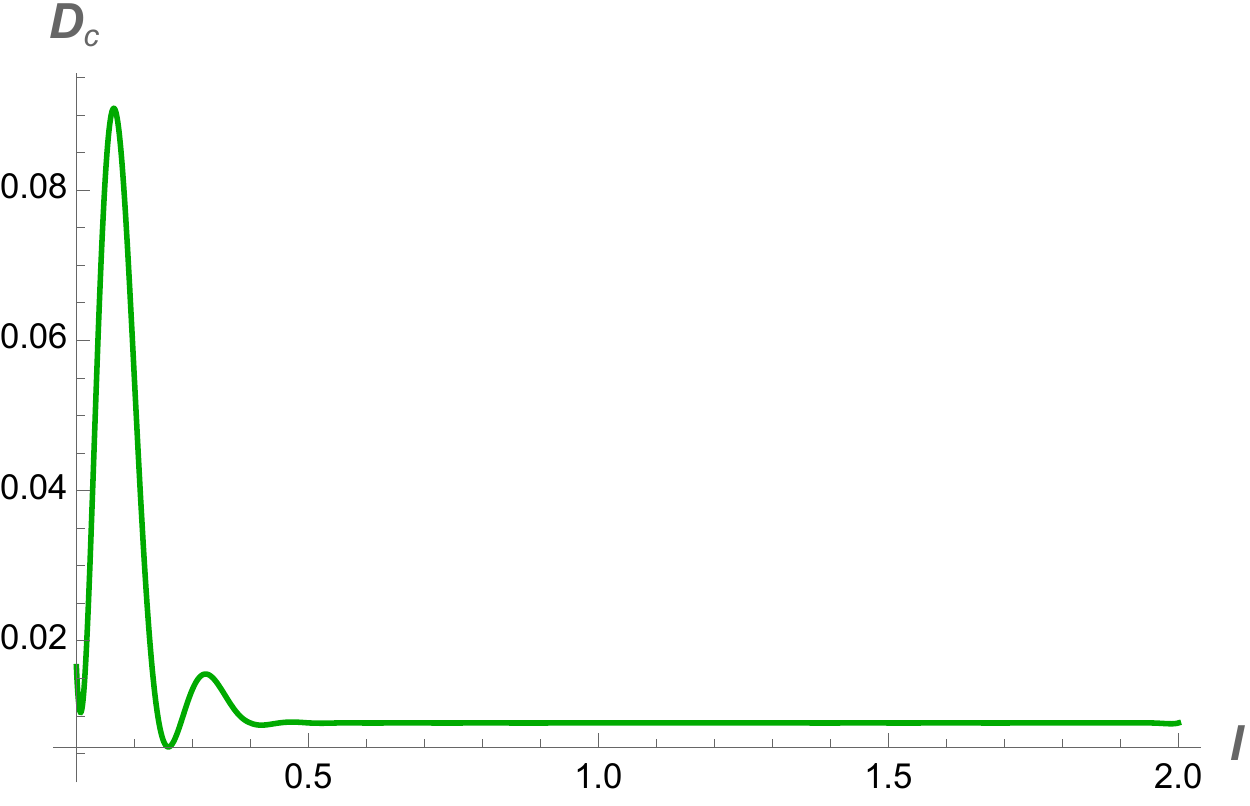} 
  \caption{$D_c$ in Nunez-Legramandi metric. }
 \label{fig:DcNunez1}
\end{figure}

Again, using the relation for the entanglement entropy in this background, the  mutual information for two strips can be calculated and then the distance $D$ where the mutual information drops to zero is the critical distance $D_c$ which its behavior versus width of the strip $l$ has been shown in figure \ref{fig:DcNunez1}. This behavior is expected since this geometry is confining and therefore $D_c$ first goes up and then comes down as $l$ being increased and then becomes constant.

\subsection{Backgrounds with smeared branes}

The previous calculations were for the cases with fully-localized D3/D7 intersection, where the branes are located at a fix location in the transverse direction. The results can also be compared with other backgrounds and also with the geometries where these branes are smeared.

The metric for those backgrounds could be written as
\begin{gather}
ds^2_{10}=h^{-1/2} \eta_{\mu \nu} dx^\mu dx^\nu+h^{1/2} \Big( d\rho^2+ \rho^2 d \Omega_3^2 + e^{-\phi} ( d\omega^2 + \omega^2 d \theta^2 ) \Big),
\end{gather}
where here the (near-core) wrap factor $h=h(\rho, \omega) $ is
\begin{gather}
h(\rho, \omega)=1+\frac{R^4}{ (\rho^2+ e^{- \phi} \omega^2  )^2 }\  ,
\end{gather}
where $R^4=4\pi g_s N \alpha'^2$ and the dilaton field $\phi$ and axion field $\chi$ can be found from the relations $
e^{- \phi(\omega) }= \beta_0 \log \frac{\omega^2_\Lambda}{\omega}$ and  $\chi(\theta)= \frac{N_f}{2\pi} \theta$, where the integration constants are $\beta_0=\frac{N_f}{4\pi}$ and $ \omega^2_\Lambda= \omega_0^2 e^{1/(g_s \beta_0) }$.

The main difference about this geometry is that the factor $h(\rho, \omega)$ is a factor of both of the coordinates $\rho$ and $\omega$ which makes the calculations more difficult and we leave it for the future works. Note that, for the small values of $\omega$, the near-core region of D3/D7 intersection corresponds to the IR region of the dual field theory. This background contains RR four-form potential of the D3-brane solution which then its effects on the correlations between the two mixed strips could be studied.

Other anisotropic holographic models for heavy quarks which consist of Einstein-dilaton-three-Maxwell action have been constructed in \cite{Arefeva:2020byn} and \cite{Arefeva:2020vae}, where the effects of magnetic field have been investigated. In these models also the mixed quantum information measures such as negativity or entanglement of purification, or Crofton form and kinematic space could be studied.

\begin{figure}[ht!]
 \centering
  \includegraphics[width=9 cm] {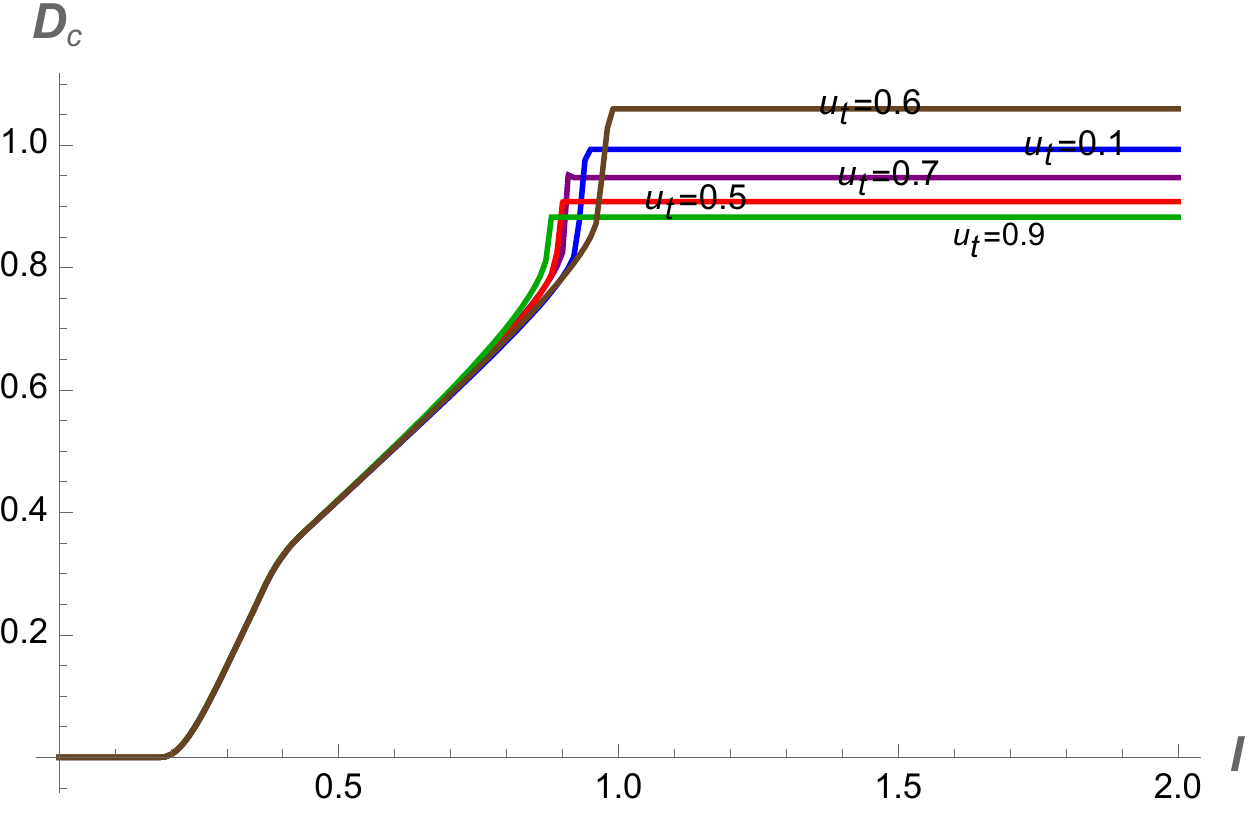} 
  \caption{The phase diagram for calculating $D_c$, coming from $S_C-S_D$. }
 \label{fig:cregularizedphases}
\end{figure}

Note also that for constructing these results we just used the connected part of the entanglement entropy, i.e, $S_C$. We could also use the regularized entropy, $S_C-S_D$, where $S_D$ is the solution for the ``disconnected part'' which is just a constant, and then in the numerical codes we could take the minimum of $S_C$ and $S_D$. For example for the Witten-QCD case, we could get somehow a different phase diagram as shown in figure \ref{fig:cregularizedphases}. But the general results would be similar and no more phases cab be detected using this quantity instead of $S_C$.

\section{Crofton form in confining backgrounds}\label{sec:Crofton}

For checking the interconnections between geometry and topology, and the quantum information measures, it would be interesting to calculate the Crofton form for these confining models as well and compare its behaviors for each case.

In \cite{Czech:2015qta}, using ideas from integral geometry, the authors found the connections between the length of a curve and the number of geodesics, or ``random'' lines it would be intersected and therefore its connection to the Crofton formula. This formula can give further intuitions about the structures of these various confining backgrounds from the perspective of the bulk reconstruction and can shed more lights on their various specific properties such as singularities, throats, bulges, walls, etc and their effects on the correlations between the two strips in the mixed setup in such confining models. Therefore, the behavior of the Crofton form for each space will be presented here.

\subsection{AdS-soliton}
\begin{figure}[ht!]
 \centering
  \includegraphics[width=8.5 cm] {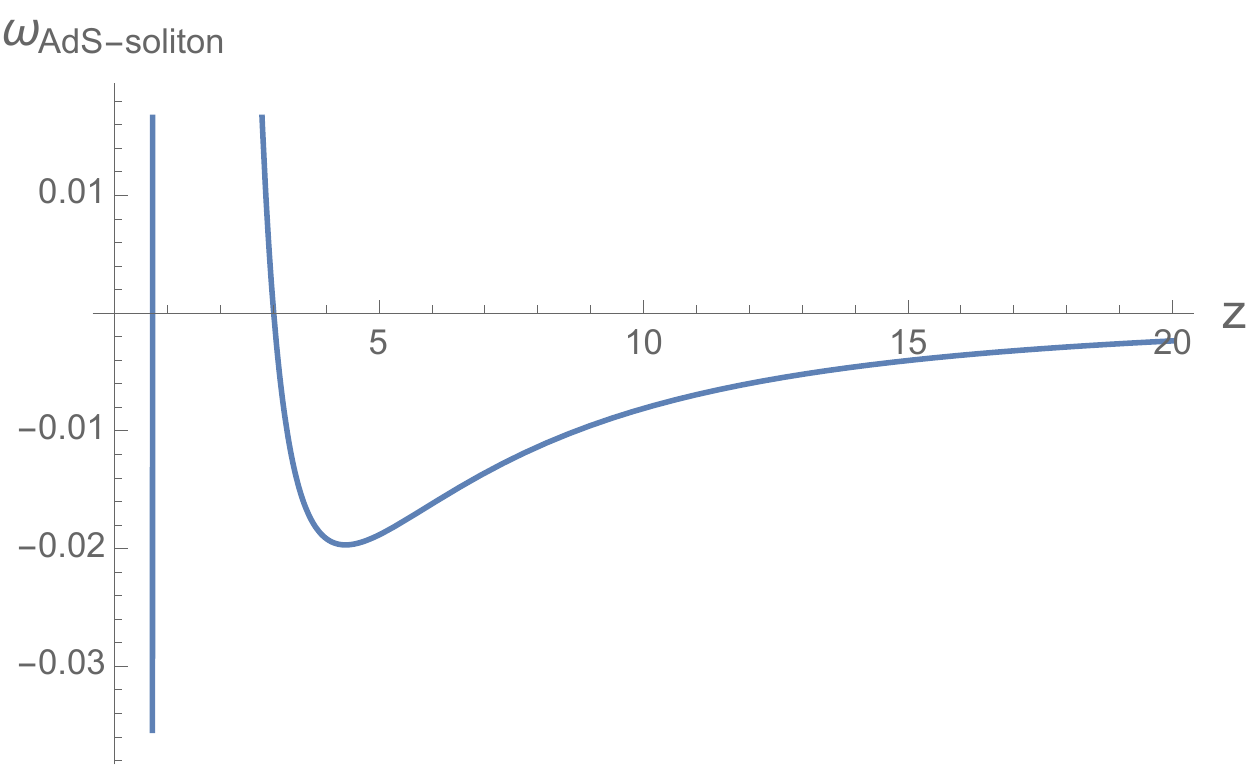} 
  \caption{The plot of Crofton form $\omega$ vs. $z$ in AdS-soliton background. }
 \label{fig:croftonsoliton}
\end{figure}

The Crofton form for the AdS-soliton geometry with the metric \ref{eq:maingeometry} would be
\begin{gather}
\omega_{\text{AdS-soliton} }= \frac{\left(2 z_0-\frac{(d-6) z^8 \left(\frac{z}{z_0}\right){}^{-d}}{z_0^7}\right) \left(1-\left(\frac{z_t}{z_0}\right){}^{8-d}\right)}{2 z_0  z_t^2 \left(1-\left(\frac{z}{z_0}\right){}^{8-d}\right)^2 \left(1-\frac{z^2 \left(1-\left(\frac{z_t}{z_0}\right){}^{8-d}\right)}{\left(1-\left(\frac{z}{z_0}\right){}^{8-d}\right) z_t^2}\right)^{\frac{3}{2} }}-\frac{1}{z^2 \sqrt{1-\frac{z^2 \left(1-\left(\frac{z_t}{z_0}\right){}^{8-d}\right)}{\left(1-\left(\frac{z}{z_0}\right){}^{8-d}\right){}^2 z_t}}},
\end{gather}
which its behavior is shown in figure \ref{fig:croftonsoliton}.

\begin{figure}[ht!]
 \centering
  \includegraphics[width=8.5 cm] {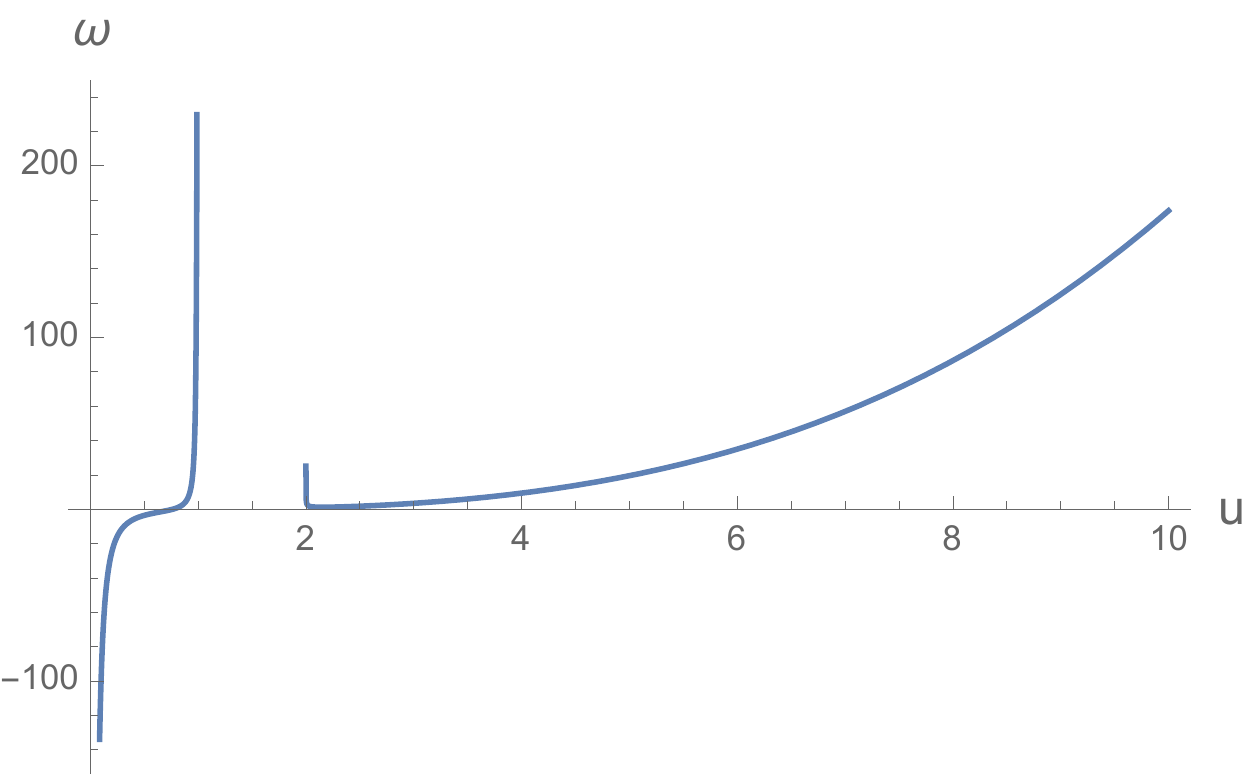} 
  \caption{The plot of Crofton form $\omega$ vs. $z$ in AdS-soliton background. }
 \label{fig:soliton1}
\end{figure}

\begin{figure}[ht!]
 \centering
  \includegraphics[width=8.5 cm] {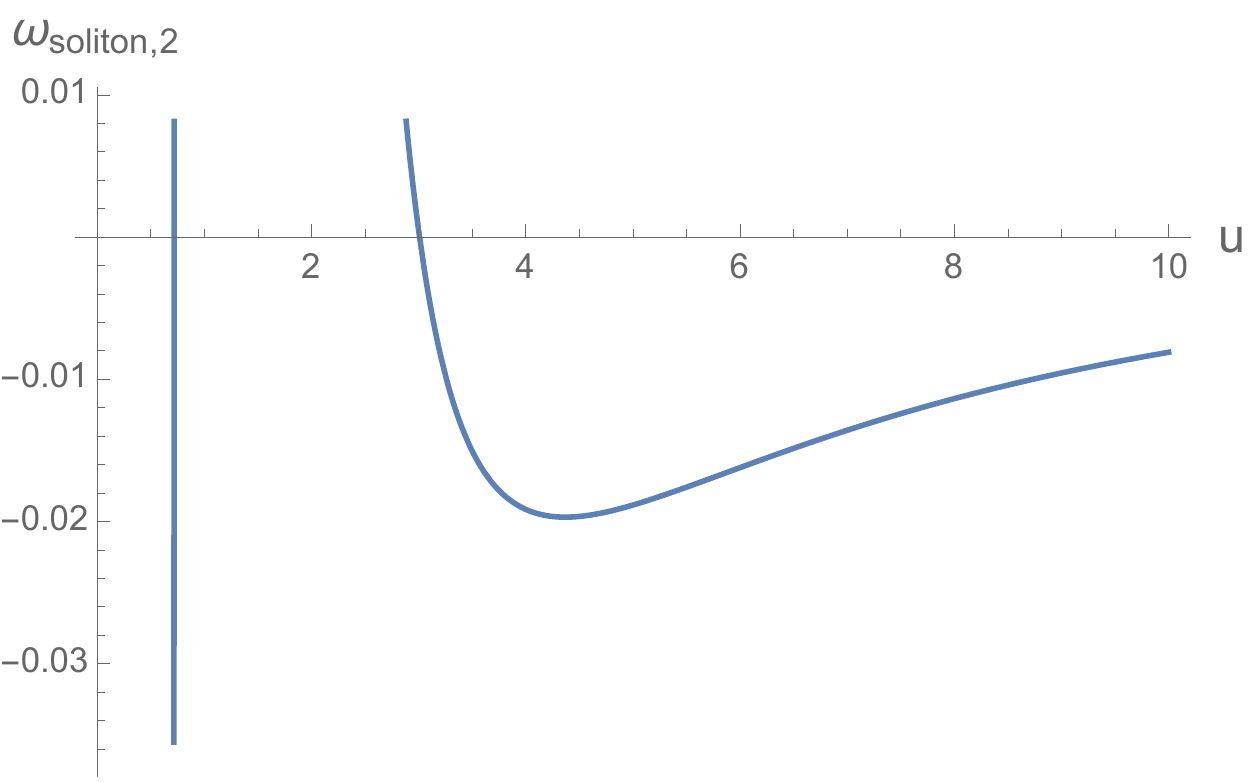} 
  \caption{The plot of Crofton form $\omega$ vs. $z$ in AdS-soliton background. }
 \label{fig:soliton2}
\end{figure}

If one uses other forms of metric for AdS soliton, for instance the form of metric of \ref{eq:solitonform2}, one could get plots with a general similar behavior as shown in figures \ref{fig:soliton1} or \ref{fig:soliton2}.

So one could see that the standard behavior is similar, excepts for some constants, as a well is being formed around the singularity, or the turning points, while in the larger AdS-radius, $\omega$ increases, with also a decreasing gradient, or as we see later, it would become constant for $u \to \infty$.

\begin{figure}[ht!]
 \centering
  \includegraphics[width=8 cm] {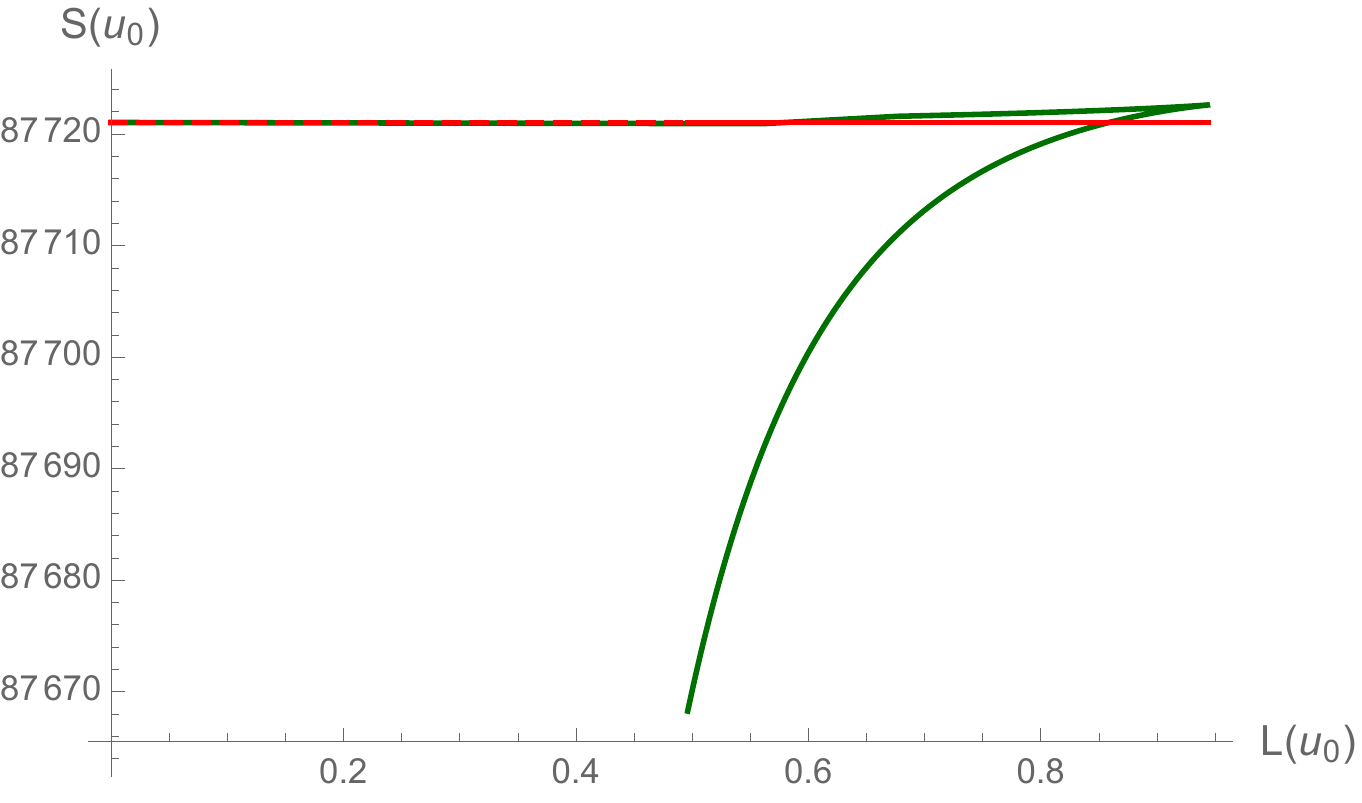} 
  \caption{The general behavior $S$ vs. $ L$ in confining models. The second derivate of the green curve (versus $L$ instead of $u$) could be used to define a new quantity similar to the Crofton form which could give further information about the interplay between geometry and information.}
 \label{fig:EEcrofton}
\end{figure}

Note that a similar quantity to Crofton form could also be defined by $\frac{\partial^2 S}{\partial L^2}$ as well, which after $L_{\text{crit}}$ it becomes zero and before that, it is the second derivative of the green curve of figure \ref{fig:EEcrofton}, and for large values of  $u$, it becomes constant. This new quantity could also give further information. 

In addition, in works such as \cite{Espindola:2018ozt}, it has been shown that not all the curves in the entanglement wedge are reconstructable from entanglement entropies and rather there are ``entanglement shades" which for their reconstruction, in addition to differential entropy, the slightly generalized version of entanglement of purification, called ``differential purification" would also be needed  \footnote{We thank the referee for this comment.}. So based on this idea, another version of Crofton form can also be defined using this differential purification, as $\frac{\partial^2 \text{(EoP)}}{\partial L^2}$, which then can be implemented in confining geometries to probe further regions of the bulk and probably find more phases. We leave such studies for the future works and continue with the usual Crofton form here.

\subsection{Sakai-Sugimoto and deformed Sakai-Sugimoto}
The Crofton form, $\omega$, for the Sakai-Sugimoto geometry would be
\begin{gather}
\omega_{\text{Sakai-Sugimoto} }=\frac{V_3 V_4 R^3_{D_4}  } {2g_s^2 G_N^{(10) } }   \frac{u^4 \Big(10 u_{\text{KK}}^3 u_t^5-5 u^5 u_{\text{KK}}^3-7 u^3 u_t^5+2 u^8\Big)}{2 \Big((u^3-u_{\text{KK}}^3) (u^5-u_t^5)  \Big)^{\frac{3}{2} }}  du \wedge d\theta.
\end{gather}

\begin{figure}[ht!]
 \centering
  \includegraphics[width=8.5 cm] {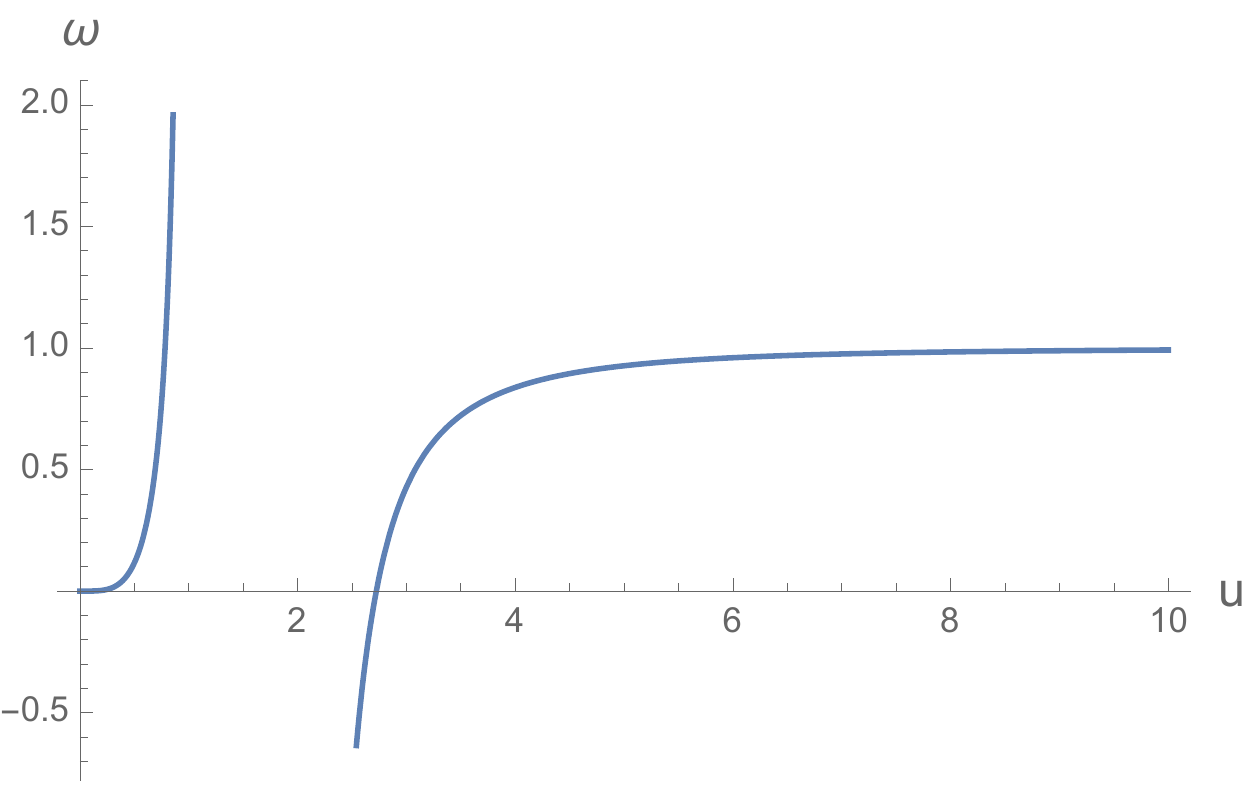} 
  \caption{The plot of Crofton form $\omega$ vs. $r$ in the Sakai-Sugimoto background.}
 \label{fig:croftonSS}
\end{figure}

The plot of $\omega$ versus $u$ for $u_{KK}=1$ and $u_t=2$ has been shown in figure \ref{fig:croftonSS}. One can again detect a well around $u_t$ and it becomes constant in the large values of $u$.

For the deformed case we get two additional term as
\begin{equation} \label{deformedcroft}
\begin{split}
\omega_{\text{deformed Sakai-Sugimoto} } &=\frac{V_3 V_4 R^3_{D_4}  } {2g_s^2 G_N^{(10) } } \Bigg \lbrack  \frac{u^4 \Big(10 u_{\text{KK}}^3 u_t^5-5 u^5 u_{\text{KK}}^3-7 u^3 u_t^5+2 u^8\Big)}{2 \Big((u^3-u_{\text{KK}}^3) (u^5-u_t^5)  \Big)^{\frac{3}{2} }}  \nonumber\\ &
+  \frac{d^2 x^4 (u) }{du^2} \sqrt{\frac{u^5 (1- \frac{u^3_{KK} }{u^3} ) }{1-\frac{u_t^5}{u^5} } }  \nonumber\\ &
+\frac{dx^4(u) }{du} \frac{u^{5/2} \left(7 u_{\text{KK}}^3 u_t^5-2 u^5 u_{\text{KK}}^3-10 u^3 u_t^5+5 u^8\right)}{2 \Big((u^3-u_{\text{KK}}^3) (u^5-u_t^5)  \Big)^{\frac{3}{2} }} \Bigg \rbrack du \wedge d\theta,
\end{split}
\end{equation}

Then, using the relation \ref{eq:deformedrel}, we get
\begin{equation} \label{deformedcroft}
\begin{split}
\omega_{\text{deformed Sakai-Sugimoto} } &=\frac{V_3 V_4 R^3_{D_4}  } {2g_s^2 G_N^{(10) } } \frac{u^4}{2 \left({\frac{u^8-u^5 u_{\text{KK}}^3}{u_0^8-u_0^5 u_{\text{KK}}^3}-1}\right)^{\frac{1}{2}} \left(u^3-u_{\text{KK}}^3\right){}^{\frac{3}{2} } \left(u^5-u_t\right){}^{\frac{3}{2} }}\times  \nonumber\\ &
\Bigg (  \frac{R_{D_4}^{\frac{3}{2}} (3 u_0^8 u^3-(2 u^5+3 u_0^5) u_{\text{KK}}^6+(13 u^8-3 u_0^5 u^3+3 u_0^8) u_{\text{KK}}^3-11 u^{11}) (u^5-u_t^5)}{u_0^5 u_{\text{KK}}^3-u^5 u_{\text{KK}}^3+u^8-u_0^8}  \nonumber\\ &
+R_{D_4}^{\frac{3}{2}} \left(7 u_{\text{KK}}^3 u_t^5-2 u^5 u_{\text{KK}}^3-10 u^3 u_t^5+5 u^8\right)  \nonumber\\ &
+ \Big(\frac{u^8-u^5 u_{\text{KK}}^3}{u_0^8-u_0^5 u_{\text{KK}}^3}-1\Big)^\frac{1}{2} \left(10 u_{\text{KK}}^3 u_t^5-5 u^5 u_{\text{KK}}^3-7 u^3 u_t^5+2 u^8\right)  \Bigg ) du \wedge d\theta,
\end{split}
\end{equation}

\begin{figure}[ht!]
 \centering
  \includegraphics[width=8.5 cm] {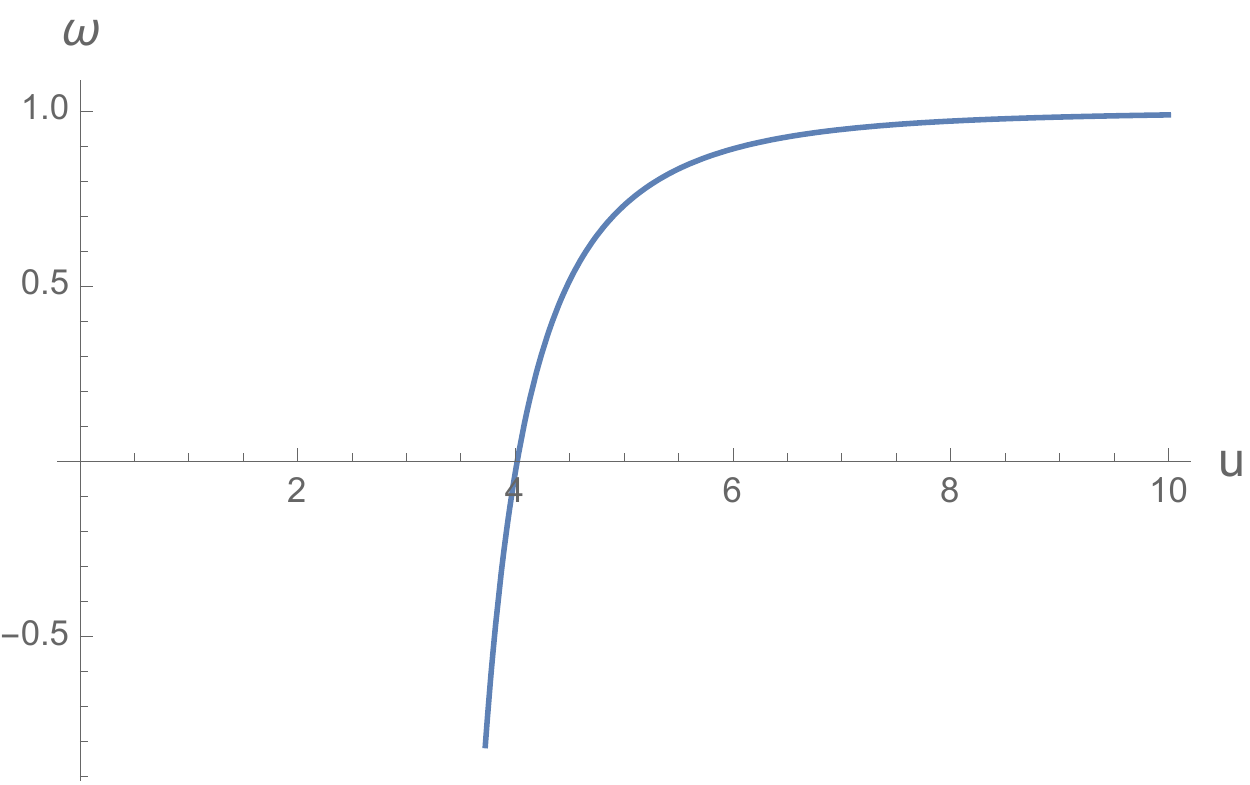} 
  \caption{The plot of Crofton form $\omega$ vs. $r$ in the \textit{deformed Sakai-Sugimoto background}. }
 \label{fig:croftondSS}
\end{figure}

The general behavior of  $\omega_{\text{deformed Sakai-Sugimoto} }$ vs. $u$ is shown in figure \ref{fig:croftondSS}.

One could see that for the intermediate and bigger values of $u$ the behavior is similar to the Sakai-Sugimoto case, as one would expect.

\subsection{Klebanov-Strassler}

The Crofton form in the KS background would be

\begin{equation} \label{deformedcroft}
\begin{split}
\omega_{\text{KS} } &= \frac{2^{2/3} \pi ^3 V_2 \epsilon ^{-8/3} \epsilon ^4 (\alpha_p {g_s} M)^2}{3 G_N} \frac{h(\tau ) \sinh (2 \tau )}{\sqrt{1-\sinh ^2\left(\tau _0\right) \text{csch}^2(\tau ) \left(\frac{\tau _0-\sinh \left(\tau _0\right) \cosh \left(\tau _0\right)}{\tau -\sinh (\tau ) \cosh (\tau )}\right){}^{\frac{2}{3} }}}-  \nonumber\\ &
\frac{h(\tau ) \sinh ^2\left(\tau _0\right) \left(2 \tau _0-\sinh \left(2 \tau _0\right)\right) (12 \tau  \coth (\tau )-5 \sinh (3 \tau ) \text{csch}(\tau )+3)}{6 (\sinh (2 \tau )-2 \tau )^2 \sqrt[3]{\frac{\tau _0-\sinh \left(\tau _0\right) \cosh \left(\tau _0\right)}{\tau -\sinh (\tau ) \cosh (\tau )}} \left(1-\sinh ^2\left(\tau _0\right) \text{csch}^2(\tau ) \left(\frac{\tau _0-\sinh \left(\tau _0\right) \cosh \left(\tau _0\right)}{\tau -\sinh (\tau ) \cosh (\tau )}\right){}^{\frac{2}{3}}\right){}^{\frac{3}{2}}}+ \nonumber\\ &
\frac{\sinh ^2(\tau ) h'(\tau )}{\sqrt{1-\sinh ^2\left(\tau _0\right) \text{csch}^2(\tau ) \left(\frac{\tau _0-\sinh \left(\tau _0\right) \cosh \left(\tau _0\right)}{\tau -\sinh (\tau ) \cosh (\tau )}\right){}^{\frac{2}{3}}}}.
\end{split}
\end{equation}

Due to the integral in the relation of $h(\tau)$ function, the analytic solution could not be found and the Crofton form for the KS could only be studied numerically. We expect its behavior would be very similar to the KT case, as in figure \ref{fig:croftonKT}.

\subsection{Witten-QCD}

The Crofton form for the Witten-QCD model is

\begin{equation} \label{wittenqcdcroft}
\begin{split}
\omega & =\frac{-1}{2} \partial^2_u S du \wedge d\theta= \nonumber\\&
 \frac{V_2}{G^{(10)}_N} \frac{4\pi^2 R^{\frac{9}{2}}}{9 g_s^2 \sqrt{u_t} }  \left(  \frac{u \left(2 \left(u^2-2 u_0^2\right) u_t^6+\left(7 u_0^2 u^3+4 u_0^5-4 u^5\right) u_t^3+2 u^8-7 u_0^5 u^3\right)}{2 \left(u^5-u_0^5 -u^2 \left(u_t^3-u_0^2\right) \right){}^{\frac{3}{2}} \sqrt{u^3-u_t^3}}  \right) du \wedge d\theta. 
\end{split}
\end{equation}
Note that again we only use the connected solution for entanglement entropy here.

\begin{figure}[ht!]
 \centering
  \includegraphics[width=8.5 cm] {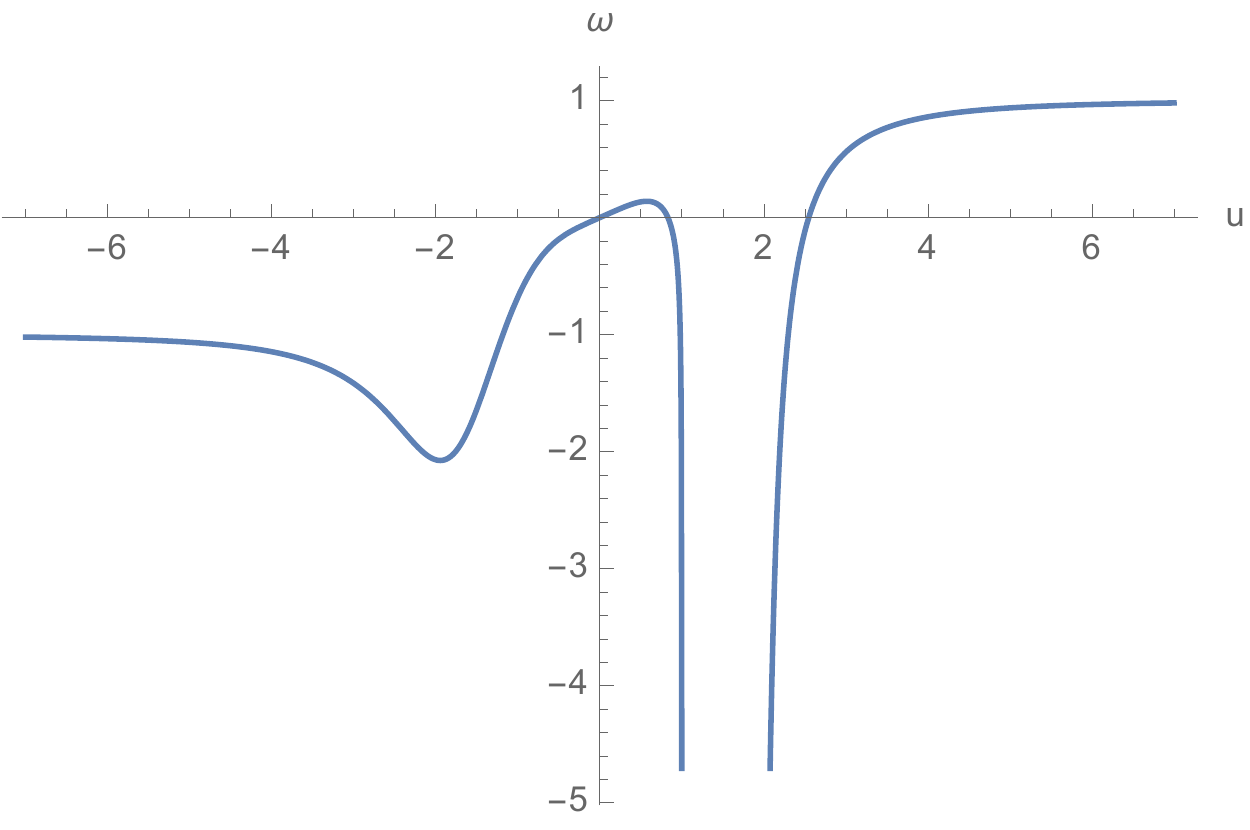} 
  \caption{The plot of Crofton form $\omega$ vs. $u$ in WQCD background. }
 \label{fig:croftonwQCD}
\end{figure}

The behavior of Crofton form for $u_0=2$ and $u_t=1$ in WQCD background has been shown in figure \ref{fig:croftonwQCD} where one can see that around $u_t$ and $u_0$ we have an infinite well.

\subsection{Maldacena-Nunez}

The Crofton form in MN background would be
\begin{gather}
\omega_{MN}= c_{MN} \frac{2 \sinh (4 r) \left(\sinh ^4(2 r)-2 \sinh ^4\left(2 r_0\right)\right)}{\left(\sinh ^4(2 r)-\sinh ^4\left(2 r_0\right)\right) \sqrt{1-\sinh ^4\left(2 r_0\right) \text{csch}^4(2 r)}},
\end{gather}
where $c_{MN}=\frac{V_2 \pi^3 e^{4\phi_0} }{G_N^{(10)} } $.

\begin{figure}[ht!]
 \centering
  \includegraphics[width=8.5 cm] {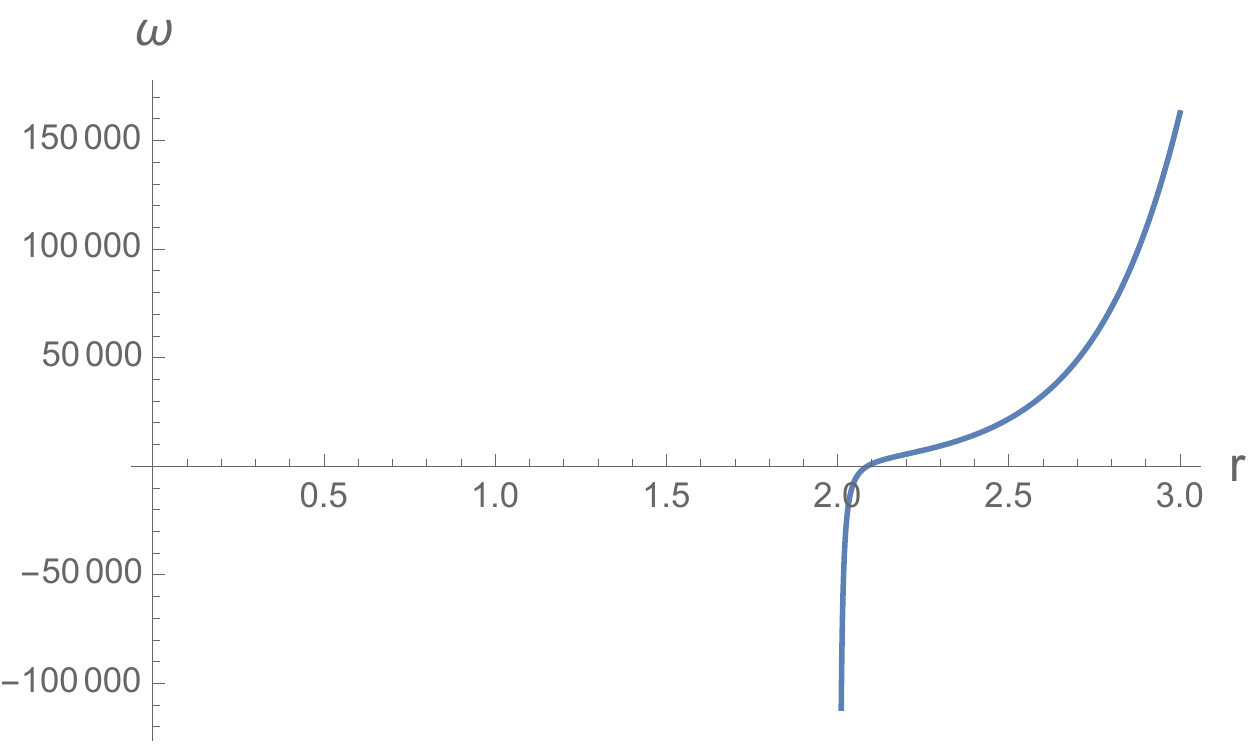} 
  \caption{The plot of Crofton form $\omega$ vs. $r$ in the MN background. }
 \label{fig:croftonMN}
\end{figure}

The plot of Crofton form for $r_0=2$ in MN case has been shown in figure \ref{fig:croftonMN}. Again a deep well could be detected.

\subsection{Klebanov-Tseytlin}

The Crofton form for the KT background would be
\begin{gather}
\omega_{KT}=c_{KT}\frac{2 r^9 \log ^{\frac{3}{2}} (\frac{r}{r_s} )  \big(\log  (\frac{r}{r_s} )+1 \big)-r_0^6 r^3 \log (\frac{r_0}{r_s} ) \log^{\frac{1}{2} }  (\frac{r}{r_s} ) \big(8 \log (\frac{r}{r_s} )+3\big )}{2  \left(r^6 \log  \big(\frac{r}{r_s} \big)-r_0^6 \log  \big (\frac{r_0}{r_s} \big) \right)^{\frac{3}{2} }},
\end{gather}
where $c_{KT}=\frac{12 V_2 \pi^3 M^2 g_s \epsilon^4}{ G_N^{(10)} }$ is just a constant.

\begin{figure}[ht!]
 \centering
  \includegraphics[width=8.5 cm] {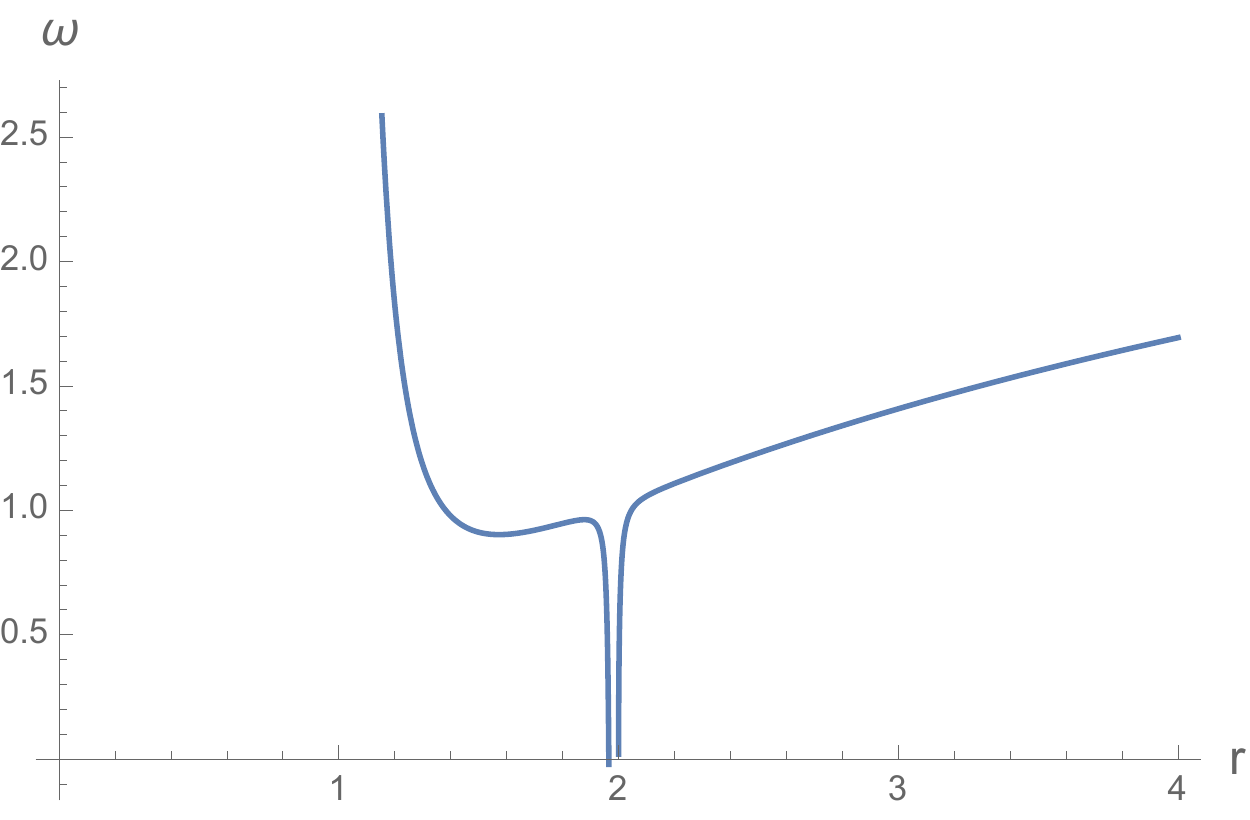} 
  \caption{The plot of Crofton form $\omega$ vs. $r$ in the KT background. }
 \label{fig:croftonKT}
\end{figure}

The plot of $\omega_{KT}$ versus $r$ for singularity point of $r_s=2$ and turning point of $r_0=1$ is shown in figure \ref{fig:croftonKT}.

Again, one can see similar to the WQCD case, for the Crofton form, there is a well around the singularity point.

\subsection{Domain Wall QCD}

The Crofton form for the D5 black brane \ref{eq:D5metric} which were used in the domain wall background \ref{sec:DomainWall} is
\begin{gather}
\omega_{\text{Domain-Wall}}= \frac{u \left(u^2 \left(3 u_t^2 u_{\Lambda }^2-3 u_t^4+u_{\Lambda }^4\right)+2 u_t^2 u_{\Lambda }^2 \left(u_t^2-u_{\Lambda }^2\right)+u^6-2 u^4 u_{\Lambda }^2\right)}{(u^2-u_{\Lambda }^2)^{\frac{1}{2}} \left(\left(u^2-u_t^2\right) \left(u_t^2+u^2-u_{\Lambda }^2\right)\right)^{\frac{3}{2}}}
\end{gather}

 \begin{figure}[ht!]
 \centering
  \includegraphics[width=8.5 cm] {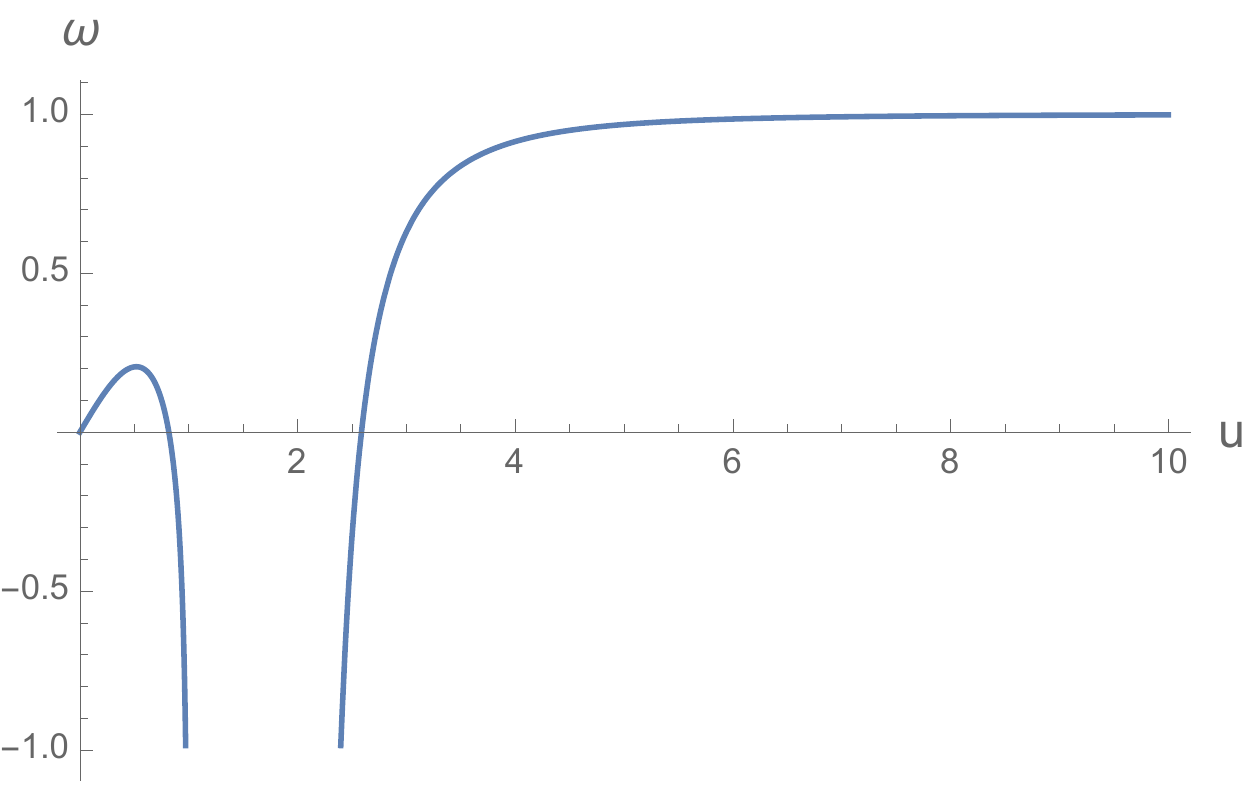}  
  \caption{The Crofron form for D5 black brane geometry for $u_t=2$ and $u_\Lambda=1$. }
 \label{fig:wallp2}
\end{figure}

Again, the hole around the $u_\Lambda$ and $u_t$ could be detected while for the larger values of $u$ it would become constant and again it would point to the structure of this background as the Crofron form can be used as a holographic tool in the bulk reconstruction.

\subsection{Nunez-Legramandi metric}
The scalar part of the Crofton for this metric would be
\begin{gather}
\omega_{NL}= \left( \frac{3 e^{3 \rho }}{\sqrt{1-\frac{\left(e^{5 \text{$\rho $0}}-\frac{32}{78125 \sqrt{5}}\right) e^{3 \text{$\rho $0}-3 \rho }}{e^{5 \rho }-\frac{32}{78125 \sqrt{5}}}}}-\frac{e^{3 \rho } \left(\frac{3 \left(e^{5 \text{$\rho $0}}-\frac{32}{78125 \sqrt{5}}\right) e^{3 \text{$\rho $0}-3 \rho }}{e^{5 \rho }-\frac{32}{78125 \sqrt{5}}}+\frac{5 \left(e^{5 \text{$\rho $0}}-\frac{32}{78125 \sqrt{5}}\right) e^{2 \rho +3 \text{$\rho $0}}}{\left(e^{5 \rho }-\frac{32}{78125 \sqrt{5}}\right)^2}\right)}{2 \left(1-\frac{\left(e^{5 \text{$\rho $0}}-\frac{32}{78125 \sqrt{5}}\right) e^{3 \text{$\rho $0}-3 \rho }}{e^{5 \rho }-\frac{32}{78125 \sqrt{5}}}\right)^{3/2}} \right).
\end{gather}

\begin{figure}[ht!]
 \centering
  \includegraphics[width=8.5 cm] {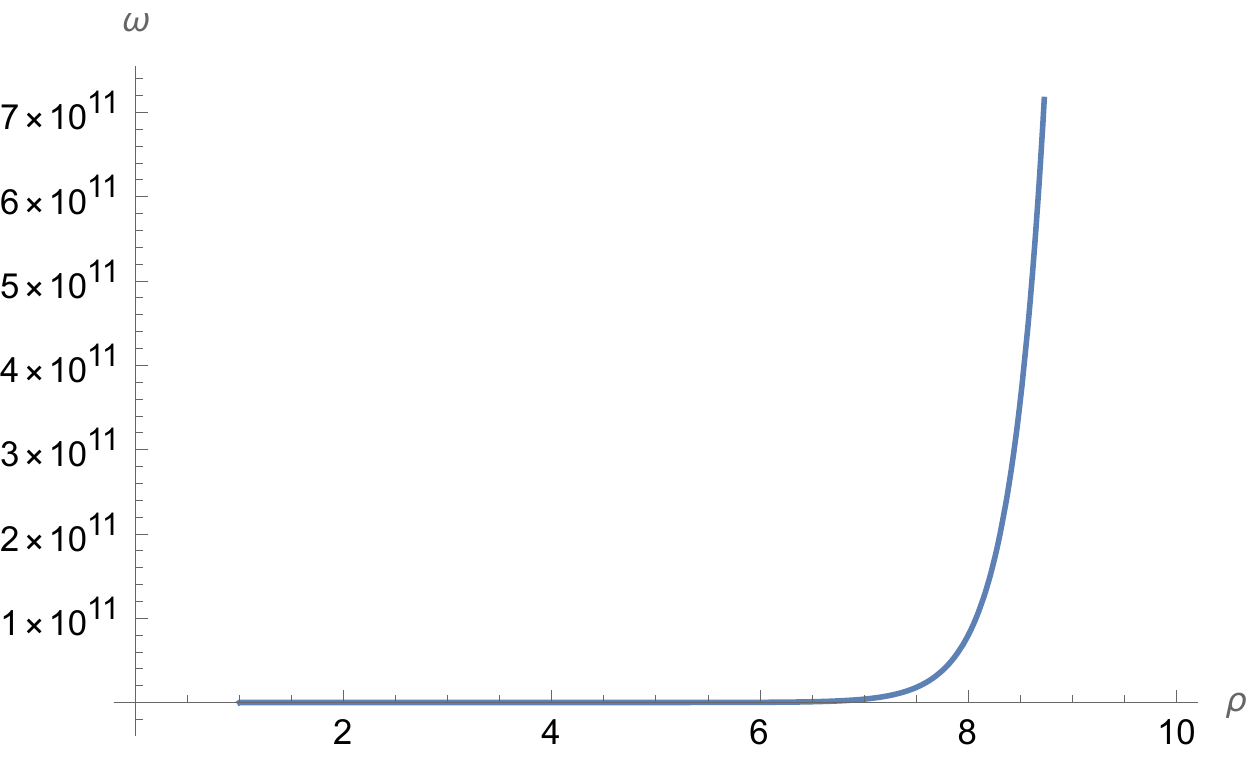}  
  \caption{The Crofron form for D5 black brane geometry for $u_t=2$ and $u_\Lambda=1$. }
 \label{fig:croftonLM}
\end{figure}

Its behavior is shown in figure \ref{fig:croftonLM}.

Again, the behavior of the Crofton form for this new confining metric for large $\omega$ is not what we expect from a good holographic confining geometry and therefore it probably need to be modified in order to satisfy the expected quantum information properties. However, a well or discontinuity can be observed around $\rho_0$ or the end of geometry or horizon as we expect and in this regard, this geometry satisfy this particular requirement.

\subsection{Other measures and applications}

Other correlation measures could also employed to detect these phases in the confining geometries. We specifically expect  measures, such as logarithmic negativity, complexity of purification or reflected entropy could also detect these four phases. For instance, recently in \cite{Dong:2021oad},  the entanglement negativity has been studied in a toy model of evaporating black hole which consists of JT gravity with an end-of-the-world (EOW) brane and similar to our confining models, four phases have been found there which were called, disconnected, cyclically connected, anti-cyclically connected and pairwise connected geometries where the last of these geometries are new replica wormholes that break the replica symmetry spontaneously. The transitions between these geometries have been studied analytically there.

Other methods such as convex optimization theory or bit threads \cite{Evslin:2007ti, Lust:2008zd, Koerber:2006hh,Bakhmatov:2017ihw,Agon:2021tia, Pedraza:2021mkh, Pedraza:2021fgp}, could also help to depict the correlation structures in the confining geometries, and specifically shows the effects of the wall in the bulk on the entanglement pattern.

Other tools and measures, such as balanced partial entanglement and the entanglement wedge cross section \cite{Wen:2021qgx, Han:2021ycp}, entanglement contours \cite{Han:2021ycp, Ageev:2021ipd}, horocycles \cite{Czech:2017zfq}, lambda lengths of Penner \cite{Penner:1987, Boldis:2021snw}, Holevo information \cite{Bao:2021ebo}, etc,  could be employed for analyzing the entanglement pattern in these backgrounds.

As some applications of these study, it worths to mention that in works such as \cite{deBoer:2009wk}, the ``holographic'' neutron stars have been studied where the boundary states were considered as degenerate composite operators which are dual to holographic degenerate stars in the AdS side. Their model is like a set of $n_F$ ``shells'' which are distinguished by the number of derivatives. The quantum correlations between these shells, then, could be probed by mixed entanglement measures and the phase structure of neutron stars could be constructed. As we showed here that the critical distance $D_c$ could distinguish the scale of chiral breaking/restoration and confinement/deconfinement, it could be a new tool to probe the phase structure of neutron stars, in addition to the entanglement entropy.

\section{Conclusion}\label{sec:conclude}

In order to analyze the mixed correlation patterns in the confining backgrounds and holographic top-down QCD models, in this work we used a symmetric setup with two equal size strips with width $l$ and infinite length, where there is a distance $D$ among them. These two subsystems are entangled with each other and with the rest of the Hilbert space, creating a mixed correlated system, and therefore entanglement entropy cannot be a correct measure of their correlations. We propose here that, in addition to other mixed correlation measures, the critical distance $D_c$ between the two strips, where the mutual information $I$ between them suddenly drops to zero, could act as a new tool for probing the deep in the bulk, and it can detect the dual QCD phase structures. Specifically, in our models, by varying the relative sizes of $l$, $D$ and $u_{KK}$, the critical distance could distinguish the scale of confinement/deconfinement phase transition versus the scale of chiral symmetry breaking/restoration.

In addition, for these confining geometries, we calculated the Crofton form, which can further exhibit the interconnections between entanglement and geometries. For these confining models, we showed that Crofton form would display a universal behavior, as it diverges around the infrared point and becomes constant in the large radiuses, which again demonstrates the effects of the IR wall and confinement on the correlation patterns.

In future works, we plan to apply other correlation measures to derive the phase structures of confining models and also solutions with black holes, and by comparing the properties of such phase diagrams, derive results about the universal properties of the pattern of mixed correlations.

\section*{Acknowledgments}
I would like to thank Matti Järvinen for supports and helpful discussions. I would also like to thank MDPI publisher for supports in the form of 2019 Universe and 2021 Galaxies travel awards which were helpful in participating various conferences to discuss this work. I am grateful to Prof. Sang Pyo Kim and the organizers of the 17th Italian-Korean symposium for relativistic astrophysics and workshop on cosmology and quantum space time (CQUeST 2021) for their kind invitation. This work has been supported by an appointment to the JRG Program at the APCTP through the Science and Technology Promotion Fund and Lottery Fund of the Korean Government. It has also been supported by the Korean Local Governments – Gyeongsangbuk do Province and Pohang City – and by the National Research Foundation of Korea (NRF) funded by the Korean government (MSIT) (grant number 2021R1A2C1010834).

 \medskip

\bibliography{jhepmixedQCDphases.bib}
\bibliographystyle{JHEP}
\end{document}